\documentclass{nature}

\usepackage{caption}
\usepackage{graphicx}
\usepackage[scaled]{helvet}
\usepackage{amsmath}
\usepackage{amssymb}
\usepackage{bm}
\newcommand*{\myfont}{\fontfamily{phv}\selectfont}
\usepackage{hyperref}
\usepackage{anyfontsize}

\usepackage{lineno}
\usepackage{color, colortbl}

\definecolor{CLBlue}{rgb}{0, .25, .8}
\definecolor{MyBlue}{rgb}{0, .24, .40}
\definecolor{MyTurquoise}{rgb}{0, .53, .49}
\definecolor{MyGreen}{rgb}{0, .35, 0}
\definecolor{MyOrange}{rgb}{.8, .46, 0}
\definecolor{MyRed}{rgb}{.57, .07, 0}
\definecolor{MyPurple}{rgb}{.46, .1, .46}
\definecolor{LightGrey}{rgb}{0.95,.95,.95}

\title{Human information processing in complex networks}

\author{Christopher W. Lynn$^1$, Lia Papadopoulos$^1$, Ari E. Kahn$^{2,3}$, \& Danielle S. Bassett$^{1,3,4,5,6,7,*}$}

\begin{document}


\maketitle

\begin{affiliations}
\item Department of Physics \& Astronomy, College of Arts \& Sciences, University of Pennsylvania, Philadelphia, PA 19104, USA
\item Department of Neuroscience, Perelman School of Medicine, University of Pennsylvania, Philadelphia, PA 19104, USA
\item Department of Bioengineering, School of Engineering \& Applied Science, University of Pennsylvania, Philadelphia, PA 19104, USA
\item Department of Electrical \& Systems Engineering, School of Engineering \& Applied Science, University of Pennsylvania, Philadelphia, PA 19104, USA
\item Department of Neurology, Perelman School of Medicine, University of Pennsylvania, Philadelphia, PA 19104, USA
\item Department of Psychiatry, Perelman School of Medicine, University of Pennsylvania, Philadelphia, PA 19104, USA
\item Santa Fe Institute, Santa Fe, NM 87501, USA
\end{affiliations}

\newpage

\begin{abstract}

Humans communicate using systems of interconnected stimuli or concepts -- from language and music to literature and science -- yet it remains unclear how, if at all, the structure of these networks supports the communication of information. Although information theory provides tools to quantify the information produced by a system, traditional metrics do not account for the inefficient ways that humans process this information. Here we develop an analytical framework to study the information generated by a system as perceived by a human observer. We demonstrate experimentally that this perceived information depends critically on a system's network topology. Applying our framework to several real networks, we find that they communicate a large amount of information (having high entropy) and do so efficiently (maintaining low divergence from human expectations). Moreover, we show that such efficient communication arises in networks that are simultaneously heterogeneous, with high-degree hubs, and clustered, with tightly-connected modules -- the two defining features of hierarchical organization. Together, these results suggest that many communication networks are constrained by the pressures of information transmission, and that these pressures select for specific structural features.

\end{abstract}

\newpage

Humans receive information in discrete chunks, which transition from one to another -- as words in a sentence or notes in a musical progression -- to create coherent messages. The networks formed by these chunks (nodes) and transitions (edges) encode the structure of allowed messages, fundamentally governing the ways that we communicate with one another. Although attempts to study the information properties of such transition networks date to the foundation of information theory itself,\cite{Shannon-01} with applications to linguistics,\cite{Bar-01,Dretske-01} music theory,\cite{Cohen-02} social and information networks,\cite{Rosvall-02,Gomez-03} the Internet,\cite{Liben-01} and transportation,\cite{Rosvall-01} fundamental questions concerning the impact of network structure on how humans process information remain unanswered.

The primary difficulty in quantifying the information content of a message is accounting for the human perspective: formally, a message's information content is not inherent, but rather depends crucially on the receiver's expectations (or estimated probabilities) of different symbols and stimuli.\cite{Shannon-01, Dretske-01, Cover-01} Whereas for computers the probabilities of different symbols are often prescribed, human expectations are biased\cite{Hilbert-01} and differ from person to person,\cite{Dretske-01} with measurable consequences for behavior\cite{Laming-01} and cognition.\cite{Koechlin-01} However, recent advances in psychology and neuroscience have shed light on how humans learn and internally estimate the structure of complex probabilistic systems.\cite{Saffran-01, Dehaene-01, Schapiro-01, Kahn-01, Lynn-06, Lynn-08} Given this progress, it is now possible and compelling to build a framework to quantify human information processing and to consider what types of networks support efficient communication.

\begin{figure}
\centering
\includegraphics[width = .7\textwidth]{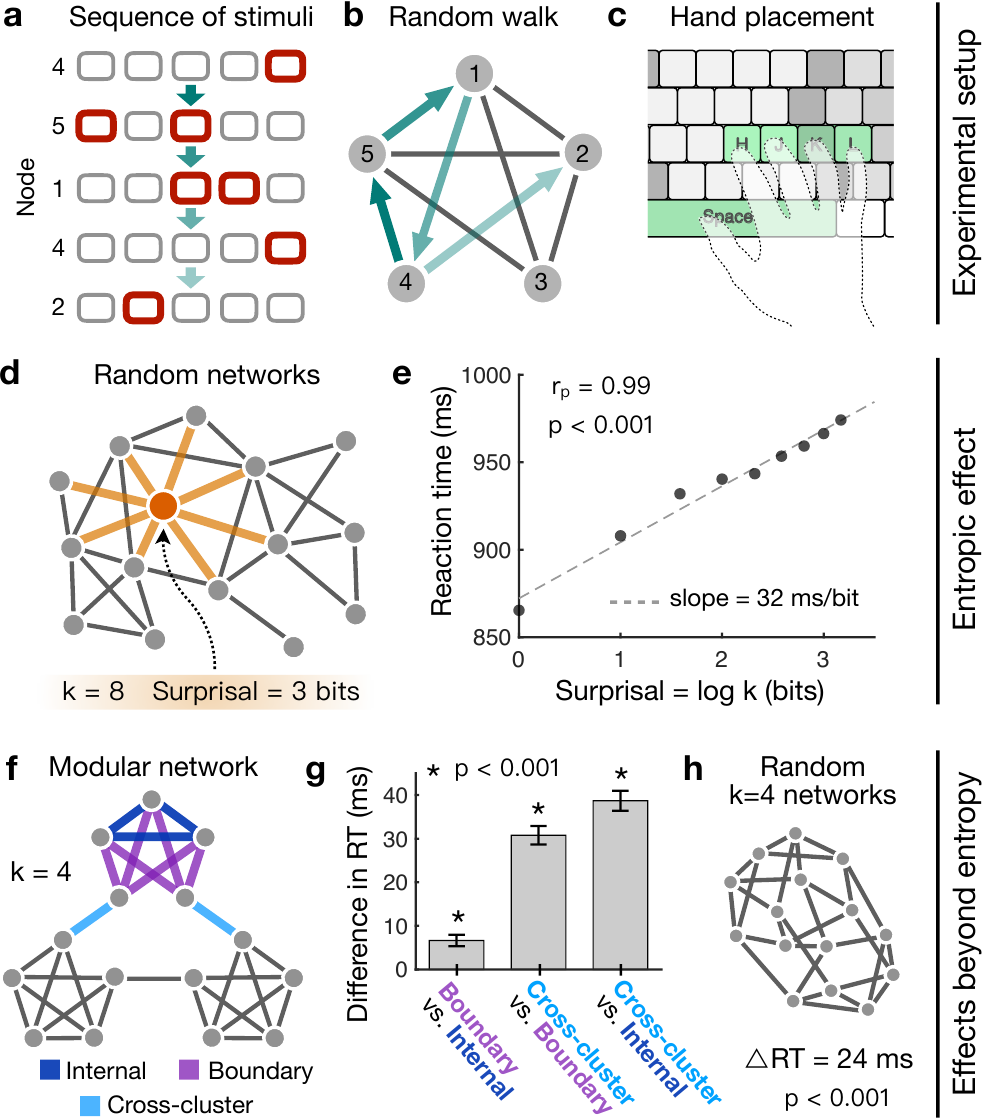} \\
\raggedright
\myfont\textbf{Fig. 1 $|$ Human behavioral experiments reveal the dependence of perceived information on network topology.}
\captionsetup{labelformat=empty}
{\spacing{1.25} \caption{\label{experiment} \myfont \textbf{a}-\textbf{c}, Experimental setup for our serial reaction time tasks. \textbf{a}, Subjects are shown sequences of 1500 stimuli, with each stimulus consisting of five squares with one or two highlighted in red. \textbf{b}, The sequential order of stimuli is determined by a random walk on an underlying network. \textbf{c}, In response to each stimulus, subjects press keys on a keyboard corresponding to the highlighted squares. We use both one- and two-button responses because they allow for networks of size up to $N = 15$. To control for the behavioral effects of the different one- and two-button responses, we (i) randomize the assignment of stimuli to nodes for each subject and (ii) regress out behavioral dependencies on individual stimuli.\cite{Kahn-01} \textbf{d}-\textbf{e}, Effect of produced information on reaction times, referred to as the entropic effect. \textbf{d}, For each subject, we draw an Erd\"{o}s-R\'{e}nyi random network with $N = 15$ nodes and $E = 30$ edges; the information produced by a transition $i\rightarrow j$ (or its surprisal) is $\log k_i$, where $k_i$ is the degree of node $i$. \textbf{e}, $\quad\quad\quad\quad\quad\quad\quad\quad\quad\quad\quad\quad\quad$}}
\end{figure}
\addtocounter{figure}{-1}
\begin{figure}
\centering
\raggedright
\captionsetup{labelformat=empty}
{\spacing{1.25} \caption{\myfont Reaction times, averaged over all transitions that begin at nodes of a given degree $k$, are significantly correlated with the produced information $\log k$ (Pearson correlation coefficient $r_p = 0.99$, $p < 0.001$, $n=177$ subjects). \textbf{f}-\textbf{h}, Effects of network topology on reaction times after controlling for produced information. \textbf{f}, We control for variations in produced information by focusing on networks of constant degree $k = 4$, such as the modular network, which contains three distinct types of transitions: those deep within clusters (dark blue), those at the boundaries of clusters (purple), and those between clusters (light blue). \textbf{g}, Each type of transition produces reaction times that are distinct from the other two; differences in reaction times and $p$-values are estimated using mixed effects models ($n = 173$ subjects; see Supplementary Sec. 5). \textbf{h}, The difference in reaction times $\Delta RT$ between random degree-4 networks and the modular network; the modular network yields consistently faster reactions ($n = 84$ subjects). In addition to the population-level results in panels \textbf{e}, \textbf{g}, and \textbf{h}, we also find significant individual variation in subjects' sensitivity to network topology (see Supplementary Sec. 8).}}
\end{figure}

\section*{Humans perceive information beyond entropy}
\vspace{-30pt}
\noindent\rule{\textwidth}{.5pt}

\vspace{-9pt}
We set out to study the amount of information a human perceives when observing a sequence of stimuli. Naturally, one might naively expect a human to perceive roughly the same amount of information as is being produced by a sequence, or its Shannon entropy.\cite{Shannon-01, Cover-01} Here, to motivate our analytic results, we carry out a set of experiments showing that these two quantities -- the information perceived by a human and the information produced by a sequence -- differ systematically. To experimentally measure perceived information, we employ a paradigm recently developed in statistical learning,\cite{Schapiro-01, Kahn-01, Lynn-06, Lynn-08} presenting subjects with sequences of stimuli on a screen (Fig. \ref{experiment}a) and asking them to respond to each stimulus by pressing the indicated keys on a keyboard (Fig. \ref{experiment}b). Although many real communication systems have long-range correlations, the production of information is traditionally modeled as a Markov process,\cite{Shannon-01, Cover-01} or equivalently, a random walk on a (possibly weighted, directed) network.\cite{Rosvall-02} Therefore, we assign each stimulus to a node in an underlying network, and we stipulate the order of stimuli within a sequence using random walks (Fig. \ref{experiment}b; Methods). By measuring subjects' reaction times and error rates, we can infer how much information they perceive: slow reactions or many errors reflect surprising transitions (with high perceived information), while fast reactions or few errors indicate expected transitions (with low perceived information).\cite{Laming-01, Kahn-01, Lynn-06}

In a random walk, the probability of transitioning from node (or stimulus) $i$ to a neighboring node $j$ is $P_{ij} = 1/k_i$, where $k_i$ is the degree of node $i$. Thus, the amount of information produced by a single transition $i\rightarrow j$ (often referred to as surprisal\cite{Shannon-01}) is given by $-\log P_{ij} = \log k_i$ (Fig. 1d).\cite{Cover-01} Indeed, subjects' behavior is remarkably well-predicted by the information surprisal, with each additional bit of produced information inducing a linear 32 ms increase in reaction times (Fig. \ref{experiment}e) and a 0.3\% increase in the number of errors (see Supplementary Sec. 6). However, even if we present subjects with networks of constant degree -- forcing each transition to produce an identical amount of information -- we still discover consistent variations in behavior that are driven by network topology.

For example, consider the modular network in Fig. \ref{experiment}f, which by symmetry only contains three types of transitions. Each transition produces reaction times and error rates that are distinct from the other two (Fig. \ref{experiment}g), with transitions between or at the boundaries of clusters generating longer reaction times and more errors (see Supplementary Sec. 6) than those deep within a cluster. In addition to differences in behavior at the level of individual transitions, we also find overall variations between different networks. Specifically, when compared to random networks of constant degree (Fig. \ref{experiment}g), the modular network yields significantly faster reactions (and swifter learning rates; see Supplementary Sec. 7), indicating a decrease in the average perceived information. Moreover, similar effects have recently been demonstrated across a range of experimental settings,\cite{Lynn-08} including networks of varying size and topology;\cite{Karuza-03, Kahn-01, Lynn-06} networks with weighted edges;\cite{Saffran-01, Dehaene-01, Meyniel-02} time-varying networks;\cite{Lynn-06, Meyniel-02} different types of stimuli;\cite{Saffran-01, Dehaene-01, Meyniel-02, Tompson-01, Schapiro-01, Karuza-03} and various behavioral and cognitive measures.\cite{Schapiro-01, Dehaene-01, Saffran-01} Together, these results reveal that humans perceive information -- beyond the information produced by a sequence -- in a manner that depends systematically on network topology.

\section*{Quantifying perceived information: Cross entropy}
\vspace{-30pt}
\noindent\rule{\textwidth}{.5pt}

\vspace{-9pt}
The differences between perceived information and produced information can be understood as stemming from the inaccuracy of human expectations. As discussed above, given a transition probability matrix $P$, a transition $i\rightarrow j$ produces $-\log P_{ij}$ bits of information. By contrast, to a person with estimated transition probabilities $\hat{P}$, the same transition will convey $-\log \hat{P}_{ij}$ bits of information.

\begin{figure}[t!]
\centering
\includegraphics[width = .8\textwidth]{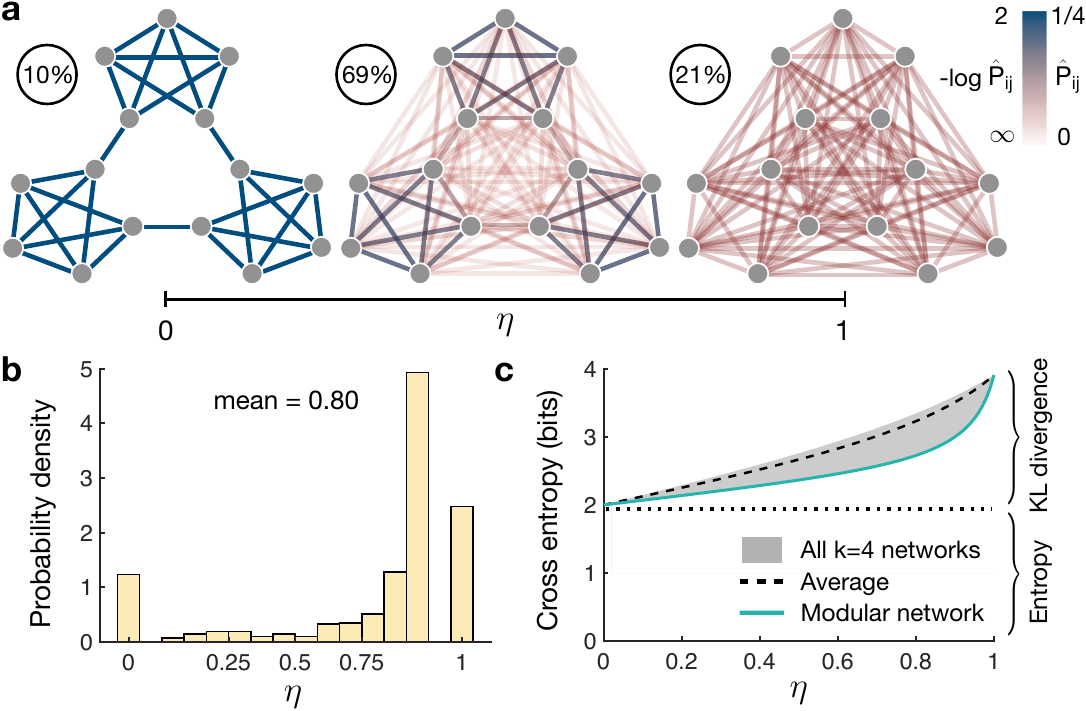} \\
\raggedright
\myfont\textbf{Fig. 2 $|$ Modeling human estimates of transition probabilities.}
\captionsetup{labelformat=empty}
{\spacing{1.25} \caption{\label{model} \myfont \textbf{a}, Illustration of the internal estimates of the transition probabilities $\hat{P}$ in the modular network. For $\eta \rightarrow 0$ (left), the estimates become exact, while for $\eta \rightarrow 1$ (right), the estimates become all-to-all, losing any resemblance to the true network. For intermediate $\eta$ (center), transitions within clusters maintain higher probabilities (and therefore lower surprisal) than transitions between clusters, thereby explaining the differences in reaction times in Fig. \ref{experiment}g. Percentages indicate the proportion of subjects, across all tasks, belonging to each category. \textbf{b}, Distribution of the accuracy parameter $\eta$ estimated from subjects' reaction times (see Supplementary Sec. 4); the distribution is over all 518 completed tasks ($n = 434$ subjects). \textbf{c}, Cross entropy $S(P,\hat{P})$ as a function of $\eta$ for all $k$-4 networks of size $N=15$ (shaded region). The modular network (solid line) maintains a lower cross entropy than the average across all $k$-4 networks (dashed line), thereby explaining the difference in reaction times in Fig. \ref{experiment}h.}}
\end{figure}

Although several models have been proposed for how humans estimate transition probabilities,\cite{Dehaene-01, Meyniel-02, Schapiro-01, Lynn-08} converging evidence indicates that humans integrate transitions over time.\cite{Howard-01, Dayan-01, Gershman-01, Lynn-06, Garvert-01} Such temporal integration yields expectations that include higher powers of the transition matrix: $\hat{P} = C \sum_{t = 0}^{\infty} f(t) P^{t+1}$, where $f(t)\ge 0$ is a decreasing function and $C=(\sum_t f(t))^{-1}$ is a normalization constant (we note that $\hat{P}$ is guaranteed to converge if $\sum_t f(t)$ converges). For example, if $f(t) = 1/t!$, then the transition probability estimates $\hat{P}$ are nearly identical to the network communicability\cite{Estrada-01, Garvert-01} (see Supplementary Sec. 4). Here, we focus on the specific choice $f(t) = \eta^t$, where $\eta\in (0,1)$ represents the inaccuracy of a person's expectations (Fig. \ref{model}a). This model can be derived from a number of different cognitive theories -- including the temporal context model of episodic memory,\cite{Howard-01} temporal difference learning and the successor representation in reinforcement learning,\cite{Dayan-01, Gershman-01} and the free energy principle from information theory.\cite{Lynn-06} Inferring $\eta$ from each subject's reaction times (Fig. \ref{model}b; see Methods), we find that 10\% of subjects hold exact estimates of the transition structure ($\eta \rightarrow 0$; Fig. \ref{model}a, left), while 21\% have expectations that are completely disordered ($\eta \rightarrow 1$; Fig. \ref{model}a, right). Importantly, most subjects have expectations that lie between these two extremes (Fig. \ref{model}a, center), yielding a decrease in the expected probability of between- versus within-cluster transitions in the modular network. This decrease in expected probability, in turn, gives rise to an increase in perceived information, thereby explaining the observed variations in subjects' reaction times and error rates for different parts of the modular network (Fig. \ref{experiment}g).

We are now prepared to study the average perceived information of an entire communication network. Averaging the perceived information of individual transitions over the random walk process, we have $\langle-\log \hat{P}_{ij}\rangle_P = -\sum_{ij} \pi_i P_{ij} \log \hat{P}_{ij}$, where $\bm{\pi}$ is the stationary distribution of $P$. Interestingly, this quantity -- known as the \textit{cross entropy} $S(P,\hat{P})$ between $P$ and $\hat{P}$ -- splits naturally into the entropy $S(P)$, or the average produced information, and the KL divergence $D_{\text{KL}}(P||\hat{P})$, or the inefficiency of the observer's expectations:
\begin{equation}
\underbrace{\vphantom{\frac{\hat{P}_{ij}}{P_{ij}}} \langle-\log \hat{P}_{ij}\rangle_P}_{\text{\normalsize $S(P,\hat{P})$}} \,=\, \underbrace{\vphantom{\frac{\hat{P}_{ij}}{P_{ij}}} \langle-\log P_{ij}\rangle_P}_{\text{\normalsize \vphantom{$\hat{P}$} $S(P)$}} \,+\, \underbrace{\langle-\log \frac{\hat{P}_{ij}}{P_{ij}}\rangle_P}_{\text{\normalsize $D_{\text{KL}}(P||\hat{P})$}}.
\end{equation}
This relationship has a number of immediate consequences, including the fact that the information a human perceives $S(P,\hat{P})$ is lower-bounded by the information that a system produces $S(P)$ (since $D_{\text{KL}}(P||\hat{P}) \ge 0$). Moreover, inefficiency is minimized when a person's expectations are exact (since $D_{\text{KL}}(P||\hat{P}) = 0$ only when $\hat{P} = P$).\cite{Cover-01} For example, consider the set of degree-4 networks from our human experiments (Fig. \ref{experiment}h). While all such networks have identical entropy, their differing topologies induce a range of cross entropies, which vary as a function of $\eta$ (Fig. \ref{model}c). Notably, the modular graph displays lower cross entropy than most other degree-4 networks (Fig. \ref{model}c), thus explaining the observed difference in subjects' behaviors (Fig. \ref{experiment}h).

\addtocounter{figure}{-1}
\begin{figure}
\centering
\myfont\textbf{Table 1 $|$ Properties of the real communication networks examined in this paper.} \\[.5em]
{\fontsize{9}{7.5}\selectfont
\begin{tabular}{l l l c c c c}
\hline
\\[-.9em]
{\footnotesize \textbf{Type}} / Name & $N$ & $E$ & $S^{\text{real}}$ (bits) & $S^{\text{rand}}$ (bits) & $D_{\text{KL}}^{\text{real}}$ (bits) & $D_{\text{KL}}^{\text{rand}}$ (bits) \\[.1em]
\hline
\hline
\\[-.9em]
{\footnotesize \textcolor{MyBlue}{\textbf{Language (noun transitions)}}} \\
Shakespeare & 11,234 & 97,892 & 6.15 & 4.16 & 1.74 & 2.17 \\
Homer & 3,556 & 23,608 & 5.25 & 3.79 & 1.75 & 2.12 \\
Plato & 2,271 & 9,796 & 4.41 & 3.19 & 1.74 & 2.04 \\
Jane Austen & 1,994 & 12,120 & 4.92 & 3.66 & 1.71 & 2.10 \\
William Blake & 370 & 781 & 2.59 & 2.24 & 1.64 & 1.77 \\
Miguel de Cervantes & 6,090 & 43,682 & 5.55 & 3.89 & 1.76 & 2.14 \\
Walt Whitman & 4,791 & 16,526 & 4.24 & 2.89 & 1.76 & 2.00 \\
{\footnotesize \textcolor{MyTurquoise}{\textbf{Semantic relationships}}} \\
Bible & 1,707 & 9,059 & 4.31 & 3.48 & 1.45 & 2.07 \\
Les Miserables & 77 & 254 & 3.25 & 2.82 & 0.84 & 1.65 \\
Edinburgh Thesaurus & 7,754 & 226,518 & 6.26 & 5.88 & 2.07 & 2.21 \\
Roget Thesaurus & 904 & 3,447 & 3.19 & 3.02 & 1.76 & 1.99 \\
Glossary terms & 60 & 114 & 2.32 & 2.09 & 1.29 & 1.55 \\
FOLDOC & 13,274 & 90,736 & 4.11 & 3.83 & 1.72 & 2.14 \\
ODLIS & 1,802 & 12,378 & 4.59 & 3.83 & 1.70 & 2.11 \\
{\footnotesize \textcolor{MyGreen}{\textbf{World Wide Web}}} \\
Google internal & 12,354 & 142,296 & 6.15 & 4.56 & 1.35 & 2.19 \\
Education & 2,622 & 6,065 & 3.01 & 2.36 & 0.92 & 1.85 \\
EPA & 2,232 & 6,876 & 3.34 & 2.74 & 1.75 & 1.95 \\
Indochina & 9,638 & 45,886 & 3.88 & 3.33 & 0.58 & 2.08 \\
2004 Election blogs & 793 & 13,484 & 5.78 & 5.11 & 1.36 & 2.01 \\
Spam & 3,796 & 36,404 & 5.30 & 4.30 & 1.66 & 2.16 \\
WebBase & 6,843 & 16,374 & 3.48 & 2.41 & 1.09 & 1.87 \\
{\footnotesize \textcolor{MyOrange}{\textbf{Citations}}} \\
arXiv Hep-Ph & 12,711 & 139,500 & 5.02 & 4.49 & 1.68 & 2.19 \\
arXiv Hep-Th & 7,464 & 115,932 & 5.56 & 4.98 & 1.64 & 2.20 \\
Cora & 3,991 & 16,621 & 3.50 & 3.14 & 1.48 & 2.04 \\
DBLP & 240 & 858 & 3.30 & 2.93 & 1.37 & 1.88 \\
{\footnotesize \textcolor{MyRed}{\textbf{Social relationships}}} \\
Facebook & 13,130 & 75,562 & 4.22 & 3.59 & 1.78 & 2.11 \\
arXiv Astr-Ph & 17,903 & 196,972 & 5.39 & 4.49 & 1.41 & 2.19 \\
Adolescent health & 2,155 & 8,970 & 3.22 & 3.14 & 1.78 & 2.03 \\
Highschool & 67 & 267 & 3.11 & 3.07 & 1.15 & 1.57 \\
Jazz & 198 & 2,742 & 5.09 & 4.81 & 0.94 & 1.61 \\
Karate club & 34 & 78 & 2.58 & 2.32 & 1.05 & 1.40 \\
{\footnotesize \textcolor{MyPurple}{\textbf{Music (note transitions)}}} \\
Thriller -- Michael Jackson & 67 & 446 & 4.03 & 3.78 & 0.76 & 1.38 \\
Hard Day's Night -- Beatles & 41 & 212 & 3.62 & 3.42 & 0.49 & 1.21 \\
Bohemian Rhapsody -- Queen & 71 & 961 & 5.01 & 4.77 & 0.55 & 0.95 \\
Africa -- Toto & 39 & 163 & 3.41 & 3.13 & 0.84 & 1.29 \\
Sonata No 11 -- Mozart & 55 & 354 & 3.91 & 3.73 & 0.83 & 1.28 \\
Sonata No 23 -- Beethoven & 69 & 900 & 4.86 & 4.72 & 0.65 & 0.96 \\
Nocturne Op 9-2 -- Chopin & 59 & 303 & 3.62 & 3.42 & 0.95 & 1.43 \\
Clavier Fugue 13 -- Bach & 40 & 143 & 3.06 & 2.92 & 0.89 & 1.37 \\
Ballade Op 10-1 -- Brahms & 69 & 670 & 4.42 & 4.31 & 0.80 & 1.18 \\
\hline
\end{tabular}} \\[-.5 em]
\captionsetup{labelformat=empty}
{\spacing{1} \caption{\myfont{\footnotesize For each network we list its type and name, number of nodes $N$ and edges $E$, entropy of the real network $S^{\text{real}}$ and after randomizing the edges $S^{\text{rand}}$, and KL divergence of the real network $D_{\text{KL}}^{\text{real}}$ and after randomization $D_{\text{KL}}^{\text{rand}}$ with $\eta$ set to the average value $0.80$ from our experiments. $S^{\text{rand}}$ and $D_{\text{KL}}^{\text{rand}}$ are averaged over 100 randomizations. For descriptions of and references for these networks, see Supplementary Sec. 15.}}}
\end{figure}

\section*{Information properties of real communication networks}
\vspace{-30pt}
\noindent\rule{\textwidth}{.5pt}

\begin{figure}[t!]
\centering
\includegraphics[width = .95\textwidth]{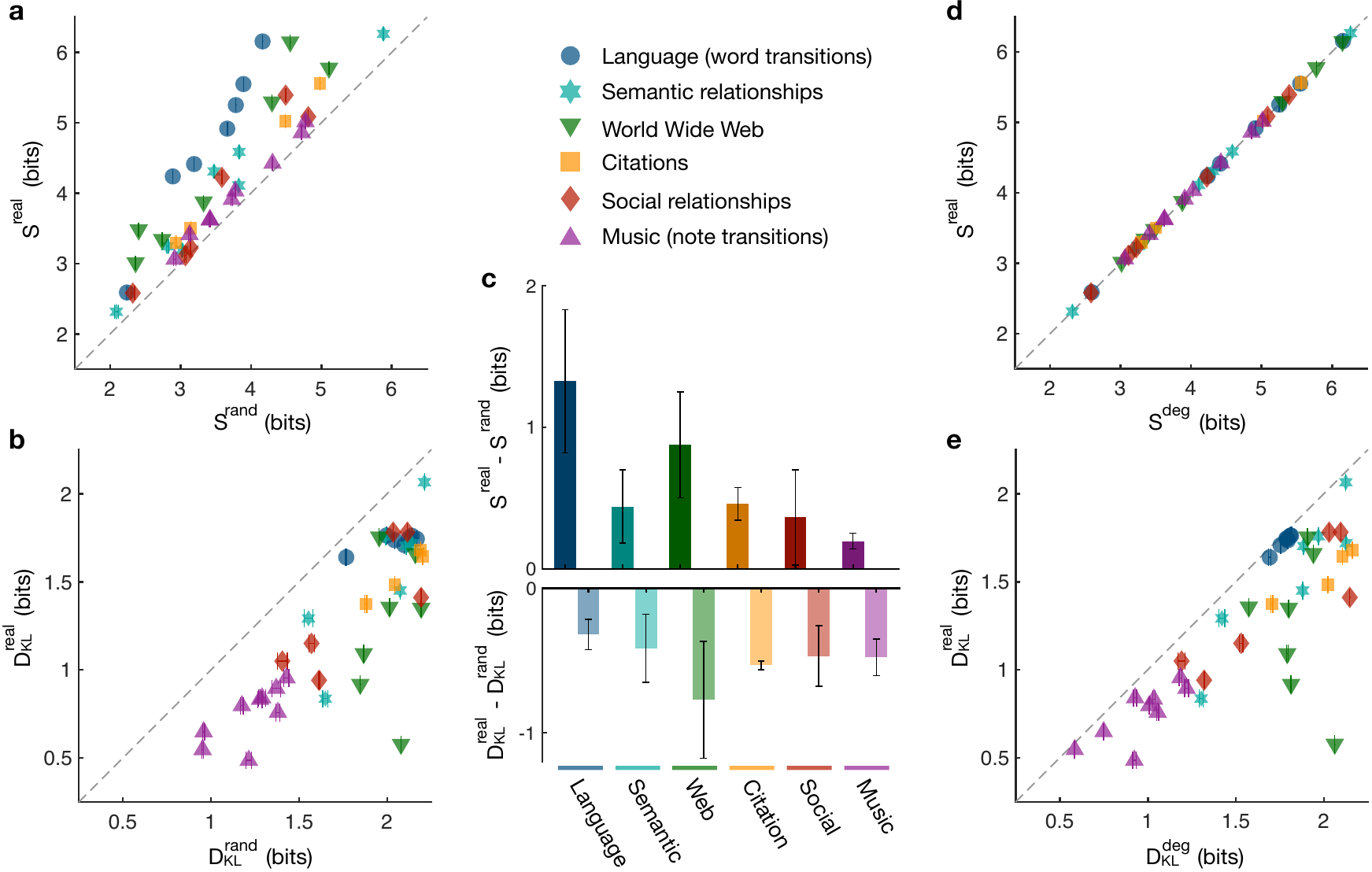} \\
\raggedright
\myfont\textbf{Fig. 3 $|$ The entropy and KL divergence of real communication networks.}
\captionsetup{labelformat=empty}
{\spacing{1.25} \caption{\label{real_nets} \myfont \textbf{a}, Entropy of fully randomized versions of the networks listed in Table 1 ($S^{\text{rand}}$) compared with the true values ($S^{\text{real}}$). \textbf{b}, KL divergence of fully randomized versions of the real networks ($D_{\text{KL}}^{\text{rand}}$) compared with the true values ($D_{\text{KL}}^{\text{real}}$). Human expectations $\hat{P}$ are calculated with $\eta$ set to the average value $0.80$ from our experiments; however, the results remain qualitatively the same across all values of $\eta$ (Supplementary Sec. 9). \textbf{c}, Difference between $S^{\text{real}}$ and $S^{\text{rand}}$ (top) and difference between $D_{\text{KL}}^{\text{real}}$ and $D_{\text{KL}}^{\text{rand}}$ (bottom) for different network types, with error bars indicating standard deviation over networks of each type. \textbf{d}, Entropy of degree-preserving randomized networks ($S^{\text{deg}}$) compared with $S^{\text{real}}$. \textbf{e}, KL divergence of degree-preserving randomized networks ($D_{\text{KL}}^{\text{deg}}$) compared with $D_{\text{KL}}^{\text{real}}$ with fixed $\eta = 0.80$. In panels \textbf{a}, \textbf{b}, \textbf{d}, and \textbf{e}, data points and error bars (standard deviations) are estimated from 100 realizations of the randomized networks. All networks are undirected; for examination of directed versions see Supplementary Sec. 9.}}
\end{figure}

\vspace{-9pt}
Using the framework developed above, we are ultimately interested in characterizing the perceived information generated by real communication systems. The networks chosen (Table 1) have all either evolved or been designed to communicate information through sequences of stimuli (such as words or musical notes) or concepts (such as scientific papers, websites, or social interactions). Strikingly, we find that the networks share two consistent properties: they produce large amounts of information (high entropy; Fig. \ref{real_nets}a), while at the same time maintaining low inefficiency (low KL divergence; Fig. \ref{real_nets}b). Specifically, these properties hold relative to completely randomized versions of the networks (Table 1) , with $\eta$ set to the average value $0.8$ from our human experiments (Fig. \ref{model}b). Interestingly, different network types exhibit these information properties to varying degrees (Fig. \ref{real_nets}c). For example, language networks have the highest entropy but also the highest KL divergence, perhaps reflecting the pressure on language to maximize information rate. Meanwhile, music networks are low in both entropy and KL divergence, possibly mirroring their role as a means for entertainment rather than rapid communication.

If we instead compare the communication networks against randomized versions that preserve node degrees,\cite{Maslov-01} we find that the entropy is unchanged (Fig. \ref{real_nets}d), indicating that produced information depends only on the degree distribution. By contrast, even compared to these entropy-preserving networks, the KL divergence of real networks remains low (Fig. \ref{real_nets}e). We verify that these results largely hold for (i) all values of $\eta$, (ii) different models of human expectations $\hat{P}$, and (iii) directed versions of the above networks (Supplementary Sec. 9). Moreover, we find that the information properties of communication networks can vary dramatically in time,\cite{Derex-01, Momennejad-02} with most networks dynamically evolving (for example, over the course of a musical piece or the growth of a social network) to optimize efficient communication -- that is, to maximize entropy and minimize divergence from human expectations (Supplementary Sec. 10).

Finally, to demonstrate that efficient communication is not required by all real communication networks, it is important to consider examples where the results in Fig. \ref{real_nets} break down. We give two such examples in Supplementary Sec. 11, showing that (i) directed citation networks have markedly low entropy and (ii) transitions between words of all parts of speech have relatively high KL divergence. However, if we allow transitions to move both forward and backward along citations (as is typical when traversing scientific literature), then citation networks regain their high entropy (Fig. \ref{real_nets}a). Similarly, if we focus on ``content" words that carry meaning (such as the nouns in Fig. \ref{real_nets}) rather than ``grammatical" words (such as articles, prepositions, and conjunctions) -- a common distinction in the study of language networks\cite{Milo-01, Foster-01} -- then word transitions regain their low KL divergence from human expectations (Figs. \ref{real_nets}b,e). Thus, even for networks that appear to have high entropy or low KL divergence,  studying the context-specific ways that they transmit information to humans often reveals that efficient communication is maintained.

\section*{Efficient communication is driven by hierarchically modular structure}
\vspace{-30pt}
\noindent\rule{\textwidth}{.5pt}

\begin{figure}
\centering
\includegraphics[width = .7\textwidth]{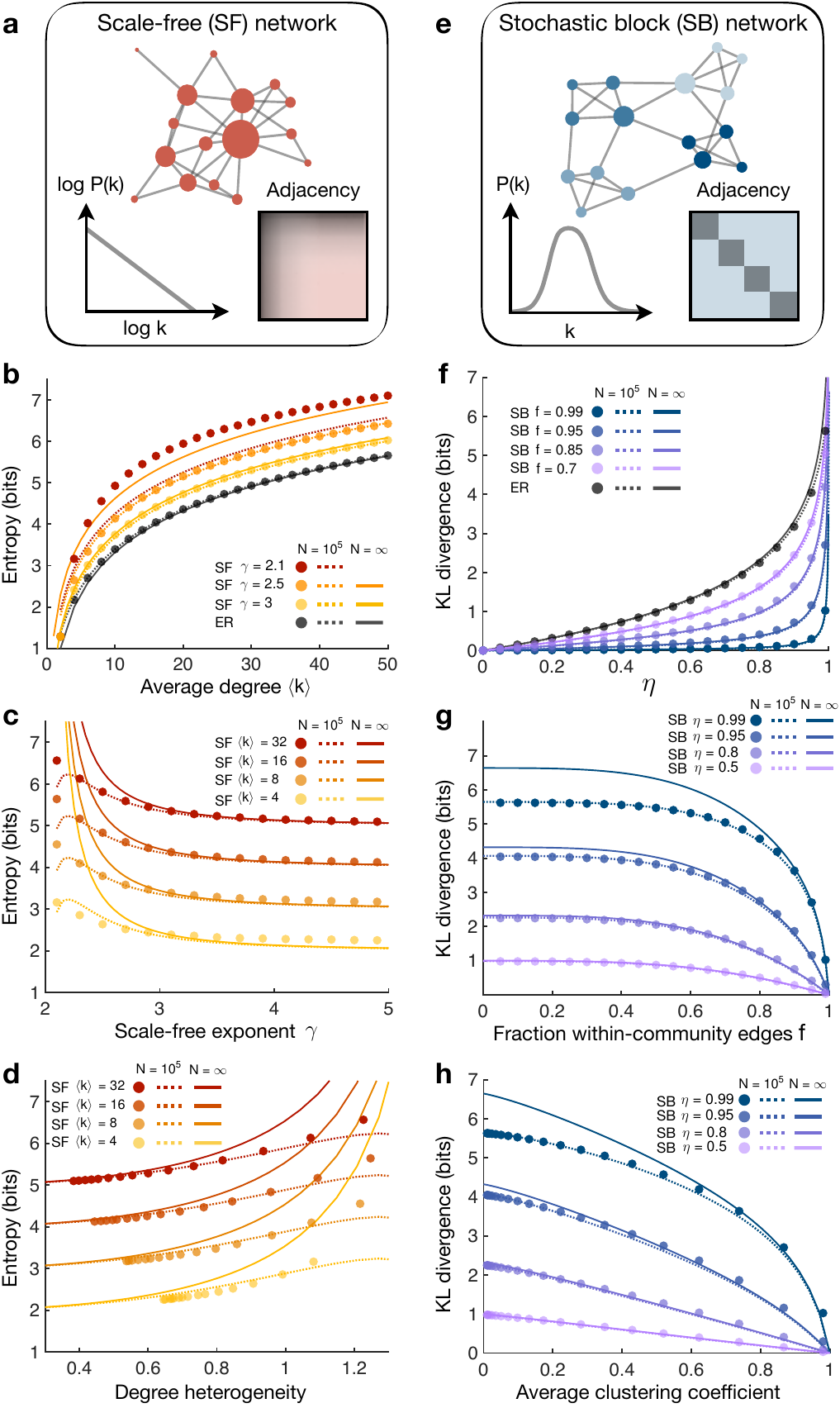} \\
\raggedright
\myfont\textbf{Fig. 4 $|$ The impact of network topology on entropy and KL divergence.}
\captionsetup{labelformat=empty}
{\spacing{1.25} \caption{\label{structure} \myfont \textbf{a}, Scale-free (SF) network, characterized by a power-law degree distribution and the presence of high-degree hub nodes. \textbf{b}, Entropy as a function of the average degree $\langle k\rangle$ for Erd\"{o}s $\,\,\,\,\,\,\,\,\,\,\,\,\,\,\,\,\,\,\,\,\,\,\,\,\,\,\,\,\,\,\,\,\,$}}
\end{figure}

\vspace{-9pt}
Given the high entropy and low KL divergence displayed by real networks, it is natural to wonder what structural features give rise to these properties. To begin, for undirected networks one can show that $S = \sum_i k_i\log k_i$, demonstrating that the entropy of a network is determined by its degree sequence (Fig. \ref{real_nets}d).\cite{Burda-01} It is clear that the entropy grows with increasing node degrees, supporting the intuition that denser networks yield more complex random walks. Moreover, since $S$ is convex in $\bm{k}$, the entropy is larger for networks with a small number of high-degree nodes and many low-degree nodes. Interestingly, such heterogeneous structure is observed in human language,\cite{Cancho-01} the Internet,\cite{Barabasi-01} social networks,\cite{Newman-02} and scale-free networks\cite{Barabasi-01} (although not all networks with heterogeneous degrees are scale-free\cite{Stumpf-01}). To investigate the relationship between a network's entropy and its degree distribution, we derive a number of analytic results in the thermodynamic limit $N\rightarrow \infty$ (Supplementary Sec. 12). For example, the entropy of an Erd\"{o}s-R\'{e}nyi network is given by $S \approx \log \langle k\rangle$ for large average degree $\langle k\rangle$. For scale-free networks with degree exponent $\gamma$ (Fig. \ref{structure}a), we find that $S = \log \langle k\rangle + \frac{1}{\gamma - 2} - \log\frac{\gamma - 1}{\gamma - 2}$, indicating that $\gamma = 2$ is a critical exponent since the entropy diverges as $\gamma \rightarrow 2$. Generating ensembles of Erd\"{o}s-R\'{e}nyi and scale-free networks, we numerically verify the logarithmic dependence of $S$ on $\langle k\rangle$ (Fig. \ref{structure}b). Moreover, we find that $S$ increases for decreasing $\gamma$ (Fig. \ref{structure}c), suggesting that the entropy grows with increasing degree heterogeneity, which we also confirm numerically (Fig. \ref{structure}d). This final result reveals that, after controlling for edge density, the entropy is largest for networks with heavy-tailed degree distributions.

\addtocounter{figure}{-1}
\begin{figure}[t!]
\centering
\raggedright
\captionsetup{labelformat=empty}
{\spacing{1.25} \caption{\myfont -R\'{e}nyi (ER) and SF networks with different scale-free exponents $\gamma$. Data points are exact calculations for ER and SF networks generated using the static model\cite{Goh-01} with size $N = 10^4$. Lines are derived from the expected degree distributions: dashed lines are numerical results for $N=10^4$ and solid lines are analytic results for $N\rightarrow\infty$ (see Supplementary Sec. 12 for derivations). Note that the thermodynamic limit for $\gamma = 2.1$ does not appear in the displayed range. \textbf{c}, Entropy as a function of $\gamma$ for SF networks with fixed $\langle k\rangle$. In the thermodynamic limit (solid lines), the entropy diverges as $\gamma\rightarrow 2$, and the analytic results are nearly exact for $\gamma > 3$. \textbf{d}, Entropy as a function of degree heterogeneity $H = \langle|k_i - k_j|\rangle/\langle k\rangle$, where $\langle|k_i - k_j|\rangle$ is the absolute difference in degrees averaged over all pairs of nodes,\cite{Liu-01} for SF networks with fixed $\langle k\rangle$ and variable $\gamma$. \textbf{e}, Stochastic block (SB) network, characterized by dense connectivity within communities and sparse connectivity between communities. \textbf{f}, KL divergence as a function of the accuracy parameter $\eta$ for ER and SB networks with communities of size $N_c=100$ and different fractions $f$ of within-community edges. Data points are exact calculations for networks with $N=10^4$ and $\langle k\rangle = 100$, and lines are analytic calculations for $N=10^4$ (dashed) and $N\rightarrow\infty$ (solid; see Supplementary Sec. 13 for derivations). \textbf{g}, KL divergence as a function of $f$ for SB networks with fixed $\eta$. The analytic results are nearly exact for $\eta < 0.8$. \textbf{h}, KL divergence as a function of the average clustering coefficient for SB networks with fixed $\eta$ and variable $f$.}}
\end{figure}

In contrast to the entropy, the KL divergence depends on the expectations of an observer. As these expectations become more accurate (that is, as $\eta$ decreases), we expect $D_{\text{KL}}(P||\hat{P})$ to decrease (as in Fig. \ref{model}c). But how does the KL divergence depend on network structure? For an undirected network with adjacency matrix $G$, we can expand in the limit of small $\eta$ to find that $D_{\text{KL}} \approx -\log(1-\eta) - \frac{\eta}{E\ln 2}\sum_i \frac{1}{k_i}\bigtriangleup_i$, where $\bigtriangleup_i = (G^3)_{ii}/2$ is the number of (possibly weighted) triangles involving node $i$ (Supplementary Sec. 13). Therefore, we see that $D_{\text{KL}}$ is smaller for networks with a large number of triangles, explaining, for instance, the low KL divergence of the modular network (Figs. \ref{experiment}h and \ref{model}c). Indeed, an abundance of triangles is typically associated with modular structure, a ubiquitous feature of real communication networks, from social and scientific interactions\cite{Girvan-01,Rosvall-02} to language\cite{Motter-01} and the Internet.\cite{Eriksen-01} To investigate the impact of modularity on the KL divergence, we derive analytic expressions for $D_{\text{KL}}$ that hold for all values of $\eta$ in the thermodynamic limit (Supplementary Sec. 13). The KL divergence of an Erd\"{o}s-R\'{e}nyi network is given by $D_{\text{KL}} = -\log (1-\eta)$. For stochastic block networks with communities of size $N_c$ and a fraction of within-community edges $f$ (Fig. \ref{structure}e), we find that $D_{\text{KL}} = -\log\left[1 - \eta\left(1 - \frac{\langle k\rangle}{N_c}\frac{(1-\eta)f^3}{1-\eta f}\right)\right]$. Generating sets of Erd\"{o}s-R\'{e}nyi and stochastic block networks, we confirm the analytic predictions that $D_{\text{KL}}$ grows with increasing $\eta$ (Fig. \ref{structure}f) and decreases for increasing modularity (Fig. \ref{structure}g) and clustering (Fig. \ref{structure}h). Therefore, even after controlling for the inaccuracy $\eta$ of human expectations, we find that modular organization serves to decrease the inefficiency of information transmission.

\begin{figure}
\centering
\includegraphics[width = .5\textwidth]{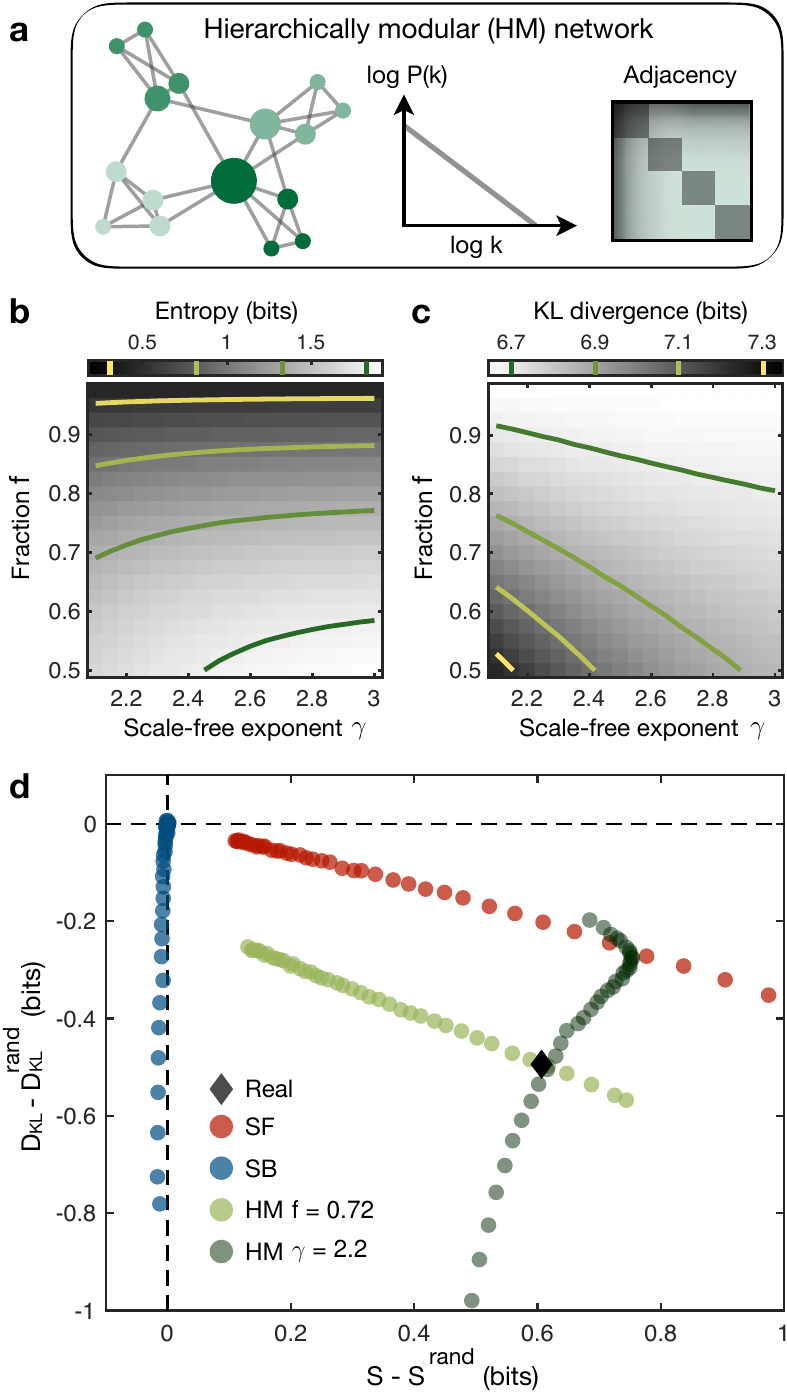} \\
\raggedright
\myfont\textbf{Fig. 5 $|$ Hierarchically modular networks support the efficient communication of information.}
\captionsetup{labelformat=empty}
{\spacing{1.25} \caption{\label{hierarchical} \myfont \textbf{a}, Hierarchically modular (HM) network, characterized by a power-law degree distribution and modular structure (Supplementary Sec. 14). \textbf{b}, Entropy as a function of the scale-free exponent $\gamma$ and the fraction of within-community edges $f$ for HM networks with size $N=10^4$, average degree $\langle k\rangle = 100$, and community size $N_c = 100$. Solid lines denote networks of equal entropy. \textbf{c}, KL divergence as a function of $\gamma$ and $f$ for HM networks with the same size and density as panel \textbf{b} and $\eta$ set to the average value $0.80$ from our experiments (Fig. \ref{model}b). Solid lines denote networks of equal KL divergence. \textbf{d}, Average entropies and KL divergences of real and model networks compared to fully randomized versions. Data points are averages over the set of networks in Table 1, where for each real network we generate SF networks with $\quad\quad\quad\quad\quad\quad\quad\quad\quad\quad\quad\quad$}}
\end{figure}
\addtocounter{figure}{-1}
\begin{figure}[t!]
\centering
\raggedright
\captionsetup{labelformat=empty}
{\spacing{1.25} \caption{\myfont variable $\gamma$ (red), SB networks with communities of size $n \approx \sqrt{N}$ and variable $f$ (blue), and HM networks with $n \approx \sqrt{N}$ and variable $\gamma$ (fixed $f = 0.72$; light green) or variable $f$ (fixed $\gamma = 2.2$; dark green), all with $N$ and $E$ equal to the real network. HM networks with $\gamma = 2.2$ and $f = 0.72$ yield the same average entropy and KL divergence as real communication networks.}}
\end{figure}

To attain both the high entropy and low KL divergence observed in real communication systems, it appears that networks must be simultaneously heterogeneous and modular, the two defining features of hierarchical organization.\cite{Ravasz-01} In order to test this hypothesis, we employ a model that combines the heterogeneous degrees of scale-free networks with the modular structure of stochastic block networks (Fig. \ref{hierarchical}a; see Supplementary Sec. 14 for an extended description). By adjusting $\gamma$ and $f$, we show that these hierarchically modular networks display both a range of entropies (Fig. \ref{hierarchical}b) and KL divergences (Fig. \ref{hierarchical}c). In fact, while scale-free networks do not exhibit the low KL divergence of real communication networks nor do stochastic block networks display their high entropy, we find that hierarchically modular networks can attain both properties (Fig. \ref{hierarchical}d). Taken together, these results indicate that heterogeneity and modularity -- precisely the features commonly observed in real communication systems\cite{Cancho-01, Newman-02, Barabasi-01, Girvan-01, Rosvall-02, Motter-01, Eriksen-01, Ravasz-01} -- are both required to achieve high information production and low inefficiency.

\section*{Conclusions and outlook}
\vspace{-30pt}
\noindent\rule{\textwidth}{.5pt}

\vspace{-9pt}
In this study, we develop tools to quantify the information humans receive from complex networks. We demonstrate experimentally that humans perceive information, beyond the information produced by a sequence, in a way that depends critically on network topology. Moreover, we find that real communication networks support the rapid and efficient transmission of information, and that this efficient communication arises from hierarchical organization. These results raise a number of questions concerning the relationship between human cognition and the structure of communication systems. For example, how have communication networks evolved over time -- or perhaps even co-evolved with the brain\cite{Deacon-01} -- to facilitate information transmission? Furthermore, how can we design communication systems, from human-technology interfaces\cite{Dix-01} to classroom lectures,\cite{Hayes-01} to optimize efficient communication? The framework presented here provides the mathematical tools to begin answering these questions.

To conclude, we highlight a number of ways that our work can be systematically generalized to analyze more realistic communication systems. First, while we model the production of information as a Markov process (equivalently, a random walk), future work should incorporate the long-range dependencies present in many real communication systems.\cite{Brown-02, Pachet-01} The primary difficulty, however, lies in understanding how humans estimate non-Markov transition structures, with most existing work in statistical learning and artificial grammars focusing on Markov processes.\cite{Saffran-01, Dehaene-01, Schapiro-01, Kahn-01, Lynn-06, Lynn-08, Meyniel-01, Karuza-03, Tompson-01, Girvan-01} Second, while we have used tools from information theory to quantify the perceived information of a network,\cite{Shannon-01, Cover-01} these methods do not incorporate the semantic information carried by individual nodes (e.g., words, notes, concepts).\cite{Bar-01, Dretske-01} Thus, in order to improve our understanding of real-world communication systems, future progress will require important interdisciplinary efforts from both cognitive scientists (to study how humans estimate non-Markov structures) and information theorists (to quantify semantic information in human contexts).

\newpage

\begin{methods}

\noindent \textbf{Experimental setup.} Subjects performed a self-paced serial reaction time task using a computer screen and keyboard. Each stimulus was presented as a horizontal row of five grey squares; all five squares were shown at all times. The squares corresponded spatially with the keys `Space', `H', `J', `K', and `L' (Fig. \ref{experiment}c). To indicate a target key or pair of keys for the subject to press, the corresponding squares would become outlined in red (Fig. \ref{experiment}a). When subjects pressed the correct key combination, the squares on the screen would immediately display the next stimulus. If an incorrect key or pair of keys was pressed, the message `Error!' was displayed on the screen below the stimulus and remained until the subject pressed the correct key(s). The order in which stimuli were presented to each subject was determined by a random walk on a network of $N=15$ nodes. For each subject, one of the 15 key combinations was randomly assigned to each node in the network.

In the first experiment, each subject was assigned an Erd\"{o}s-R\'{e}nyi network with $E=30$ edges. In the second experiment, all subjects responded to sequences of stimuli drawn from the modular network (Fig. \ref{experiment}f), which has the same number of nodes and edges. We remark that each node in the modular network is connected to four other nodes, so the entropy of each transition was a constant $-\log \frac{1}{4} = 2$ bits. Some subjects performed both of the first two experiments in back to back stages, with the order of the experiments counterbalanced across subjects. In the third experiment, subjects underwent two stages. In one stage subjects responded to stimuli drawn from the modular network, while in the other stage each subject was assigned a random $k$-4 network. The order of the two stages was counterbalanced. For each stage of each experiment, subjects responded to sequences of 1500 stimuli.

\noindent \textbf{Experimental procedures.} All participants provided informed consent in writing and experimental methods were approved by the Institutional Review Board of the University of Pennsylvania. In total, we recruited 363 unique participants to complete our studies on Amazon's Mechanical Turk: 106 completed just the first experiment, 102 completed just the second experiment, 71 completed both the first and second experiments in back-to-back stages, and 84 completed the third experiment. Worker IDs were used to exclude duplicate participants between experiments, and all participants were financially remunerated for their time. In the first two experiments, subjects were paid \$3-\$11 for up to an estimated 30-60 minutes: \$3 per network for up to two networks, \$2 per network for correctly responding on at least 90\% of the trials, and \$1 for completing two stages. In the third experiment, subjects were paid up to \$9 for an estimated 60 minutes: \$5 for completing the experiment and \$2 for correctly responding on at least 90\% of the trials on each stage.

\noindent \textbf{Data analysis.} To make inferences about subjects' internal expectations based upon their reaction times, we excluded all trials in which subjects responded incorrectly. We also excluded reaction times that were implausible, either three standard deviations from a subject's mean reaction time, below 100 ms, or over 3500 ms.

\noindent \textbf{Measuring the effects of topology on reaction times.} In order to estimate the effects of network topology on subjects' reaction times, one must overcome large inter-subject variability. To do so, we used linear mixed effects models, which have become prominent in human research where many measurements are made for each subject.\cite{Schall-01} Compared with standard linear models, mixed effects models allow for differentiation between effects that are subject-specific and those that are representative of the prototypical individual in our experiments. Here, all models were fit using the \texttt{fitlme} function in MATLAB (R2018a), and random effects were chosen as the maximal structure that (i) allowed the model to converge and (ii) did not include effects whose 95\% confidence intervals overlapped with zero. In what follows, when referring to our mixed effects models, we employ the standard R notation.

For the first experiment, in order to measure the impact of entropy on reaction times (Fig. \ref{experiment}e), we regressed out a number of biomechanical dependencies: (i) variability due to the different button combinations, (ii) the natural quickening of reactions with trial number, and (iii) the change in reaction times between stages. We also regressed out the effects of recency on subjects' reaction times. Specifically, we fit a mixed effects model with the formula `$\text{RT} \sim \log(\text{Trial})*\text{Stage} + \text{Target} + \text{Recency} + (1 + \log(\text{Trial})*\text{Stage} + \text{Recency} \,|\, \text{ID})$', where RT is the reaction time, Trial is the trial number (it is common to consider $\log(\text{Trial})$ rather than the trial number itself\cite{Kahn-01, Lynn-06}), Stage is the stage of the experiment, Target is the target button combination, Recency is the number of trials since the last instance of the current stimulus, and ID is each subject's unique ID.

For the second experiment, to measure differences in reaction times between transitions in the modular network (Fig. \ref{experiment}g), we fit a mixed effects model of the form `$\text{RT} \sim \log(\text{Trial})*\text{Stage} + \text{Target} + \text{Recency} + \text{Trans\_Type} + (1 + \log(\text{Trial})*\text{Stage} + \text{Recency} \,|\, \text{ID})$', where Trans\_Type is a dummy variable representing the type of transition (Fig. \ref{experiment}g) and the other variables are defined above. The three models for the three different comparisons are summarized in Tables S2-S4.

For the third experiment, to measure the difference in reaction times between the modular network and random $k$-4 networks (Fig. \ref{experiment}h), we fit a mixed effects model of the form `$\text{RT} \sim \log(\text{Trial})*\text{Stage} + \text{Target} + \text{Recency} + \text{Graph} + (1 + \log(\text{Trial})*\text{Stage} + \text{Recency} \,|\, \text{ID})$', where Graph is a dummy variable representing the type of network (either modular or random $k$-4). This model is summarized in Table S5.

\noindent \textbf{Estimating $\eta$ values.} Given a choice for the parameter $\eta$, and given a sequence of past nodes $x_1,\hdots,x_{t-1}$, the internal expectation of the next node $x_t$ is predicted to be $\hat{P}_{x_{t-1},x_t}$. We predict subjects' reaction times $r(t)$ using the linear model $\hat{r}(t) = r_0 - r_1\log \hat{P}_{x_{t-1},x_t}$, where $-\log \hat{P}_{x_{t-1},x_t}$ is the predicted perceived information at time $t$. Before estimating $\eta$, $r_0$, and $r_1$, we regress out subjects' biomechanical dependencies using the mixed effects model `$\text{RT} \sim \log(\text{Trial})*\text{Stage} + \text{Target} + \text{Recency} + (1 + \log(\text{Trial})*\text{Stage} + \text{Recency} \,|\, \text{ID})$', where all variables are defined above. Then, to estimate the model parameters that best describe a subject's reactions, we minimize the root-mean-square error (RMSE) with respect to each subject's reaction times. We note that, given a choice for $\eta$, the linear parameters $r_0$ and $r_1$ can be calculated analytically. Thus, the estimation problem can be restated as a one-dimensional minimization problem; that is, minimizing RMSE with respect to $\eta$. To find the global minimum, we began by calculating RMSE along 101 values for $\eta$ between $0$ and $1$. Then, starting at the minimum value of this search, we performed gradient descent until the gradient $\frac{\partial \text{RMSE}}{\partial \eta}$ fell below an absolute value of $10^{-6}$. The resulting distribution for $\eta$ over subjects are shown in Fig. \ref{model}b. For more details, see Supplementary Sec. 4. 

\end{methods}

\section*{Data Availability}

The data that support the findings of this study are available from the corresponding author upon request.





\newpage
\begin{addendum}

\item[Supplementary Information] Supplementary text and figures accompany this paper.

\item[Acknowledgements] We thank Eric Horsley, Harang Ju, David Lydon-Staley, Shubhankar Patankar, Pragya Srivastava, and Erin Teich for feedback on earlier versions of the manuscript. We thank Dale Zhou for providing the code used to parse the texts. D.S.B., C.W.L., and A.E.K. acknowledge support from the John D. and Catherine T. MacArthur Foundation, the Alfred P. Sloan Foundation, the ISI Foundation, the Paul G. Allen Family Foundation, the Army Research Laboratory (W911NF-10-2-0022), the Army Research Office (Bassett-W911NF-14-1-0679, Grafton-W911NF-16-1-0474, DCIST- W911NF-17-2-0181), the Office of Naval Research, the National Institute of Mental Health (2-R01-DC-009209-11, R01-MH112847, R01-MH107235, R21-M MH-106799), the National Institute of Child Health and Human Development (1R01HD086888-01), National Institute of Neurological Disorders and Stroke (R01 NS099348), and the National Science Foundation (BCS-1441502, BCS-1430087, NSF PHY-1554488 and BCS-1631550). L.P. is supported by an NSF Graduate Research Fellowship. The content is solely the responsibility of the authors and does not necessarily represent the official views of any of the funding agencies.
 
\item[Author Contributions] C.W.L. and D.S.B. conceived the project. C.W.L. designed the framework and performed the analysis. C.W.L. and A.E.K. performed the human experiments. C.W.L. wrote the manuscript and Supplementary Information. L.P., A.E.K., and D.S.B. edited the manuscript and Supplementary Information.
 
\item[Competing Interests] The authors declare no competing financial interests.
 
\item[Corresponding Author] Correspondence and requests for materials should be addressed to D.S.B. \\ (dsb@seas.upenn.edu).
 
\end{addendum}

\newpage

\noindent {\Large \myfont \textbf{Supplementary Information}}

\noindent {\large \myfont \textbf{\textit{Human information processing in complex networks}}}

\section{Introduction}

In this Supplementary Information, we provide extended analysis and discussion to support the results presented in the main text. In Sec. \ref{previous_work}, we clarify the fundamental differences between our work and previous research on human information processing and complex networks. In Sec. \ref{perceived_info}, we give a brief introduction to information theory and provide explicit definitions for the quantities discussed in the main text. In Sec. \ref{human_expectations}, we introduce existing research studying how humans form expectations about complex transition networks. In Sec. \ref{reaction_time}, we present the effects of graph topology on human reaction times measured in our serial response experiments. We begin in Sec. \ref{entropic_effect} by demonstrating the impact of entropy on reaction times and then proceed to describe effects beyond entropy (Sec. \ref{cluster_effect}, \ref{modular_effect}). In Sec. \ref{real_networks}, we verify that our conclusions concerning the information properties of real networks hold for (i) various values of $\eta$ (Sec. \ref{vary_eta}), (ii) different models of internal representations (Sec. \ref{diff_representations}), and (iii) directed versions of the real networks (Sec. \ref{directed_networks}). In Sec. \ref{entropy}, we derive analytic results for the entropies of various canonical network families. In Sec. \ref{KL_divergence}, we derive a number of analytic results concerning the KL divergence between random walks and human expectations. In Sec. \ref{hierarchically_modular}, we develop a generative model of hierarchically modular networks that combines the heterogeneity of scale-free networks with the community structure of stochastic block networks. Finally, in Sec. \ref{network_datasets}, the real networks analyzed in this work are listed and briefly described.

\section{Previous work}
\label{previous_work}

Our work builds on a long record of research in information theory,\cite{Shannon-01, Cover-01} network science,\cite{Strogatz-01, Albert-02} and cognitive science.\cite{Cleeremans-01, Saffran-01, Fiser-01}  Here, we clarify the relationships and differences between our work and earlier research in these areas. In particular, we emphasize two main points:

\begin{enumerate}
\item In the study of complex networks, traditional definitions of network complexity focus on the structure of a network itself.\cite{Strogatz-01, Albert-02, Rosvall-02, Gomez-03, Liben-01, Rosvall-01} While characterizing the inherent complexity of a network is a fascinating problem with numerous applications, many complex systems -- from language and music to social networks and literature -- exist for the sole purpose of communicating information with and between humans. Therefore, to fully understand the structure of these communication networks, one must consider the perspective of a human observer. In this work, we show that this shift in perspective from inherent complexity to perceived complexity can be formally defined using information theory and provides critical insights into the structure of real communication networks.

\item Significant research in cognitive science and statistical learning has studied how humans build internal expectations about the world around them,\cite{Cleeremans-01, Saffran-01, Fiser-01, Newport-01, Dehaene-01, Schapiro-01, Kahn-01, Lynn-06, Meyniel-02} generating deep insights about human learning and behavior. Building upon this work, we consider a complimentary problem that has received far less attention: Given a model of human expectations, what types of structures support efficient human communication? The answer to this question may shed light on the organization of real communication systems and help us to design new systems with desirable properties.
\end{enumerate}

\subsection{Definitions of network entropy}

Information theory has been linked with network science since its inception, when Shannon estimated the entropy rate of the English language by studying a random walk on the network of word transitions in a book.\cite{Shannon-01} Since then, information theory has been used extensively to characterize the structure and function of complex networks.\cite{Strogatz-01, Albert-02, Rosvall-02, Rosvall-03, Gomez-03, Liben-01, Rosvall-01, Newman-01, Sinatra-01} Of particular interest are ongoing efforts studying the entropies of random walks on complex networks. For example, the entropies of a number of canonical network families have been derived, including constant-degree networks\cite{Cover-01} and power-law distributed networks.\cite{Gomez-03} Meanwhile, researchers have developed strategies for maximizing the entropy of random walks by tuning the edge weights in a network,\cite{Demetrius-01, Burda-01, Sinatra-01, Coghi-01} and it is now known that temporal regularities in random walks reveal key aspects of modularity and community structure.\cite{Rosvall-03, Rosvall-02}

Our work extends these efforts by taking into account human expectations. Specifically, we consider the cross entropy (or perceived information) of random walks relative to human expectations, which can be broken down into network entropy (or produced information) and KL divergence (or the inefficiency of human expectations). Importantly, we discover that the entropy and KL divergence characterize distinct aspects of network structure: while entropy is driven by degree heterogeneity, the KL divergence is determined by a network's modular organization. Additionally, we provide a number of novel results concerning network entropy and KL divergence that may be of independent interest. These include analytic approximations for the entropies of networks with Poisson and exponential degree distributions as well as static model networks (see Supplementary Sec. \ref{entropy}) and the KL divergences of Erd\"{o}s-R\'{e}nyi and stochastic block networks (see Supplementary Sec. \ref{KL_divergence}).

\subsection{Human information processing}

Efforts to relate human cognition to information theory have a rich history, spanning the fields of cognitive science, psychology, and neuroscience. For example, information theory has been used to study linguistics,\cite{Bar-01,Dretske-01} decision-making,\cite{Hilbert-01, Wickelgren-01} Bayesian learning,\cite{Ortega-01} neural coding,\cite{Rieke-01} and vision.\cite{Delgado-01} In fact, the relativity of information -- the notion that the amount of information conveyed by a message depends not just on the inherent complexity of the message, but also on the expectations of a receiver -- was previously studied in linguistics to understand the dependence of meaning in language on context.\cite{Dretske-01} To quantify perceived information, however, one requires a mathematical model of human expectations. 

Here, we employ recent models from cognitive science and statistical learning to quantitatively study perceived information. In particular, our experimental results build upon a long line of research in cognitive science linking human reaction times to information processing\cite{Laming-01,Hyman-01} as well as efforts in statistical learning investigating the relationship between human expectations and the network structure of probabilistic transitions.\cite{Cleeremans-01, Saffran-01, Fiser-01, Newport-01, Dehaene-01, Schapiro-01, Karuza-01, Kahn-01, Karuza-03, Lynn-06, Meyniel-02} Additionally, our analytical results leverage mathematical models of human expectations that have roots in temporal context and temporal difference learning\cite{Howard-01, Gershman-01} and also appear in reinforcement learning\cite{Dayan-01,Momennejad-01} and statistical learning.\cite{Schapiro-01, Lynn-06, Meyniel-02} Using these models of human expectations $\hat{P}$, we are able to quantify the amount of information $\langle -\log \hat{P}\rangle $ that a human perceives when observing a sequence of stimuli.

\section{Perceived information}
\label{perceived_info}

We introduce a specific definition for the information of a sequence of stimuli as perceived by a human observer. We assume that the sequence is generated according to a Markov process with transition probability matrix $P$. The amount of information produced by a transition from one stimulus $i$ to another stimulus $j$ is $-\log P_{ij}$.\cite{Shannon-01} To quantify the amount of information produced by the entire sequence (per stimulus), one averages this quantity over the Markov process,\cite{Cover-01}
\begin{equation}
\label{S}
\langle  -\log P_{ij}\rangle _P = -\sum_i \pi_i \sum_j P_{ij} \log P_{ij},
\end{equation}
where $\bm{\pi}$ is the stationary distribution defined by the stationary condition $\bm{\pi}^{\intercal} = \bm{\pi}^{\intercal}P$. The average quantity in Eq. (\ref{S}) is known as the \emph{entropy rate} of the sequence, although it is often referred to simply as the \emph{entropy}, and it is denoted by $S(P)$.

While the entropy rate quantifies the amount of information produced by a sequence, we are interested in studying the amount of information that a human perceives when observing such a sequence. Consider a human observer with expectations based on an internal estimate of the transition probabilities $\hat{P}$. When observing a transition from one stimulus $i$ to another stimulus $j$, the observer perceives $-\log \hat{P}_{ij}$ bits of information, which, when averaged over the Markov process, takes the form
\begin{equation}
\label{CS}
\langle -\log \hat{P}_{ij}\rangle _P = -\sum_i \pi_i \sum_j P_{ij} \log \hat{P}_{ij}.
\end{equation}
This quantity is the \emph{cross entropy rate} (or simply the \emph{cross entropy}) $S(P,\hat{P})$ between the Markov process $P$ and the observer's expectations $\hat{P}$.

\subsection{Cross entropy}

If the observer's expectations are exact (that is, if $\hat{P} = P$), then the cross entropy (Eq. (\ref{CS})) reduces to the entropy (Eq. (\ref{S})); in other words, if the observer correctly anticipates the frequency of stimuli, then the amount of information they perceive equals the amount of information produced by the sequence itself. However, if the observer's expectations differ from reality (that is, if $\hat{P} \neq P$), then the observer perceives additional information. To see this relationship, we consider the simple identity,
\begin{equation}
S(P,\hat{P}) = S(P) + D_{\text{KL}}(P||\hat{P}),
\end{equation}
where $D_{\text{KL}}(P||\hat{P})$ is the Kullback-Leibler (KL) divergence between $P$ and $\hat{P}$, defined by
\begin{equation}
\label{DKL}
D_{\text{KL}}(P||\hat{P}) = \langle -\log \frac{\hat{P}_{ij}}{\hat{P}_{ij}} \rangle _P = -\sum_i \pi_i \sum_j P_{ij} \log\frac{\hat{P}_{ij}}{P_{ij}}.
\end{equation}
Gibbs' inequality\cite{Cover-01} states that $D_{\text{KL}}(P||\hat{P}) \ge 0$ for all $P$ and $\hat{P}$, and that $D_{\text{KL}}(P||\hat{P}) = 0$ only if $\hat{P} = P$. Therefore, we see that the perceived information (or cross entropy) is lower-bounded by the produced information (or entropy).

\subsection{Random walks on a network}

Every stationary Markov process is equivalent to a random walk on an underlying (possibly weighted, directed) network, where each state is encoded as a node in the network. Specifically, given a transition probability matrix $P$, one can choose an adjacency matrix $G$ such that
\begin{equation}
\label{Pij}
P_{ij} = \frac{1}{k^{\text{out}}_i}G_{ij},
\end{equation}
where $k_i^{\text{out}} = \sum_j G_{ij}$ is the out-degree of node $i$. To develop a number of analytic results, we briefly consider the special case of an undirected network. In this case, the out-degree of a node $i$ is referred to simply as its degree $k_i$. If $G$ is connected, then there exists a unique stationary distribution over nodes, and it is proportional to the degree vector, such that $\bm{\pi} = \frac{1}{2E}\bm{k}$, where $E = \frac{1}{2}\sum_{ij} G_{ij}$ is the number of edges in the network. Therefore, for random walks on a connected, undirected network, we find that the cross entropy can be written as
\begin{equation}
S(P,\hat{P}) = -\frac{1}{2E}\sum_{ij} G_{ij}\log \hat{P}_{ij},
\end{equation}
reflecting a weighted average of $-\log\hat{P}_{ij}$ over the edges in the network. Moreover, if we further restrict our focus to unweighted networks, then the entropy takes a particularly simple form:\cite{Gomez-03}
\begin{equation}
\label{S(k)}
S(P) = \frac{1}{2E}\sum_i k_i \log k_i.
\end{equation}
In this case, it is clear that the entropy of a random walk is uniquely defined by the degree sequence of the network,\cite{Cover-01} a result that is verified numerically for real networks in Fig. 2d.

\section{Human expectations}
\label{human_expectations}

When observing sequences of stimuli, humans constantly rely on their internal estimate of the transition structure to anticipate what is coming next.\cite{Nobre-01, Hyman-01, Sternberg-01, Johnson-01,Winograd-01} Indeed, building expectations about probabilistic relationships allows humans to perform abstract reasoning,\cite{Bousfield-01} produce language,\cite{Friederici-01} develop social intuition,\cite{Gopnik-01, Tompson-01} and segment streams of stimuli into self-similar parcels.\cite{Reynolds-01} Moreover, as discussed above, a person's internal expectations, defined by the estimated transition probability matrix $\hat{P}$, determine the amount of information $S(P,\hat{P})$ that they receive from a transition structure defined by $P$. To study the cross entropy $S(P,\hat{P})$, we require a model $\hat{P} = F(P)$ of how humans internally estimate transition structures in the world around them.

\subsection{Temporal integration of stimuli}
\label{temporal_integration}

Models describing how humans learn and estimate transition structures typically stem from Bayesian inference\cite{Gopnik-01, Piantadosi-01, Tenenbaum-01, Meyniel-02} or notions of hierarchical learning.\cite{Meyniel-01, Dehaene-01, Newport-01, Cleeremans-01} A common thread across many models is that humans relate stimuli that are not directly adjacent in time.\cite{Schapiro-01,Nobre-01} These non-adjacent relationships have been hypothesized to reflect planning for the future,\cite{Dayan-01, Gershman-01} context-dependent memory effects,\cite{Howard-01, Jazayeri-01} and even errors in optimal Bayesian learning.\cite{Lynn-06, Meyniel-02} Independent of the underlying mechanisms, the fact that humans relate non-adjacent stimuli results in a common functional form for the expectations $\hat{P}$ where the true transition structure $P$ is integrated over time. Mathematically, this means that $\hat{P}$ includes higher powers of $P$:
\begin{equation}
\label{model1}
\hat{P} = C(f(0)P + f(1)P^2 + \hdots) = C\sum_{t = 0}^{\infty} f(t)P^{t+1},
\end{equation}
with progressively higher powers down-weighted by a decreasing function $f(t)\ge 0$, where $C = (\sum_{t = 0}^{\infty} f(t))^{-1}$ is a normalization constant. We remark that $\hat{P}$ in Eq. \ref{model1} is guaranteed to converge as long as the sum $\sum_{t = 0}^{\infty} f(t)$ converges.

There exist a number of simple choices for the function $f(t)$. For example, if people's integration of the transition structure drops off as a power law, then we have $f(t) = (t+1)^{-\alpha}$ with power-law exponent $\alpha > 1$. Instead, if the integration drops off with the factorial of $t$ (that is, if $f(t) = 1/t!$), then $\hat{P} = e^{-1}Pe^P$, where $e^P$ is the matrix exponential. We remark that this model for $\hat{P}$ is nearly equivalent to the communicability of $P$ from graph theory,\cite{Estrada-01} which has recently been used to model human expectations.\cite{Girvan-01} In Sec. \ref{diff_representations} we study the information properties of real networks under these alternative models of human expectations, finding qualitatively equivalent results to those described in the main text.

\subsection{Exponential model}
\label{exponential_model}

\begin{figure}[t!]
\centering
\includegraphics[width = \textwidth]{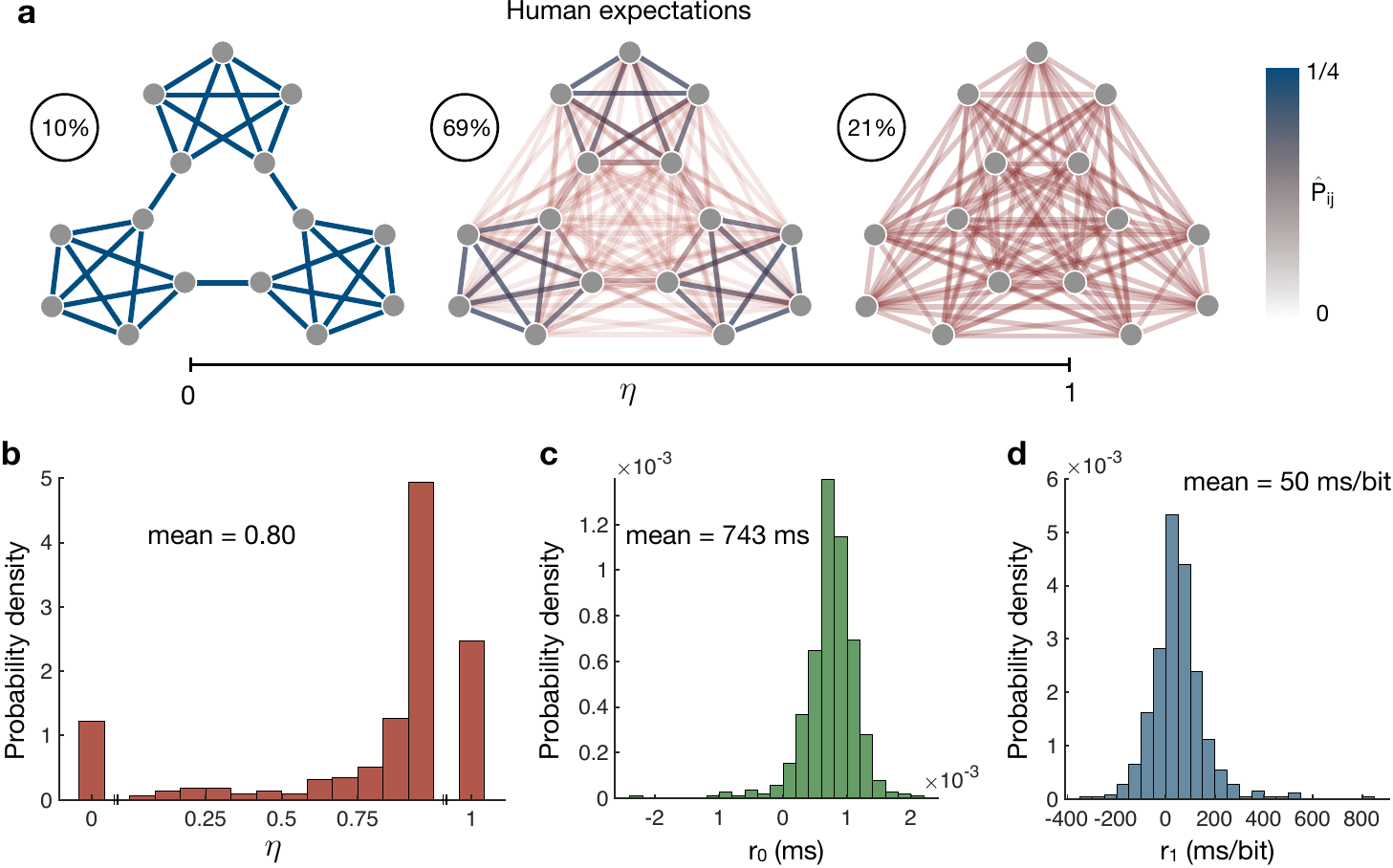} \\
\raggedright
\myfont\textbf{Fig. S\ref{expectations} $|$ Estimated model parameters relating human expectations to reaction times.}
\captionsetup{labelformat=empty}
{\spacing{1.25} \caption{\label{expectations} \myfont \textbf{a}, Human expectations $\hat{P}$ for the modular network. For $\eta\rightarrow 0$, expectations become exact (left; 10\% of subjects), while for $\eta\rightarrow 1$, expectations become all-to-all, losing any resemblance to the true structure (right; 21\% of subjects). At intermediate values of $\eta$, the communities maintain probability weight, while expectations for cross-cluster transitions weaken (center; 69\% of subjects). \textbf{b}-\textbf{d}, Distributions of model parameters estimated from subjects' reaction times. Distributions are over all 518 completed sequences. For the integration parameter $\eta$ (\textbf{b}), 53 subjects were best described as having exact representations ($\eta\rightarrow 0$) and 107 lacked any notion of the transition structure ($\eta\rightarrow 1$), while across all subjects the average value was $\eta = 0.80$. The intercept $r_0$ is mostly positive (\textbf{b}), with an average value of $743$ ms. The slope $r_1$ is also mostly positive (\textbf{d}), with an average value of $50$ ms/bit.}}
\end{figure}

Throughout the main text, we focus on a specific model for $\hat{P}$ in which the integration of the transition structure drops off exponentially, such that $f(t) = \eta^t$, where $\eta \in (0,1)$ is the integration constant. This model is closely related to the successor representation from reinforcement learning,\cite{Dayan-01,Momennejad-01} which can be derived from temporal context and temporal difference learning,\cite{Gershman-01} and can independently be shown to arise from errors in human cognition.\cite{Lynn-06} The model takes the following concise analytic form,
\begin{align}
\label{model2}
\hat{P} &= \left(\sum_{t=0}^{\infty}\eta^t\right)^{-1} \sum_{t=0}^{\infty}\eta^tP^{t+1} \nonumber \\
&= (1-\eta)P\sum_{t=0}^{\infty}(\eta P)^t \\
&= (1-\eta)P(I - \eta P)^{-1}, \nonumber
\end{align}
where the second equality follows by noticing that $\sum_{t=0}^{\infty}\eta^t = 1/(1-\eta)$ and the third equality follows from the fact that $\sum_{t=0}^{\infty}(\eta P)^t$ converges to $(I - \eta P)^{-1}$ since the spectral radius $\rho(\eta P) = \eta$ is less than one. In the limit $\eta\rightarrow 0$, we see that $\hat{P}\rightarrow P$, and hence the estimate becomes equivalent to the true transition structure $P$ (Fig. S\ref{expectations}a). By contrast, in the limit $\eta\rightarrow 1$, we find that $\hat{P}\rightarrow \bm{1}\bm{\pi}^{\intercal}$, where $\bm{1}$ is the vector of all ones and $\bm{\pi}$ is the stationary distribution, such that the expectations lose all resemblance to the true structure (Fig. S\ref{expectations}a). For intermediate values of $\eta$, higher-order features of the network, such as communities of densely-connected nodes, maintain much of their probability weight, while some of the fine-scale features, like the edges between communities, fade away (Fig. S\ref{expectations}a). This strengthening of expectations for transitions within communities relative to transitions between communities is precisely the effect we observe in human reaction times (Fig. 1e).

In order to make quantitative predictions for the KL divergence $D_{\text{KL}}(P||\hat{P})$, it is useful to have an estimate for the integration parameter $\eta$ based on real human data. We estimate $\eta$ by making predictions for subjects' reaction times and then minimizing the prediction error with respect to $\eta$. Given a sequence of nodes $x_1,\hdots,x_{t-1}$, we note that the reaction to the next node $x_t$ is determined by the perceived information of the transition from $x_{t-1}$ to $x_t$, with expectations calculated at time $t-1$. Formally, this perceived information is given by $-\log \hat{P}_{x_{t-1},x_t}$, and we make the following linear prediction for the reaction time,
\begin{equation}
\hat{r}(t) = r_0 - r_1 \log \hat{P}_{x_{t-1},x_t},
\end{equation}
where the intercept $r_0$ represents a person's minimum average reaction time (with perfect anticipation of the next stimulus, $\hat{P}_{x_{t-1},x_t} = 1$) and the slope $r_1$ quantifies the strength of the relationship between a person's reactions and their perceived information, measured in units of time per bit. Before estimating the model parameters, we first regress out the dependencies of each subject's reaction times on the button combinations, trial number, experimental stage, and recency using a mixed effects model of the form `$RT\sim \log(Trial)*Stage + Target + Recency + (1 + \log(Trial)*Stage + Recency \,|\, ID)$', where $RT$ is the reaction time, $Trial$ is the trial number between 1 and 1500 (we found that $\log(Trial)$ was far more predictive of subjects' reaction times than the trial number itself), $Stage$ is the stage of the experiment (either one or two), $Target$ is the target button combination, $Recency$ is the number of trials since the last instance of the current stimulus, and $ID$ is each subject's unique ID. Then, to estimate the parameters $\eta$, $r_0$, and $r_1$ that best describe a subjects' reaction times, we minimize the RMS error $\sqrt{\frac{1}{T}\sum_t(r(t) - \hat{r}(t))^2}$, where $r(t)$ is the reaction time on trial $t$ after regressing out the above dependencies and $T$ is the number of trials in the experiment. The distributions of the estimated parameters are shown in Fig. S\ref{expectations}b-d. Among the 518 completed sequences (across 363 unique subjects), 53 were best described as having expectations that exactly matched the transition structure ($\eta\rightarrow 0$) and 107 seemed to lack any notion of the transition structure whatsoever ($\eta\rightarrow 1$), with an overall average value of $\eta = 0.80$.

\begin{figure}[t!]
\centering
\includegraphics[width = \textwidth]{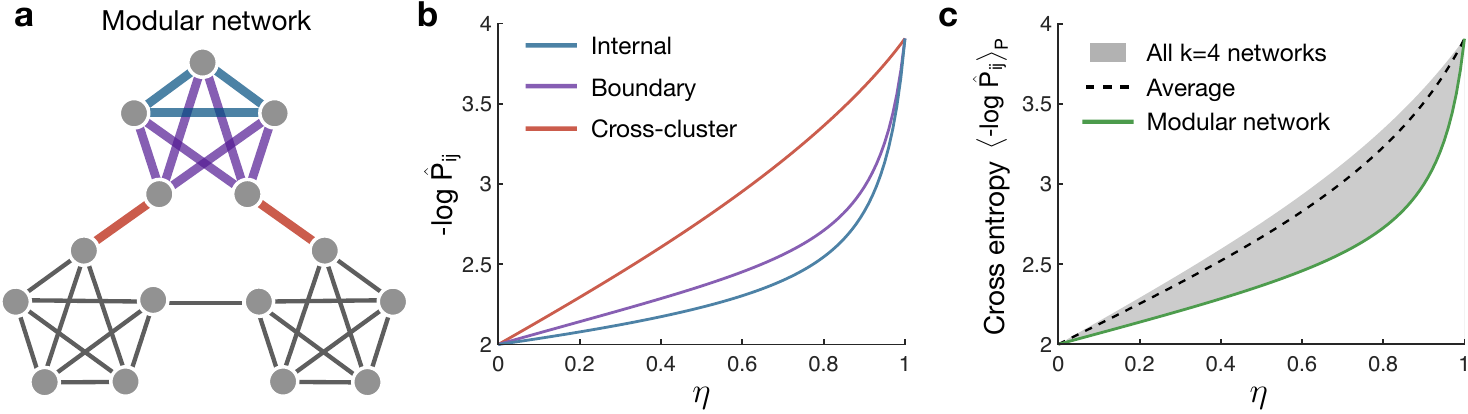} \\
\raggedright
\myfont\textbf{Fig. S\ref{network_effects} $|$ Network effects on human reaction times beyond entropy.}
\captionsetup{labelformat=empty}
{\spacing{1.25} \caption{\label{network_effects} \myfont \textbf{a}, Modular network with three modules of five nodes each. By symmetry the network contains three distinct types of edges: those deep within communities (blue), those at the boundaries of communities (purple), and those between communities (red). \textbf{b}, Perceived information $-\log \hat{P}_{ij}$ for the three edge types as a function of $\eta$. Across all values of $\eta$, the perceived information is highest for cross-cluster edges, followed by boundary edges, and lowest for internal edges, thus explaining the observed differences in human reaction times (Fig. 1e). \textbf{c}, Cross entropy (or network-averaged perceived information) $\langle -\log \hat{P}_{ij}\rangle _P$ as a function of $\eta$ for the modular network (green) and all k-4 networks (the grey region denotes the range and the dashed line denotes the mean). The modular network maintains nearly the lowest cross entropy among k-4 networks across all values of $\eta$, thereby explaining the overall decrease in reaction times in the modular network relative to random k-4 networks (Fig. 1f).}}
\end{figure}

Equipped with the model of human expectations in Eq. (\ref{model2}), we can make quantitative predictions for the perceived information of different transition structures. For example, considering the three types of transitions in the modular network (Fig. S\ref{network_effects}a), we find across all values of $\eta$ that the perceived information $-\log \hat{P}_{ij}$ is highest for transitions between communities, followed by transitions at the boundaries of communities, and lowest for transitions deep within communities (Fig. S\ref{network_effects}b). This prediction precisely matches the variations in reaction times for the different transitions observed in our human experiments (Fig. 1e). Furthermore, we find that the average perceived information (or cross entropy) $\langle  -\log \hat{P}_{ij}\rangle _P$ is lower in the modular network than almost any other network of the same entropy across all values of $\eta$ (Fig. S\ref{network_effects}c). This final prediction explains the observed decrease in reaction times in the modular network relative to random entropy-preserving networks (Fig. 1f).

\section{Network effects on reaction times}
\label{reaction_time}

In order to directly probe the information that humans perceive, we employ an experimental framework recently developed in statistical learning.\cite{Schapiro-01, Karuza-01, Kahn-01, Karuza-03, Lynn-06} Specifically, we present human subjects with sequences of stimuli on a computer screen, each stimulus depicting a row of five grey squares with one or two of the squares highlighted in red (Fig. 1a, left). In response to each stimulus, subjects are asked to press one or two computer keys mirroring the highlighted squares (Fig. 1a, right). Each of the 15 different stimuli represents a node in an underlying transition network, upon which a random walk stipulates the sequential order of stimuli (Fig. 1b). By measuring the speed with which a subject responds to each stimulus, we can infer how much information they are processing -- a fast reaction reflects an unsurprising (or uninformative) transition, while a slow reaction reflects a surprising (or informative) transition.\cite{Laming-01, Hyman-01, Sternberg-01, McCarthy-01, Kahn-01, Lynn-06}

In order to extract the effects of network structure on subjects' reaction times, we use linear mixed effects models, which have become prominent in human research where many measurements are made for each subject.\cite{Schall-01,Baayen-01} To fit our mixed effects models and to estimate the statistical significance of each effect we use the \texttt{fitlme} function in MATLAB (R2018a). In what follows, when referring to our mixed effects models, we adopt the standard R notation.\cite{Bates-01}

\subsection{Entropic effect}
\label{entropic_effect}

We first investigate the effect of produced information on subjects' reaction times. For undirected and unweighted networks, the produced information (or surprisal) for a single transition from a node $i$ to one of $i$'s neighbors is $\log k_i$, where $k_i$ is the degree of node $i$. To study a range of surprisal values, we consider completely random networks in which the node degrees are allowed to vary (specifically, we consider random networks with $N=15$ nodes and $E = 30$ edges). We regress out the dependencies of each subject's reaction times on the button combinations, trial number, experimental stage, and recency using a mixed effects model with the formula `$RT\sim \log(Trial)*Stage + Target + Recency + (1 + \log(Trial)*Stage + Recency \,|\, ID)$', where $RT$ is the reaction time, $Trial$ is the trial number between 1 and 1500, $Stage$ is the stage of the experiment (either one or two), $Target$ is the target button combination, $Recency$ is the number of trials since last observing a node,\cite{Baddeley-01} and $ID$ is each subject's unique ID. After regressing out these biomechanical dependencies, we find that subjects' average reaction times following nodes of a given degree are accurately predicted by the produced information (Fig. 1c), with a Pearson correlation of $r_p = 0.99$ ($p < 0.001)$ and a slope of 32 ms/bit.

Additionally, to take into account variations in subjects' reaction times rather than simply studying average reaction times, we employ a mixed effects model of the form `$RT\sim \log(Trial)*Stage + Target + Recency + Surprisal + (1 + \log(Trial)*Stage + Recency + Surprisal \,|\, ID)$', where $Surprisal$ is the logarithm of the degree of the preceding node. The mixed effects model is summarized in Table S1, reporting a 26 ms increase in reaction times for each additional bit of produced information. We remark that this bit rate is close to that estimated from subjects' average reaction times in random graphs (32 ms/bit; Fig. 1c) and is also comparable to the bit rate estimated from our linear prediction of subjects' reaction times in constant-degree graphs (50 ms/bit; Fig. S\ref{expectations}d).

\addtocounter{figure}{-1}
\begin{figure}[t!]
\centering
{\fontsize{12}{12} \myfont
\begin{tabular}{|c|c|c|c|c|}
\hline
Effect & Estimate (ms) & t-value & Pr$(>$$|$t$|)$ & Significance \\
\hline
\hline
(Intercept) & $1324.8 \pm 49.6$ & $26.73$ & $< 0.001$ & $***$ \\
\hline
log(Trial) & $-89.6\pm 5.8$ & $-15.41$ & $< 0.001$ & $***$ \\
\hline
Stage & $-538.9 \pm 54.1$ & $-9.96$ & $< 0.001$ & $***$ \\
\hline
Recency & $1.9 \pm 0.1$ & $21.63$ & $< 0.001$ & $***$ \\
\hline
\rowcolor{LightGrey}
Surprisal & $26.1 \pm 4.1$ & $6.39$ & $< 0.001$ & $***$ \\
\hline
log(Trial):Stage & $78.2 \pm 6.6$ & $11.91$ & $< 0.001$ & $***$ \\
\hline
\end{tabular}}
\vskip 12pt
\raggedright
\captionsetup{labelformat=empty}
{\spacing{1.25} \caption{\myfont \textbf{Table S1 $|$ Mixed effects model measuring the effect of produced information on human reaction times.} We find a significant 26 ms increase in reaction times ($n = 177$) for each additional bit of produced information, or surprisal (grey). All effects are significant with $p$-values less than 0.001 ($***$).}}
\end{figure}

\subsection{Extended cross-cluster effect}
\label{cluster_effect}

We next investigate reaction time patterns that are driven by perceived information beyond the information produced by a sequence. To experimentally control for the information produced by transitions, we focus on networks of constant degree 4 ($N = 15$ and $E = 30$). Specifically, we consider the modular network shown in Fig. S\ref{network_effects}a, consisting of three communities or clusters comprised of five nodes each. Recent research has shown that people can detect transitions between the clusters\cite{Schapiro-01} and that cross-cluster transitions yield increases in reaction times relative to within-cluster transitions.\cite{Kahn-01,Lynn-06} These behaviors are surprising in light of the fact that all edges in the network have identical transition probabilities and therefore produce identical amounts of information. Here, we extend these results to include all three of the distinct types of transitions in the modular network (Fig. S\ref{network_effects}a): those deep within communities (internal transitions), those at the boundaries of communities (boundary transitions), and those between communities (cross-cluster transitions).

We use a mixed effects model with the formula `$RT\sim \log(Trial)*Stage + Target + Recency + Trans\_Type + (1 + \log(Trial)*Stage + Recency \,|\, ID)$', where $Trans\_Type$ represents the type of transition (either internal, boundary, or cross-cluster). We find a 39 ms increase in reaction times for cross-cluster transitions relative to internal transitions within clusters (Table S2), a 31 ms increase in reaction times for cross-cluster transitions relative to boundary transitions within clusters (Table S3), and a 7 ms increase in reaction times for boundary transitions relative to internal transitions within clusters (Table S4). Notably, this hierarchy of reaction times (Fig. 1g) is the same as that predicted by our cross entropy framework (Fig. S\ref{network_effects}b).

\addtocounter{figure}{-1}
\begin{figure}[t!]
\centering
{\fontsize{12}{12} \myfont
\begin{tabular}{|c|c|c|c|c|}
\hline
Effect & Estimate (ms) & t-value & Pr$(>$$|$t$|)$ & Significance \\
\hline
\hline
(Intercept) & $1365.6 \pm 46.8$ & $29.15$ & $< 0.001$ & $***$ \\
\hline
log(Trial) & $-86.9\pm 5.2$ & $-16.75$ & $< 0.001$ & $***$ \\
\hline
Stage & $-549.2 \pm 52.9$ & $-10.38$ & $< 0.001$ & $***$ \\
\hline
Recency & $1.5 \pm 0.1$ & $18.40$ & $< 0.001$ & $***$ \\
\hline
\rowcolor{LightGrey}
Trans$\_$Type & $38.7 \pm 2.3$ & $16.99$ & $< 0.001$ & $***$ \\
\hline
log(Trial):Stage & $63.5 \pm 5.8$ & $11.01$ & $< 0.001$ & $***$ \\
\hline
\end{tabular}}
\vskip 12pt
\raggedright
\captionsetup{labelformat=empty}
{\spacing{1.25} \caption{\myfont \textbf{Table S2 $|$ Mixed effects model measuring the difference in reaction times between internal and cross-cluster transitions.} We find a significant 39 ms increase in reaction times ($n = 173$) for cross-cluster transitions relative to internal transitions within communities (grey). All effects are significant with $p$-values less than 0.001 ($***$).}}
\end{figure}

\addtocounter{figure}{-1}
\begin{figure}[t!]
\centering
{\fontsize{12}{12} \myfont
\begin{tabular}{|c|c|c|c|c|}
\hline
Effect & Estimate (ms) & t-value & Pr$(>$$|$t$|)$ & Significance \\
\hline
\hline
(Intercept) & $1349.3 \pm 45.8$ & $29.48$ & $< 0.001$ & $***$ \\
\hline
log(Trial) & $-86.0\pm 5.2$ & $-16.39$ & $< 0.001$ & $***$ \\
\hline
Stage & $-495.41 \pm 49.6$ & $-9.98$ & $< 0.001$ & $***$ \\
\hline
Recency & $1.6 \pm 0.1$ & $23.28$ & $< 0.001$ & $***$ \\
\hline
\rowcolor{LightGrey}
Trans$\_$Type & $30.8 \pm 2.1$ & $14.50$ & $< 0.001$ & $***$ \\
\hline
log(Trial):Stage & $62.1 \pm 5.8$ & $10.76$ & $< 0.001$ & $***$ \\
\hline
\end{tabular}}
\vskip 12pt
\raggedright
\captionsetup{labelformat=empty}
{\spacing{1.25} \caption{\myfont \textbf{Table S3 $|$ Mixed effects model measuring the difference in reaction times between boundary and cross-cluster transitions.} We find a significant 31 ms increase in reaction times ($n = 173$) for cross-cluster transitions relative to boundary transitions within communities (grey). All effects are significant with $p$-values less than 0.001 ($***$).}}
\end{figure}

\addtocounter{figure}{-1}
\begin{figure}[t!]
\centering
{\fontsize{12}{12} \myfont
\begin{tabular}{|c|c|c|c|c|}
\hline
Effect & Estimate (ms) & t-value & Pr$(>$$|$t$|)$ & Significance \\
\hline
\hline
(Intercept) & $1333.3 \pm 44.3$ & $30.13$ & $< 0.001$ & $***$ \\
\hline
log(Trial) & $-84.0\pm 4.9$ & $-17.11$ & $< 0.001$ & $***$ \\
\hline
Stage & $-464.8 \pm 47.2$ & $-9.84$ & $< 0.001$ & $***$ \\
\hline
Recency & $1.5 \pm 0.1$ & $24.55$ & $< 0.001$ & $***$ \\
\hline
\rowcolor{LightGrey}
Trans$\_$Type & $6.6 \pm 1.3$ & $4.96$ & $< 0.001$ & $***$ \\
\hline
log(Trial):Stage & $60.0 \pm 5.4$ & $11.12$ & $< 0.001$ & $***$ \\
\hline
\end{tabular}}
\vskip 12pt
\raggedright
\captionsetup{labelformat=empty}
{\spacing{1.25} \caption{\myfont \textbf{Table S4 $|$ Mixed effects model measuring the difference in reaction times between internal and boundary transitions within clusters.} We find a significant 7 ms increase in reaction times ($n = 173$) for boundary transitions relative to internal transitions within communities (grey). All effects are significant with $p$-values less than 0.001 ($***$).}}
\end{figure}

\subsection{Modular effect}
\label{modular_effect}

We finally investigate the effects of perceived information averaged over all transitions in a network, defined by the cross entropy in Eq. (\ref{CS}). To do so, we compare reaction times in the modular network with reaction times in random $k$-4 networks. We remark that the entropy (defined in Eq. (\ref{S})) is identical across all graphs considered. We use a mixed effects model of the form `$RT\sim \log(Trial)*Stage + Target + Recency + Network\_Type + (1 + \log(Trial)*Stage + Recency \,|\, ID)$', where $Network\_Type$ represents the type of network (either modular of random $k$-4). The estimated mixed effects model is summarized in Table S5, reporting a 24 ms increase in reaction times for random degree-preserving networks relative to the modular network. Notably, this effect is predicted by our cross entropy framework (Fig. S\ref{network_effects}c). Moreover, this result provides direct evidence that, even after controlling for the entropy of a network, modular structure reduces the total amount of information that humans perceive when observing a sequence of stimuli.

\addtocounter{figure}{-1}
\begin{figure}[t!]
\centering
{\fontsize{12}{12} \myfont
\begin{tabular}{|c|c|c|c|c|}
\hline
Effect & Estimate (ms) & t-value & Pr$(>$$|$t$|)$ & Significance \\
\hline
\hline
(Intercept) & $1195.0 \pm 48.8$ & $24.49$ & $< 0.001$ & $***$ \\
\hline
log(Trial) & $-71.9\pm 4.9$ & $-14.61$ & $< 0.001$ & $***$ \\
\hline
Stage & $-405.3 \pm 36.9$ & $-10.98$ & $< 0.001$ & $***$ \\
\hline
Recency & $1.7 \pm 0.1$ & $19.65$ & $< 0.001$ & $***$ \\
\hline
\rowcolor{LightGrey}
Network$\_$Type & $23.5 \pm 6.9$ & $3.39$ & $< 0.001$ & $***$ \\
\hline
log(Trial):Stage & $49.0 \pm 5.1$ & $9.61$ & $< 0.001$ & $***$ \\
\hline
\end{tabular}}
\vskip 12pt
\raggedright
\captionsetup{labelformat=empty}
{\spacing{1.25} \caption{\myfont \textbf{Table S5 $|$ Mixed effects model measuring the difference in reaction times between the modular network and random $k$-4 networks.} We find a significant 24 ms increase in reaction times ($n = 84$) for random $k$-4 networks (that is, networks of equal entropy) relative to the modular network (grey). All effects are significant with $p$-values less than 0.001 ($***$).}}
\end{figure}

\section{Network effects on errors}
\label{errors}

In addition to measuring the effects of network structure on subjects' reaction times, we can also investigate variations in subjects' error rates. Here, we study the same entropic, extended cross-cluster, and modular effects as in Supplementary Sec. \ref{reaction_time} above, but on error rates instead of reaction times.

\subsection{Entropic effect}
\label{error_entropic}

We first investigate the effect of produced information (or surprisal) $\log k_i$ on subjects' error rates. Specifically, we consider the same random networks as in Supplementary Sec. \ref{entropic_effect}. To measure the effect of produced information on error rates, we estimate a mixed effects model of the form `$Error \sim \log(Trial)*Stage + Target + Recency + Surprisal + (1 + \log(Trial)*Stage + Recency + Surprisal \,|\, ID)$', where $Error$ equals one for error trials and zero for correct trials, and the other variables have been defined previously. The estimated model is summarized in Table S6, with a significant 0.3\% increase in errors for each additional bit of produced information.

\addtocounter{figure}{-1}
\begin{figure}[t!]
\centering
{\fontsize{12}{12} \myfont
\begin{tabular}{|c|c|c|c|c|}
\hline
Effect & Estimate & t-value & Pr$(>$$|$t$|)$ & Significance \\
\hline
\hline
(Intercept) & $0.078 \pm 0.009$ & $8.73$ & $< 0.001$ & $***$ \\
\hline
log(Trial) & $-0.007\pm 0.001$ & $-5.57$ & $< 0.001$ & $***$ \\
\hline
Stage & $-0.037 \pm 0.011$ & $-3.30$ & $< 0.001$ & $***$  \\
\hline
Recency & $0.001 \pm 0.000$ & $14.01$ & $< 0.001$ & $***$ \\
\hline
\rowcolor{LightGrey}
Surprisal & $0.003 \pm 0.001$ & $2.74$ & $0.006$ & $**$ \\
\hline
log(Trial):Stage & $0.005 \pm 0.002$ & $3.34$ & $< 0.001$ & $***$  \\
\hline
\end{tabular}}
\vskip 12pt
\raggedright
\captionsetup{labelformat=empty}
{\spacing{1.25} \caption{\myfont \textbf{Table S6 $|$ Mixed effects model measuring the effect of produced information on error rates.} We find a significant 0.3\% increase in errors ($n = 177$) for each additional bit of produced information, or surprisal (grey). The significance column indicates $p$-values less than 0.001 ($***$), less than 0.01 ($**$), and less than 0.05 ($*$).}}
\end{figure}

\subsection{Extended cross-cluster effect}
\label{error_cluster}

\begin{figure}[h!]
\centering
\includegraphics[width = .6\textwidth]{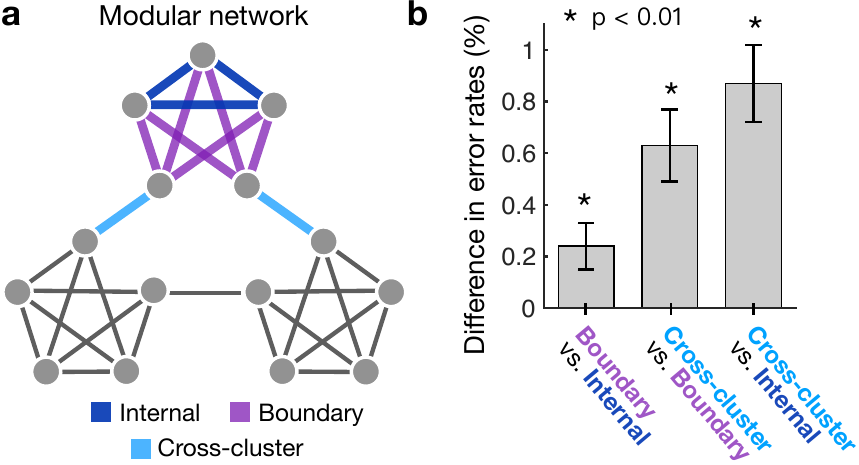} \\
\raggedright
\myfont \textbf{Fig. S\ref{errors} $|$ Effects of modular topology on error rates.}
\captionsetup{labelformat=empty}
{\spacing{1.25} \caption{\label{errors} \myfont \textbf{a}, Modular network with three types of edges: internal edges within communities (dark blue), boundary edges within communities (purple), and cross-cluster edges between communities (light blue). \textbf{b}, Differences in error rates between the different types of transitions; we find significant differences in error rates between all three types of transitions ($n = 173$ subjects).}}
\end{figure}

We next study variations in error rates that are driven by perceived information after controlling for information produced by a sequence. Considering once again the modular network in Fig. \ref{errors}a, we measure the differences in error rates between the different types of transitions. We use a mixed effects model of the form `$Error \sim \log(Trial)*Stage + Target + Recency + Trans\_Type + (1 + \log(Trial)*Stage + Recency \,|\, ID)$', where $Trans\_Type$ denotes the type of transition (either internal, boundary, or cross-cluster). We find a significant 0.9\% increase in errors for cross-cluster transitions relative to internal transitions within clusters (Table S7), a significant 0.6\% increase in errors for cross-cluster transitions relative to boundary transitions within clusters (Table S8), and a significant 0.2\% increase in errors for boundary transitions relative to internal transitions within clusters (Table S9). Notably, we find the same hierarchy of effects on error rates (Fig. S\ref{errors}) as for reaction times (Fig. 1g) and as predicted by our cross entropy framework (Fig. S\ref{network_effects}b).

\addtocounter{figure}{-1}
\begin{figure}[t!]
\centering
{\fontsize{12}{12} \myfont
\begin{tabular}{|c|c|c|c|c|}
\hline
Effect & Estimate & t-value & Pr$(>$$|$t$|)$ & Significance \\
\hline
\hline
(Intercept) & $0.068 \pm 0.010$ & $7.04$ & $< 0.001$ & $***$ \\
\hline
log(Trial) & $-0.005\pm 0.002$ & $-3.30$ & $< 0.001$ & $***$ \\
\hline
Stage & $-0.035 \pm 0.013$ & $-2.62$ & $0.009$ & $**$ \\
\hline
Recency & $0.000 \pm 0.000$ & $9.73$ & $< 0.001$ & $***$ \\
\hline
\rowcolor{LightGrey}
Trans$\_$Type & $0.009 \pm 0.002$ & $5.69$ & $< 0.001$ & $***$ \\
\hline
log(Trial):Stage & $0.006 \pm 0.002$ & $3.03$ & $0.002$ & $**$ \\
\hline
\end{tabular}}
\vskip 12pt
\raggedright
\captionsetup{labelformat=empty}
{\spacing{1.25} \caption{\myfont \textbf{Table S7 $|$ Mixed effects model measuring the difference in error rates between internal and cross-cluster transitions.} We find a significant 0.9\% increase in errors ($n = 173$) for cross-cluster transitions relative to internal transitions within communities (grey). The significance column indicates $p$-values less than 0.001 ($***$), less than 0.01 ($**$), and less than 0.05 ($*$).}}
\end{figure}

\addtocounter{figure}{-1}
\begin{figure}[t!]
\centering
{\fontsize{12}{12} \myfont
\begin{tabular}{|c|c|c|c|c|}
\hline
Effect & Estimate & t-value & Pr$(>$$|$t$|)$ & Significance \\
\hline
\hline
(Intercept) & $0.075 \pm 0.008$ & $8.91$ & $< 0.001$ & $***$ \\
\hline
log(Trial) & $-0.006\pm 0.001$ & $4.68$ & $< 0.001$ & $***$ \\
\hline
Stage & $-0.044 \pm 0.013$ & $-3.53$ & $< 0.001$ & $***$ \\
\hline
Recency & $0.000 \pm 0.000$ & $12.82$ & $< 0.001$ & $***$ \\
\hline
\rowcolor{LightGrey}
Trans$\_$Type & $0.006 \pm 0.001$ & $4.38$ & $< 0.001$ & $***$ \\
\hline
log(Trial):Stage & $0.008 \pm 0.002$ & $3.88$ & $< 0.001$ & $***$ \\
\hline
\end{tabular}}
\vskip 12pt
\raggedright
\captionsetup{labelformat=empty}
{\spacing{1.25} \caption{\myfont \textbf{Table S8 $|$ Mixed effects model measuring the difference in error rates between boundary and cross-cluster transitions.} We find a significant 0.6\% increase in errors ($n = 173$) for cross-cluster transitions relative to boundary transitions within communities (grey). All effects are significant with $p$-values less than 0.001 ($***$).}}
\end{figure}

\addtocounter{figure}{-1}
\begin{figure}[t!]
\centering
{\fontsize{12}{12} \myfont
\begin{tabular}{|c|c|c|c|c|}
\hline
Effect & Estimate & t-value & Pr$(>$$|$t$|)$ & Significance \\
\hline
\hline
(Intercept) & $0.070 \pm 0.008$ & $8.63$ & $< 0.001$ & $***$ \\
\hline
log(Trial) & $-0.006\pm 0.001$ & $-4.11$ & $< 0.001$ & $***$ \\
\hline
Stage & $-0.034 \pm 0.011$ & $-3.00$ & $0.003$ & $**$ \\
\hline
Recency & $0.000 \pm 0.000$ & $13.81$ & $< 0.001$ & $***$ \\
\hline
\rowcolor{LightGrey}
Trans$\_$Type & $0.002 \pm 0.001$ & $2.66$ & $0.008$ & $**$ \\
\hline
log(Trial):Stage & $0.006 \pm 0.002$ & $3.45$ & $< 0.001$ & $***$ \\
\hline
\end{tabular}}
\vskip 12pt
\raggedright
\captionsetup{labelformat=empty}
{\spacing{1.25} \caption{\myfont \textbf{Table S9 $|$ Mixed effects model measuring the difference in error rates between internal and boundary transitions within clusters.} We find a significant 0.2\% increase in errors ($n = 173$) for boundary transitions relative to internal transitions within communities (grey). The significance column indicates $p$-values less than 0.001 ($***$), less than 0.01 ($**$), and less than 0.05 ($*$).}}
\end{figure}

\subsection{Modular effect}
\label{error_modular}

Finally, we investigate the effect of the average perceived information (or cross entropy) of a network, while still controlling for produced information. Specifically, we compare subjects' reaction times in the modular network with reaction times in random $k$-4 networks, noting that the average produced information (or entropy) is identical across all graphs considered. We employ a mixed effects model of the form `$Error\sim \log(Trial)*Stage + Target + Recency + Network\_Type + (1 + \log(Trial)*Stage + Recency \,|\, ID)$', where $Network\_Type$ indicates the type of network (either modular of random $k$-4). Although we find a 0.2\% increase in errors for random $k$-4 networks relative to the modular network, this difference is not significant (Table S10).

\addtocounter{figure}{-1}
\begin{figure}[t!]
\centering
{\fontsize{12}{12} \myfont
\begin{tabular}{|c|c|c|c|c|}
\hline
Effect & Estimate & t-value & Pr$(>$$|$t$|)$ & Significance \\
\hline
\hline
(Intercept) & $0.038 \pm 0.010$ & $3.83$ & $< 0.001$ & $***$ \\
\hline
log(Trial) & $-0.004\pm 0.001$ & $-2.49$ & $0.013$ & $*$ \\
\hline
Stage & $-0.034 \pm 0.011$ & $-2.96$ & $0.003$ & $**$ \\
\hline
Recency & $0.000 \pm 0.000$ & $9.98$ & $< 0.001$ & $***$ \\
\hline
\rowcolor{LightGrey}
Network$\_$Type & $0.002 \pm 0.002$ & $0.96$ & $0.329$ & \\
\hline
log(Trial):Stage & $0.005 \pm 0.002$ & $2.82$ & $0.005$ & $**$ \\
\hline
\end{tabular}}
\vskip 12pt
\raggedright
\captionsetup{labelformat=empty}
{\spacing{1.25} \caption{\myfont \textbf{Table S10 $|$ Mixed effects model measuring the difference in error rates between the modular network and random $k$-4 networks.} We do not find a significant difference in error rates ($n = 84$) between the modular network and random $k$-4 networks (grey). The significance column indicates $p$-values less than 0.001 ($***$), less than 0.01 ($**$), and less than 0.05 ($*$).}}
\end{figure}

\section{Modular effect on learning rate}
\label{rate}

In the previous two sections, we investigated the effects of network structure on human reaction times and error rates, without considering the learning dynamics. Here we study the effect of network structure on learning rate, or how quickly subjects' reaction times decrease for a given increase in the number of trials. Specifically, we seek to determine which type of network is \textit{faster} to learn: the modular network (Fig. S\ref{network_effects}a) or the random $k$-4 networks (Fig. 1h). To do so, we estimate a mixed effects model of the form `$RT\sim \log(Trial)*Stage + \log(Trial)*Network\_Type + Target + Recency + (1 + \log(Trial)*Stage + Recency \,|\, ID)$'. We note that the only difference between this model and that used in Supplementary Sec. \ref{modular_effect} is the interaction term between $\log(Trial)$ and $Network\_Type$, which tells us how the network type impacts the effect of $\log(Trial)$ on reaction times (or the learning rate). The estimated model is summarized in Table S11, reporting a significant 9 ms increase in reaction times for each $e$-fold increase in $Trial$ for the random $k$-4 networks relative to the modular network. Intuitively, this means that the learning rate is faster (that is, reaction times decrease more for each increase in $Trial$) for the modular network than for the $k$-4 networks.

\addtocounter{figure}{-1}
\begin{figure}[h]
\centering
{\fontsize{12}{12} \myfont
\begin{tabular}{|c|c|c|c|c|}
\hline
Effect & Estimate (ms) & t-value & Pr$(>$$|$t$|)$ & Significance \\
\hline
\hline
(Intercept) & $1222.6 \pm 50.9$ & $24.00$ & $< 0.001$ & $***$ \\
\hline
log(Trial) & $-76.0\pm 5.3$ & $-14.21$ & $< 0.001$ & $***$ \\
\hline
Stage & $-401.1 \pm 36.6$ & $-10.95$ & $< 0.001$ & $***$ \\
\hline
Recency & $1.7 \pm 0.1$ & $19.65$ & $< 0.001$ & $***$ \\
\hline
Network$\_$Type & $-35.9 \pm 30.9$ & $-1.16$ & $0.245$ & \\
\hline
log(Trial):Stage & $48.4 \pm 5.1$ & $9.58$ & $< 0.001$ & $***$ \\
\hline
\rowcolor{LightGrey}
log(Trial):Network$\_$Type & $8.8 \pm 4.4$ & $1.98$ & $0.048$ & $*$ \\
\hline
\end{tabular}}
\vskip 12pt
\raggedright
\captionsetup{labelformat=empty}
{\spacing{1.25} \caption{\myfont \textbf{Table S11 $|$ Mixed effects model measuring the difference in learning rates between the modular network and random $k$-4 networks.} For each $e$-fold increase in the number of trials, we find a significant 9 ms increase in reaction times ($n = 84$) for random $k$-4 networks relative to the modular network (grey). The significance column indicates $p$-values less than 0.001 ($***$), less than 0.01 ($**$), and less than 0.05 ($*$).}}
\end{figure}

\section{Individual differences in network effects}
\label{individual_effects}

\begin{figure}[t]
\centering
\includegraphics[width = \textwidth]{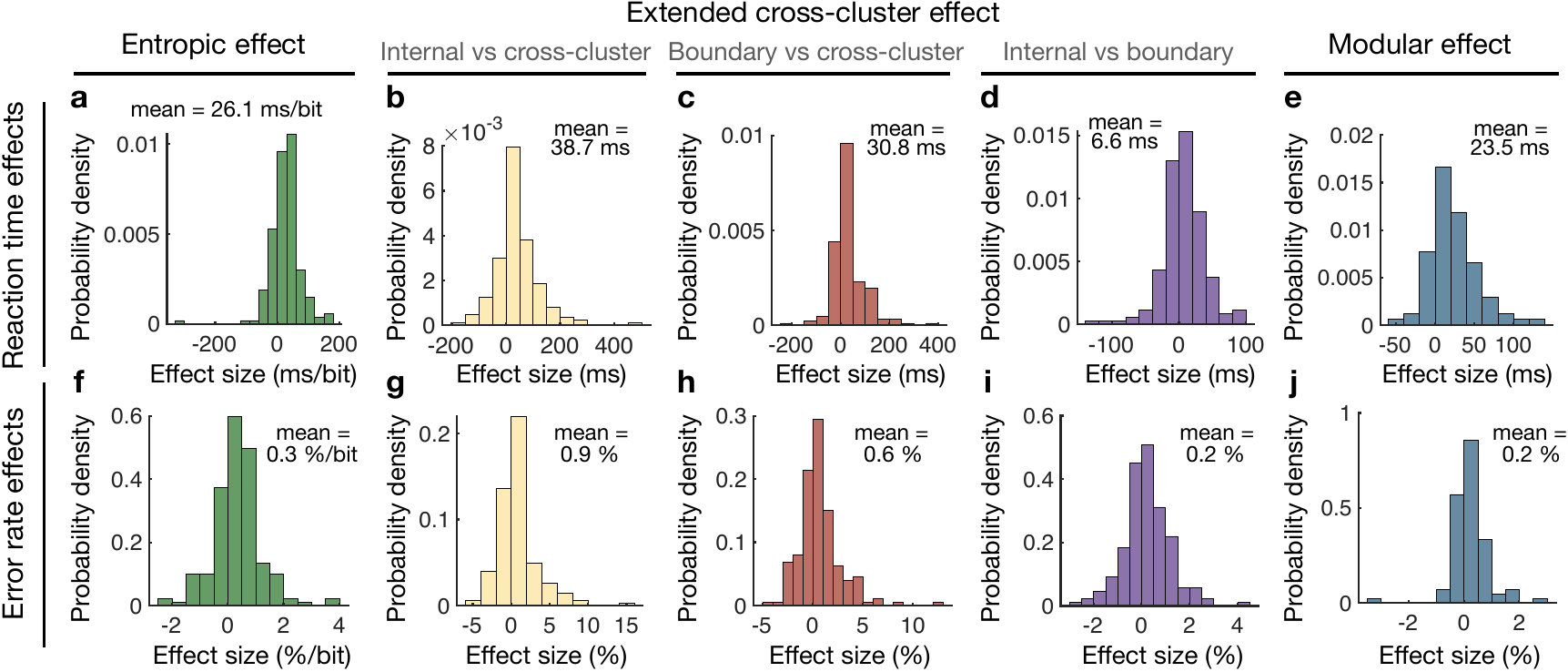} \\
\raggedright
\myfont \textbf{Fig. S\ref{individual} $|$ Distributions of network effects over individual subjects.}
\captionsetup{labelformat=empty}
{\spacing{1.25} \caption{\label{individual} \myfont \textbf{a}-\textbf{e}, Distributions over subjects of the different reaction time effects: the entropic effect ($n=177$), or the increase in reaction times for increasing produced information (\textbf{a}); the extended cross-cluster effects ($n = 173$), or the difference in reaction times between internal and cross-cluster transitions (\textbf{b}), between boundary and cross-cluster transitions (\textbf{c}), and between internal and boundary transitions (\textbf{d}) in the modular graph; and the modular effect ($n = 84$), or the difference in reaction times between the modular network and random $k$-4 networks (\textbf{e}). \textbf{f}-\textbf{j}, Distributions over subjects of the different effects on error rates: the entropic effect (\textbf{f}), the extended cross-cluster effects (\textbf{g}-\textbf{i}), and the modular effect (\textbf{j}).}}
\end{figure}

\begin{figure}[t]
\centering
\includegraphics[width = \textwidth]{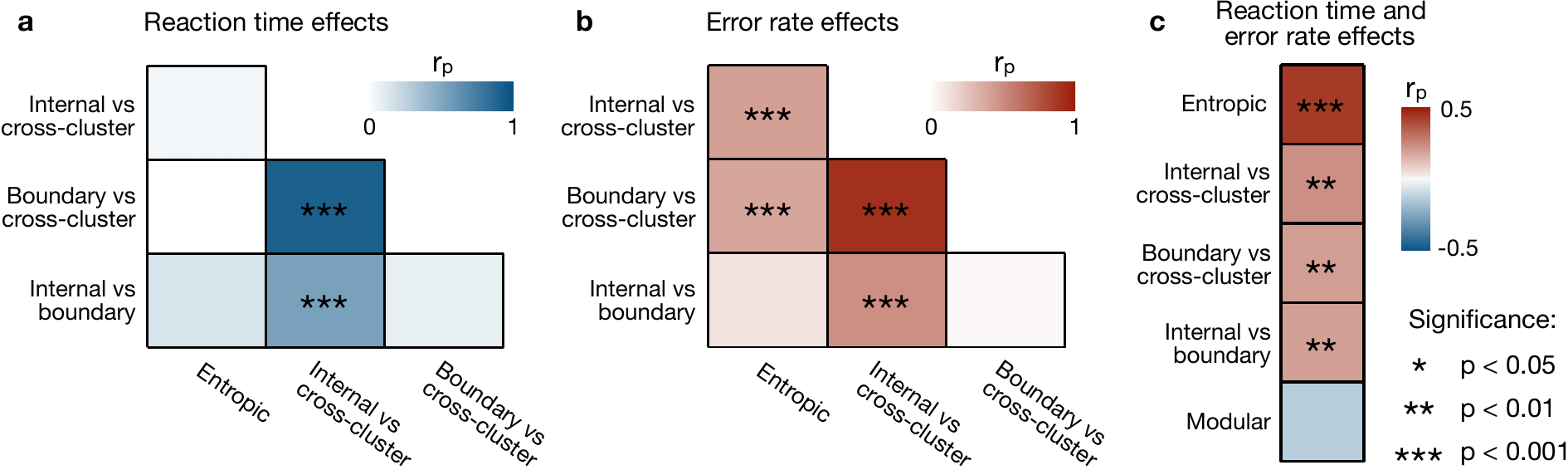} \\
\raggedright
\myfont \textbf{Fig. S\ref{correlations} $|$ Correlations between different network effects across subjects.}
\captionsetup{labelformat=empty}
{\spacing{1.25} \caption{\label{correlations} \myfont \textbf{a}, Pearson correlations between the entropic and extended cross-cluster effects on reaction times. \textbf{b}, Pearson correlations between the entropic and extended cross-cluster effects on error rates. In \textbf{a} and \textbf{b}, the modular effects on reaction times and error rates are not shown because they were measured in a different population of subjects. \textbf{c}, For each network effect, we show the Pearson correlation between the corresponding reaction time effect and error rate effect. Statistically significant correlations are indicated by $p$-values less than 0.001 ($***$), less than 0.01 ($**$), and less than 0.05 ($*$).}}
\end{figure}

In the previous three sections, we have discussed the \textit{fixed} effects of network structure on human behavior, which do not vary from person to person. However, for each network effect, we also find a significant amount of variation between individuals. Specifically, for each of the reaction time effects in Supplementary Sec. \ref{reaction_time} and error rate effects in Supplementary Sec. \ref{errors}, we fit a mixed effects model that includes a \textit{random} (or \textit{mixed}) effect term that differs for each subject. In this way, we are able to estimate the effect size for each participant in our experiments. For all of the reaction time effects (Fig. S\ref{individual}a-e) and all of the error rate effects (Fig. S\ref{individual}f-j), we find a significant standard deviation in the distribution of network effects ($p < 0.05$), indicating that each network effect exhibits significant inter-subject variability. Moreover, we find that many of the network effects on reaction times and error rates are significantly correlated across subjects (Fig. S\ref{correlations}), indicating that they are likely to be driven by common underlying mechanisms.

To understand what might be driving these individual differences in behavior, it helps to recall our linear predictions of subjects' reaction times $\hat{r}(t) = r_0 - r_1 \log \hat{P}_{x_{t-1},x_t}$, where $\hat{r}(t)$ is the predicted reaction time on trial $t$ and $\hat{P}_{x_{t-1},x_t}$ is the model of human transition probability estimates, where $x_{t-1}$ and $x_t$ are the stimuli on trials $t-1$ and $t$ (see Supplementary Sec. \ref{exponential_model}). The predictions contain three parameters, which are estimated separately for each subject: the inaccuracy parameter $\eta$, which is included in $\hat{P}$ (Fig. S\ref{expectations}b), the intercept $r_0$ (Fig. S\ref{expectations}c), and the slope $r_1$ (Fig. S\ref{expectations}d). Among these three parameters, the inaccuracy $\eta$ has drawn the most attention in the literature, having been shown to correlate with working memory performance,\cite{Lynn-06} drive differences in behaviors in reinforcement learning tasks,\cite{Gershman-03}, and determine the time-scale of episodic memories in the temporal context model.\cite{Gershman-01}

Here, we consider the possible role of $\eta$ in driving the individual differences in behaviors observed in Fig. S\ref{individual}. We first note that we should not expect a monotonic relationship between $\eta$ and any of the extended cross cluster effects (Figs. S\ref{individual}b-d and g-i) or the modular effects (Figs. S\ref{individual}e and j). Indeed, all of these effects disappear in both the high- and low-$\eta$ limits (Fig. S\ref{network_effects}b,c); for low $\eta$, humans have exact representations of the transition network and there will be no difference in the estimated probabilities of different transitions in the modular network or any other $k$-4 network, while for high $\eta$, human estimates of the transition probabilities become completely disordered and, yet again, there is no difference in the estimated transition probabilities. However, for random networks with non-uniform degrees (Fig. 1d), as $\eta$ increases the estimate $\hat{P}$ of the transition network will become less accurate, and therefore the entropic effect (Fig. 1e) should become weaker. Indeed, we find a significant negative correlation between $\eta$ and the entropic effect on reaction times (Spearman correlation $r_s = -0.25$; $p < 0.001$); we note that we use the Spearman correlation coefficient because $\eta$ is far from normally distributed (Fig. S\ref{expectations}b). Together, these results demonstrate that there are individual differences in sensitivity to network structure (Fig. S\ref{individual}), and that these differences may be related to variations in the accuracy of people's estimates of transition networks.

\section{Real networks}
\label{real_networks}

In the main text, we show that real networks exhibit two consistent information properties: they have high entropy and low KL divergence from human expectations. When calculating the KL divergence, we use the model $\hat{P}$ defined in Eq. (\ref{model2}) with $\eta$ set to the average value from our human experiments (Fig. S\ref{expectations}b). Additionally, in order to draw on our analytical results (see Supplementary Secs. \ref{entropy} and \ref{KL_divergence}), we focused on undirected versions of the real networks. Here, we show that the central conclusions in the main text concerning the information properties of real networks are robust to variations in these choices. Specifically, we verify that the KL divergence of real networks remains low for different values of $\eta$ and different models for $\hat{P}$ altogether, and we confirm that the entropy remains high and the KL divergence remains low for directed versions of the real networks.

\subsection{Varying $\eta$}
\label{vary_eta}

\begin{figure}[t!]
\centering
\includegraphics[width = .7\textwidth]{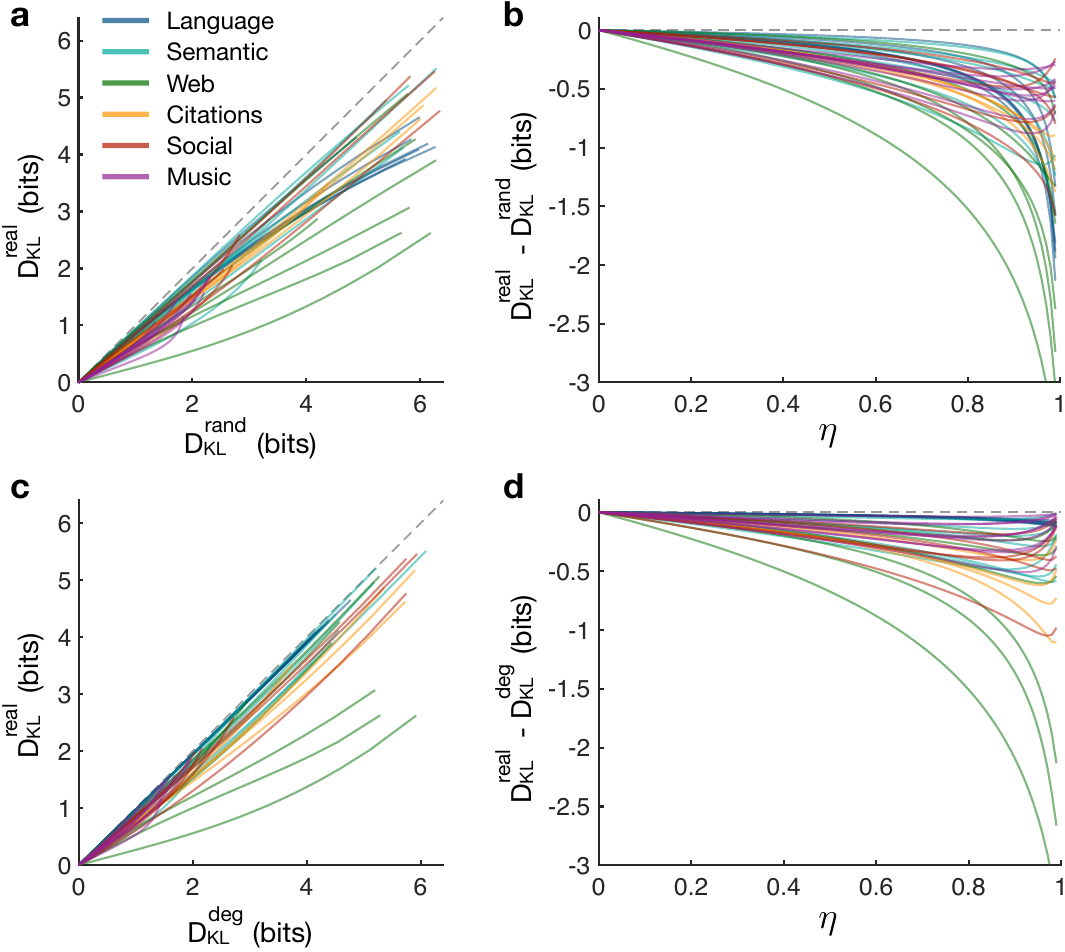} \\
\raggedright
\myfont\textbf{Fig. S\ref{DKL_eta} $|$ KL divergence of real networks for different values of $\eta$.}
\captionsetup{labelformat=empty}
{\spacing{1.25} \caption{\label{DKL_eta} \myfont \textbf{a}, KL divergence of fully randomized versions of the real networks listed in Table S12 ($D_{\text{KL}}^{\text{rand}}$) compared with the true value ($D_{\text{KL}}^{\text{real}}$) as $\eta$ varies from zero to one. Every real networks maintains lower KL divergence than the corresponding randomized network across all values of $\eta$. \textbf{b}, Difference between the KL divergence of real and fully randomized networks as a function of $\eta$. \textbf{c}, KL divergence of degree-preserving randomized versions of the real networks ($D_{\text{KL}}^{\text{deg}}$) compared with $D_{\text{KL}}^{\text{real}}$ as $\eta$ varies from zero to one. The real networks display lower KL divergence than the degree-preserving randomized versions across all values of $\eta$. \textbf{d}, Difference between the KL divergence of real and degree-preserving randomized networks as a function of $\eta$. All networks are undirected, and each line is calculated using one randomization of the corresponding real network.}}
\end{figure}

We first investigate how the KL divergence varies as a function of the inaccuracy parameter $\eta$. To recall, the KL divergence, defined in Eq. (\ref{DKL}), represents the inefficiency due to a person's expectations $\hat{P}$. We consider the model of expectations used in the main text, $\hat{P} = (1-\eta)P(I-\eta P)^{-1}$, while varying the parameter $\eta$ between zero and one. We find that all of the real networks considered maintain a lower KL divergence than fully randomized versions of the networks across all values of $\eta$ (Fig. S\ref{DKL_eta}a). In the limit $\eta\rightarrow 0$, the KL divergence of both real and randomized networks tends toward zero (Fig. S\ref{DKL_eta}a), as expected. As $\eta$ increases, the difference in efficiency between the real and fully randomized networks grows (Fig. S\ref{DKL_eta}b). We also generate randomized versions of the real networks that maintain identical entropies by preserving the degree distribution. Even when compared against random networks with the same entropy, all of the real networks attain lower KL divergence across all values of $\eta$ (Fig. S\ref{DKL_eta}c). Just as for the fully randomized networks, the difference in efficiency between real and entropy-preserving random networks grows as $\eta$ increases (Fig. S\ref{DKL_eta}d). These results confirm that our conclusions in the main text are robust to variations in the inaccuracy parameter $\eta$.

\subsection{Different internal representations}
\label{diff_representations}

\begin{figure}[t!]
\centering
\includegraphics[width = .7\textwidth]{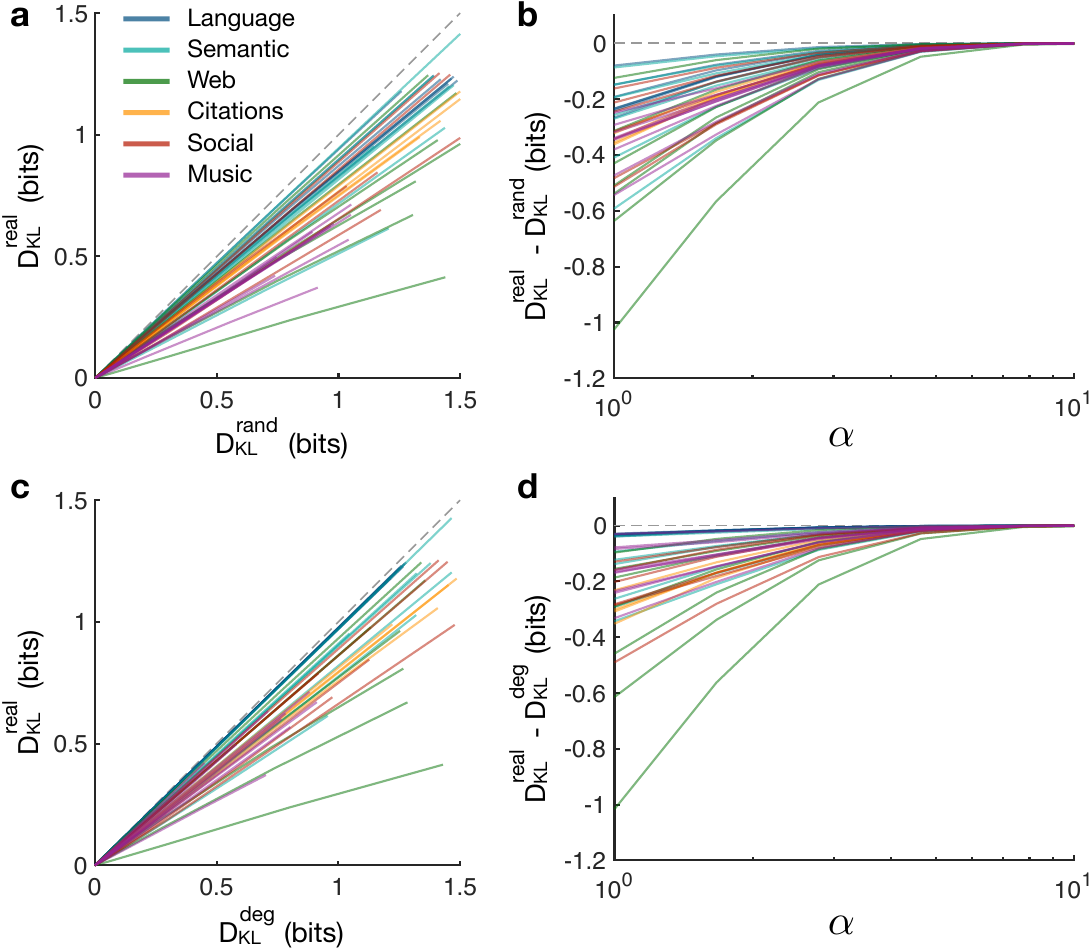} \\
\raggedright
\myfont\textbf{Fig. S\ref{DKL_power_law} $|$ KL divergence of real networks under the power-law model of human expectations.}
\captionsetup{labelformat=empty}
{\spacing{1.25} \caption{\label{DKL_power_law} \myfont \textbf{a}, KL divergence of fully randomized versions of the real networks listed in Table S12 ($D_{\text{KL}}^{\text{rand}}$) compared with the true value ($D_{\text{KL}}^{\text{real}}$). Expectations $\hat{P}$ are defined as in Eq. (\ref{model1}) with $f(t) = (t+1)^{-\alpha}$, and we allow $\alpha$ to vary between 1 and 10. The real networks maintain lower KL divergence than the randomized network across all values of $\alpha$. \textbf{b}, Difference between the KL divergence of real and fully randomized networks as a function of $\alpha$. \textbf{c}, KL divergence of degree-preserving randomized versions of the real networks ($D_{\text{KL}}^{\text{deg}}$) compared with $D_{\text{KL}}^{\text{real}}$ as $\alpha$ varies from 1 to 10. The real networks display lower KL divergence than the degree-preserving randomized versions across all values of $\alpha$. \textbf{d}, Difference between the KL divergence of real and degree-preserving randomized networks as a function of $\alpha$. All networks are undirected, and each line is calculated using one randomization of the corresponding real network.}}
\end{figure}

Here, we study the KL divergence for different models of the human expectations $\hat{P}$. First, we consider the power-law model, defined by Eq. (\ref{model1}) with integration function $f(t) = (t+1)^{-\alpha}$, where $\alpha\in (1,\infty)$ is the single parameter. Varying $\alpha$ between 1 and 10, we find that all of the real networks display lower KL divergence than fully randomized versions for all values of $\alpha$ (Fig. S\ref{DKL_power_law}a). Moreover, this difference in efficiency grows as $\alpha$ decreases (Fig. S\ref{DKL_power_law}b); that is, the difference in KL divergence increases as the expectations $\hat{P}$ integrate over longer time scales, which is analogous to $\eta$ increasing. Even when compared with random versions that preserve the entropy, the real networks still exhibit lower KL divergence across all values of $\alpha$ (Fig. S\ref{DKL_power_law}c,d).

\begin{figure}[t!]
\centering
\includegraphics[width = .7\textwidth]{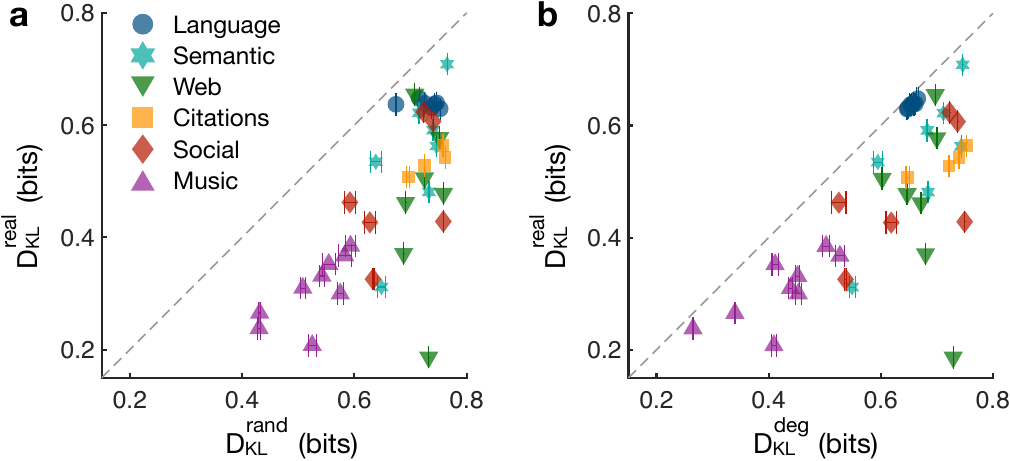} \\
\raggedright
\myfont\textbf{Fig. S\ref{DKL_factorial} $|$ KL divergence of real networks under the factorial model of human expectations.}
\captionsetup{labelformat=empty}
{\spacing{1.25} \caption{\label{DKL_factorial} \myfont \textbf{a}, KL divergence of fully randomized versions of the real networks listed in Table S12 ($D_{\text{KL}}^{\text{rand}}$) compared with the exact value ($D_{\text{KL}}^{\text{real}}$). Expectations $\hat{P}$ are defined as in Eq. (\ref{model1}) with $f(t) =1/t!$. \textbf{b}, KL divergence of degree-preserving randomized versions of the real networks ($D_{\text{KL}}^{\text{deg}}$) compared with $D_{\text{KL}}^{\text{real}}$. In both cases, the real networks maintain lower KL divergence than the randomized versions. Data points and error bars (standard deviations) are estimated from 10 realizations of the randomized networks.}}
\end{figure}

Second, we consider the factorial model for $\hat{P}$, defined by Eq. (\ref{model1}) with integration function $f(t) = 1/t!$. As discussed in Supplementary Sec. \ref{temporal_integration}, this model takes the analytic form $\hat{P} = e^{-1}Pe^P$, where $e^P$ is the matrix exponential, which is closely related to the communicability of $P$.\cite{Estrada-01, Girvan-01} Calculating the KL divergence, we find qualitatively the same results as for the previous two models. Namely, when compared against both fully randomized and entropy-preserving (i.e., degree-preserving) randomized versions, all of the real networks studied maintain a lower KL divergence (Fig. S\ref{DKL_factorial}). Taken together, the results of this and the previous subsections indicate that the low KL divergence observed in real networks is robust to different choices for the specific model of human expectations.

\subsection{Directed networks}
\label{directed_networks}

\begin{figure}[t!]
\centering
\includegraphics[width = .7\textwidth]{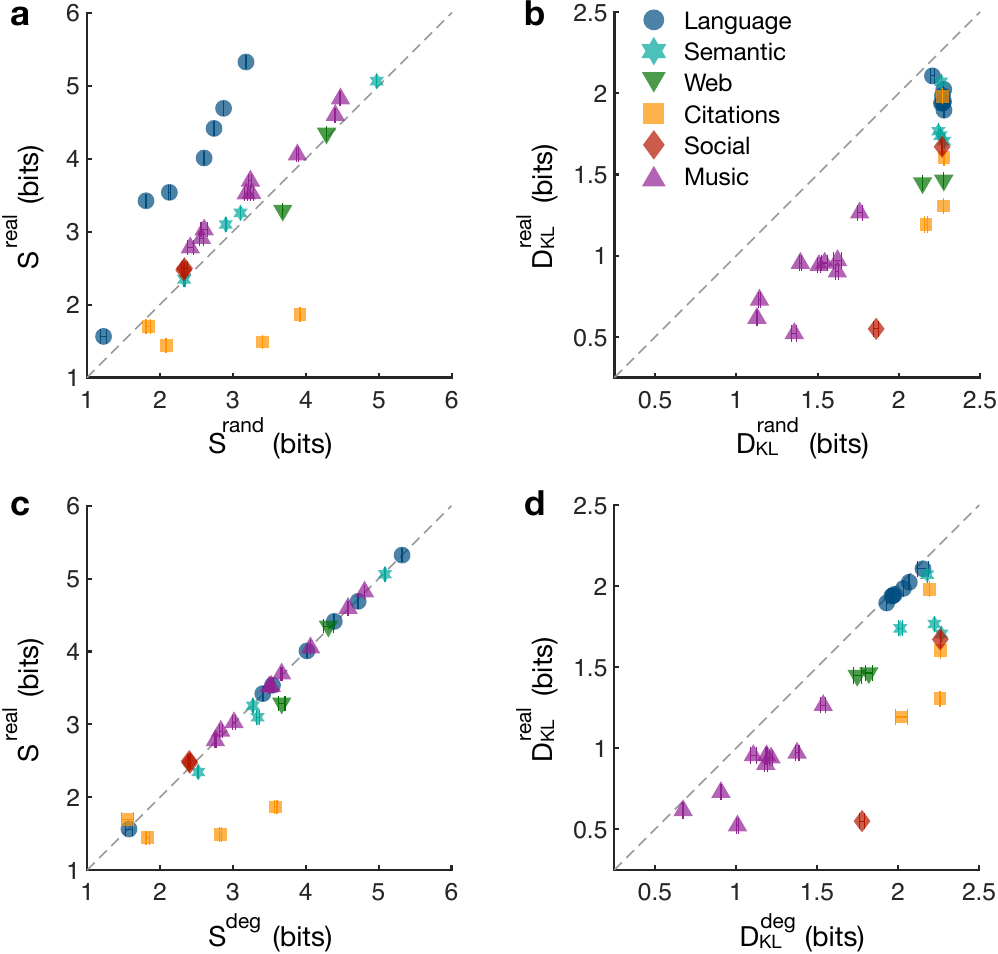} \\
\raggedright
\myfont\textbf{Fig. S\ref{directed} $|$ Entropy and KL divergence of directed versions of real networks.}
\captionsetup{labelformat=empty}
{\spacing{1.25} \caption{\label{directed} \myfont \textbf{a}, Entropy of directed versions of the real networks listed in Table S12 ($S^{\text{real}}$) compared with fully randomized versions ($S^{\text{rand}}$). Entropy is calculated directly from Eq. (\ref{S}) with the stationary distribution $\bm{\pi}$ calculated numerically. \textbf{b} KL divergence of directed versions of the real networks ($D_{\text{KL}}^{\text{real}}$) compared with fully randomized versions ($D_{\text{KL}}^{\text{rand}}$). Expectations $\hat{P}$ are defined as in Eq. (\ref{model2}) with $\eta$ set to the average value $0.80$ from our human experiments. \textbf{c}, Entropy of randomized versions of directed real networks with in- and out-degrees preserved ($S^{\text{deg}}$) compared with $S^{\text{real}}$. \textbf{d}, KL divergence of degree-preserving randomized versions of directed real networks ($D_{\text{KL}}^{\text{deg}}$) compared with $D_{\text{KL}}^{\text{real}}$. Data points and error bars (standard deviations) are estimated from 100 realizations of the randomized networks.}}
\end{figure}

We now consider directed versions of the real networks. Among the 40 networks chosen for analysis, 28 have directed versions (see Table S12). Analysis of directed networks follows in much the same way as our previous analysis of undirected networks; the only difference is that, when computing the entropy (Eq. \ref{S}) and KL divergence (Eq. \ref{DKL}), we calculate the stationary distribution $\bm{\pi}$ numerically by solving the eigenvector equation $\bm{\pi}^{\intercal} = \bm{\pi}^{\intercal}P$. We find that most of the directed real networks have higher entropy than completely randomized versions (Fig. S\ref{directed}a); the main exceptions are the citation networks, which we discuss in further detail in Supplementary Sec. \ref{inefficient}. We also find that all of the directed real networks have lower KL divergence than completely randomized versions (Fig. S\ref{directed}b), where the expectations $\hat{P}$ are calculated using the model in Eq. (\ref{model2})

If we instead compare against randomized versions that preserve both the in- and out-degrees of nodes, we see that the entropy of real networks remains relatively unchanged (Fig. S\ref{directed}c); again, the citation networks as a group represent the strongest exception to this result. Even when compared with degree-preserving randomized versions, all of the directed real networks attain a lower KL divergence (Fig. S\ref{directed}d). Generally, these results demonstrate that our conclusions regarding the information properties of real networks also apply to directed networks: (i) their entropy is higher than completely randomized versions and is primarily driven by the degree distribution, and (ii) their KL divergence is lower than both completely randomized and degree-preserving randomized versions.

\section{Temporally evolving networks}

In the main text, we studied the information properties of static communication networks. However, many of these networks are inherently temporal in nature, evolving over time to arrive at the final form that we observe today.\cite{Holme-01} This observation raises a number of interesting questions: How does the temporal nature of communication networks affect their ability to communicate information? Moreover, do communication networks evolve over time to optimize efficient communication?

To answer these questions, we consider temporally evolving versions of the real networks studied in the main text. Among the 40 networks chosen for analysis, 19 have temporal versions (see Table S12), including all of the language (noun transition) and music (note transition) networks, as well as the Facebook network and the two arXiv citation networks. For each network, we record a sequence of up to 100 subnetworks representing different snapshots in the network's evolution. For example, in the language and music networks, each subnetwork represents the transitions between nouns or notes up to a given point in the text or musical piece. Similarly, each subnetwork for the Facebook and citation networks defines the social relationships or scientific citations at a given point in the growth of the corresponding network.

\begin{figure}[t!]
\centering
\includegraphics[width = .75\textwidth]{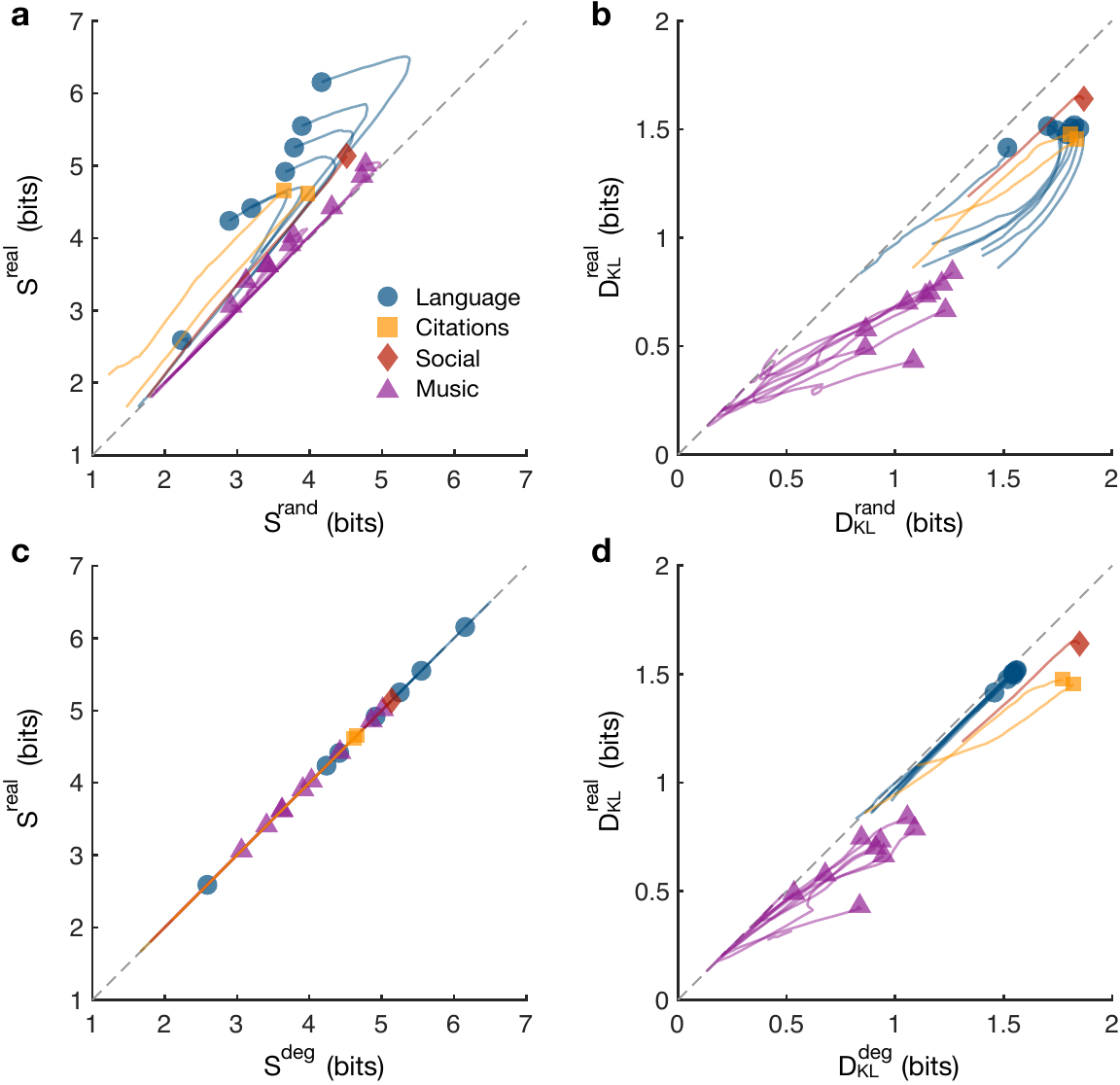} \\
\raggedright
\myfont\textbf{Fig. S\ref{grow} $|$ Entropy and KL divergence of temporally evolving versions of real networks.}
\captionsetup{labelformat=empty}
{\spacing{1.25} \caption{\label{grow} \myfont \textbf{a}, Entropy of temporally evolving versions of the real networks listed in Table S12 ($S^{\text{real}}$) compared with fully randomized versions ($S^{\text{rand}}$). Each line represents a sequence of growing networks and each symbol represents the final version of the network. \textbf{b}, KL divergence of evolving versions of the real networks ($D_{\text{KL}}^{\text{real}}$) compared with fully randomized versions ($D_{\text{KL}}^{\text{rand}}$). Expectations $\hat{P}$ are defined as in Eq. (\ref{model2}) with $\eta$ set to the average value $0.80$ from our human experiments. \textbf{c}, Entropy of temporally evolving versions of real networks ($S^{\text{real}}$) compared with degree-preserving randomized versions ($S^{\text{deg}}$). \textbf{d}, KL divergence of temporally evolving versions of real networks ($D_{\text{KL}}^{\text{real}}$) compared with degree-preserving randomized versions ($D_{\text{KL}}^{\text{deg}}$). Across all panels, each point along the lines represents an average over five realizations of the randomized networks.}}
\end{figure}

We find that the communication networks maintain higher entropy (Fig. S\ref{grow}a) and lower KL divergence (Fig. S\ref{grow}b) than completely randomized versions along almost the entirety of their evolutionary processes. Additionally, when compared against degree-preserving randomized versions, we find that the temporally evolving networks have the same entropy (Fig. S\ref{grow}c), as expected, and still maintain lower KL divergence along nearly the entire growth process (Fig. S\ref{grow}d). These results indicate that, even from the earliest stages in their development, real communication networks are organized to communicate large amounts of information (having high entropy) and to do so efficiently (having low KL divergence from human expectations).

\begin{figure}[t!]
\centering
\includegraphics[width = .8\textwidth]{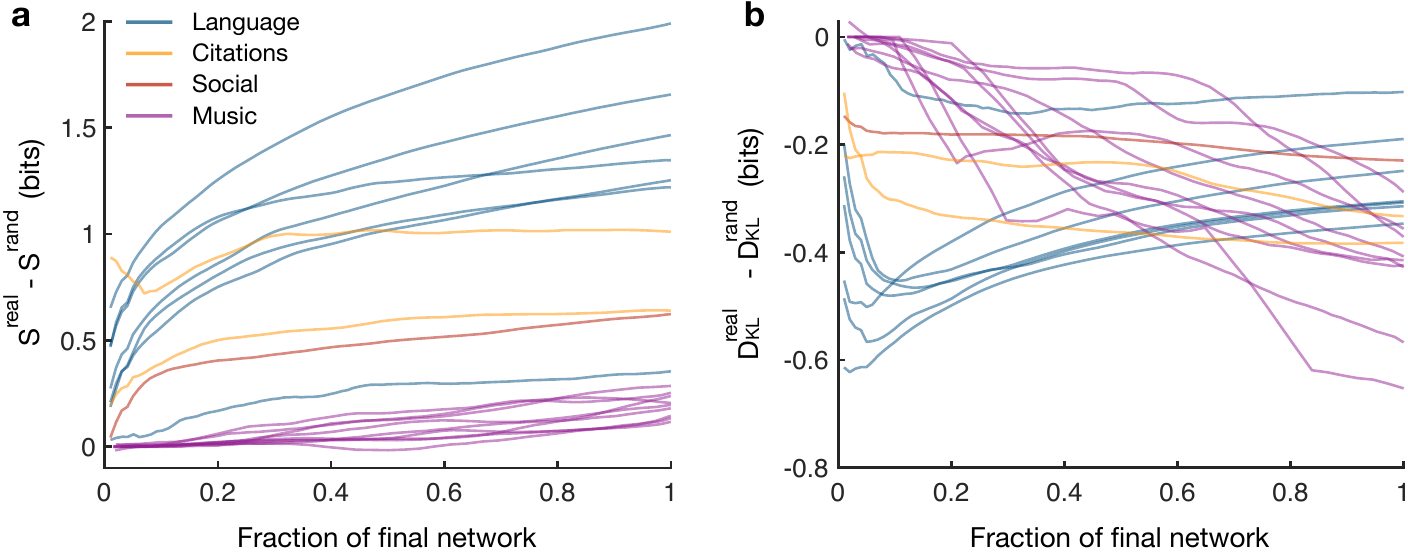} \\
\raggedright
\myfont\textbf{Fig. S\ref{grow_size} $|$ Evolution of the difference in entropy and KL divergence between real networks and randomized versions.}
\captionsetup{labelformat=empty}
{\spacing{1.25} \caption{\label{grow_size} \myfont \textbf{a}, Difference between the entropy of temporally evolving real networks ($S^{\text{real}}$) and the entropy of fully randomized versions of the same networks ($S^{\text{rand}}$) plotted as a function of the fraction of the final network size. Each line represents a sequence of growing networks that culminates in one of the communication networks studied in the main text. \textbf{b}, Difference between the KL divergence of temporally evolving real networks ($D_{\text{KL}}^{\text{real}}$) and that of fully randomized versions ($D_{\text{KL}}^{\text{rand}}$) plotted as a function of the fraction of the final network size. When calculating the KL divergences, the expectations $\hat{P}$ are defined as in Eq. (\ref{model2}) with $\eta$ set to the average value $0.80$ from our human experiments. Across both panels, each point along the lines represents an average over five realizations of the randomized networks.}}
\end{figure}

Yet, it remains unclear whether networks evolve over time to optimize efficient communication. To answer this question, we first investigate how the difference between the entropy of real networks and that of fully randomized versions changes over the course of a network's evolution (Fig. S\ref{grow_size}a). Interestingly, across all of the networks considered, we find that this difference in information production increases nearly monotonically as the networks grow, indicating that real communication networks evolve over time to transmit larger and larger amounts of information. Second, we study how the difference between the KL divergence of real networks and that of completely randomized versions changes over the evolution of a network (Fig. S\ref{grow_size}b). Notably, the music, social, and citation networks all evolve over time to minimize this difference, thereby becoming more efficient. However, language networks display a markedly different trajectory, minimizing their KL divergence (relative to randomized versions) until about 10\% of the way into their development, and then slowly growing to become less efficient. This pattern indicates that transitions between nouns communicate information most efficiently at the beginning of a text, and then become less efficient (while communicating larger amounts of information) as the text progresses. Together, these results suggest that communication networks evolve to (i) maximize the amount of information being communicated and (ii), with the exception of language networks, minimize the inefficiency of their communication.

\section{Real networks that do not support efficient communication}
\label{inefficient}

One of the central results of the paper is that real communication networks tend to have two properties: (i) high entropy and (ii) low KL divergence from human expectations. Specifically, these results tend to hold relative to fully randomized and degree-preserving randomized versions of the networks. However, it is useful to consider instances when these general results break down; that is, examples of real communication networks that either have low entropy or high KL divergence. Such examples are important for two reasons: First, they illustrate that efficient communication (defined by high entropy and low KL divergence) is not a necessary property of all real-world communication networks; and second, studying their properties reveals how efficient communication can break down. In what follows, we present two examples of real communication networks that do not support the efficient communication of information, either by having low entropy (low information production) or high KL divergence from human expectations (high inefficiency).

\subsection{Directed citation networks}

\begin{figure}[t!]
\centering
\includegraphics[width = .7\textwidth]{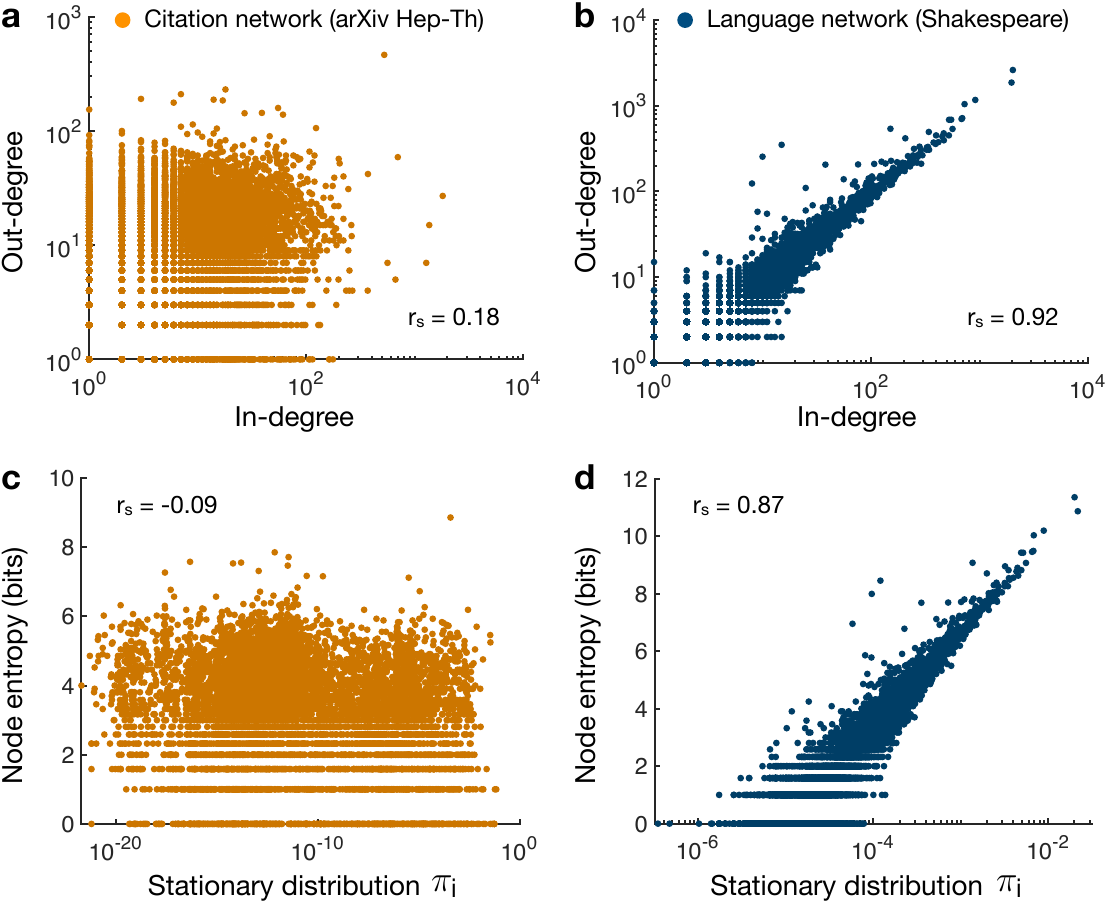} \\
\raggedright
\myfont\textbf{Fig. S\ref{citation} $|$ Comparison of directed citation and language networks.}
\captionsetup{labelformat=empty}
{\spacing{1.25} \caption{\label{citation} \myfont \textbf{a}, Out-degrees $k_i^{\text{out}} = \sum_j G_{ij}$ of nodes in the arXiv Hep-Th citation network compared with the in-degrees $k_i^{\text{in}} = \sum_j G_{ji}$ of the same nodes; we find a weak Spearman's correlation of $r_s = 0.18$. \textbf{b}, Out-degrees compared with in-degrees of nodes in the Shakespeare language (noun transition) network; we find a strong correlation $r_s = 0.92$. \textbf{c}, Entries in the stationary distribution $\pi_i$ for different nodes in the citation network compared with the node-level entropy $S_i$; we find a weakly negative correlation $r_s = -0.09$. \textbf{d}, Entries in the stationary distribution compared with node-level entropies in the language network; we find a strong correlation $r_s = 0.87$.}}
\end{figure}

First, we consider directed versions of the citation networks studied in the main paper. In the Supplementary Sec. \ref{directed_networks}, we found that the directed versions of citation networks have lower entropy than both fully randomized and degree-preserving randomized versions (Fig. S\ref{directed}a,c), contradicting our general observation that real communication networks have high entropy. Here we show that this contradiction stems from the inherently temporal nature of citation networks; namely, the fact that directed edges tend to flow backwards in time as more recent papers cite older papers. This temporal feature causes newer papers to have a lower in-degree than older papers, thereby disrupting the natural correlation between in- and out-degree in other real networks. For example, we see in the arXiv Hep-Th citation network that the in- and out-degrees are only weakly correlated (Fig. S\ref{citation}a), while for the Shakespeare language network, the in- and out-degrees are tightly correlated (Fig. S\ref{citation}b).

Since the in-degree of a node $i$ roughly corresponds to the frequency with which random walks visit $i$, we can think of the in-degrees $\bm{k}^{\text{in}}$ as approximately determining the stationary distribution $\bm{\pi}$. By contrast, the node-level entropy $S_i = -\sum_j P_{ij} \log P_{ij}$ is determined by the out-degree of node $i$, since $P_{ij} = \frac{1}{k_i^{\text{out}}} G_{ij}$ from Eq. (\ref{Pij}). Since the network-averaged entropy is simply an inner product of the stationary distribution and the node-level entropy, $S = \sum_i\pi_i S_i$, this quantity is maximized in networks for which $\pi_i$ and $S_i$ are correlated. Returning to our previous examples, we find that the stationary distribution and node-level entropy are weakly negatively correlated in the citation network (Fig. S\ref{citation}c), whereas in the language network, the stationary distribution and node entropy are tightly correlated (Fig. S\ref{citation}d). Thus, the apparent contradiction between directed citation networks and our general result that real networks have high entropy is primarily driven by the temporal nature of directed edges in citation networks. Indeed, if one instead allows random walks to flow along either direction of each edge, as in the undirected versions studied in the main text, we find that citation networks do have high entropy (Fig. 2a). Therefore, the capacity of citation networks to communicate large amounts of information depends critically on the ability of walks to hop both forward and backward along citations.

\subsection{Language networks including all parts of speech}

\begin{figure}[t!]
\centering
\includegraphics[width = .8\textwidth]{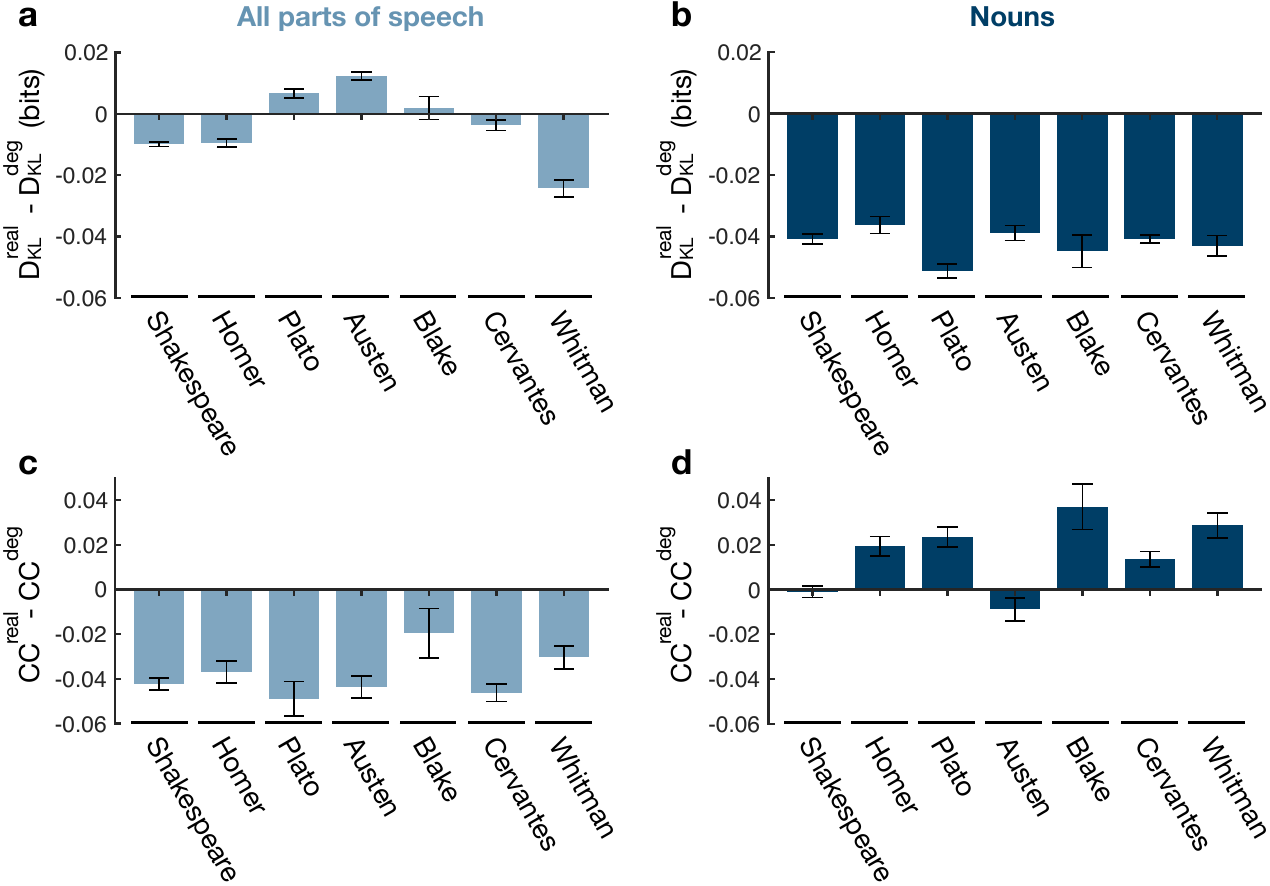} \\
\raggedright
\myfont\textbf{Fig. S\ref{language} $|$ Comparison of all-word transition networks and noun transition networks.}
\captionsetup{labelformat=empty}
{\spacing{1.25} \caption{\label{language} \myfont \textbf{a}-\textbf{b}, Difference between the KL divergence of language (word transition) networks ($D_{\text{KL}}^{\text{real}}$) and degree-preserving randomized versions of the same networks ($D_{\text{KL}}^{\text{deg}}$). We consider networks of transitions between all words (\textbf{a}) and networks of transitions between nouns (\textbf{b}). \textbf{c}-\textbf{d}, Difference between the average clustering coefficient of language networks ($CC^{\text{real}}$) and degree-preserving randomized versions of the same networks ($CC^{\text{deg}}$), where transitions are considered between all words (\textbf{c}) or only nouns (\textbf{d}). In all panels, data points and error bars (standard deviations) are estimated from 100 realizations of the randomized networks, and the networks are undirected.}}
\end{figure}

In the main text, we focus on language networks consisting of the transitions between nouns in a given text. This choice to focus on nouns follows from existing literature that distinguishes ``content" words (such as nouns), which contain meaning, from ``grammatical" words (such as articles, conjunctions, and prepositions), which define the structure of a sentence.\cite{Milo-01, Foster-01}. If we instead consider language (word transition) networks that include all parts of speech, we find that these all-word transition networks have both higher entropy and lower KL divergence than fully randomized versions, aligning with the results from the main text (Fig. 3a,b). However, when compared to degree-preserving randomized versions, we find that the all-word transition networks have nearly the same KL divergence (Fig. S\ref{language}a), with three of the seven networks exhibiting KL divergences that are either higher or statistically indistinguishable from the degree-preserving randomized versions. By contrast, the networks of transitions between nouns studied in the main text all exhibit lower KL divergence than degree-preserving randomized versions (Fig. S\ref{language}b).

From the analytic and numerical results presented in the main text (Fig. 4e-h) and Supplementary Sec. \ref{KL_divergence}, we know that decreases in KL divergence are largely driven by increases in clustering. Indeed, for the all-word transition networks, we find that the average clustering coefficients are consistently lower than for degree-preserving randomized versions (Fig. S\ref{language}c), thereby explaining their relatively high KL divergences (Fig. S\ref{language}a). To understand the low clustering (and therefore the high KL divergence) of the all-word transition networks, it is helpful to consider the fact that words typically transition from content words to grammatical words in order to maintain grammatical structure. This hopping between content and grammatical words yields transition networks with disassortative community structure,\cite{Milo-01, Foster-01} wherein words from the same class are less likely to form edges than words in different classes, which, in turn, decreases the clustering. By contrast, if we restrict our attention to content words (such as the nouns studied in the main text), we find that the transition networks exhibit high clustering (Fig. S\ref{language}d) and therefore low KL divergence (Fig. S\ref{language}b).

\section{Entropy of random walks}
\label{entropy}

Given the high entropy and low KL divergence from human expectations observed in real networks, it is natural to wonder what topological features give rise to these properties. We note that there has been a large amount of recent research studying maximum entropy random walks, wherein the topology of the network is fixed but the edge weights are tuned to maximize the entropy rate.\cite{Demetrius-01, Burda-01, Sinatra-01, Coghi-01} By contrast, here we are interested in understanding how, for fixed edge weights, different network topologies either increase or decrease the entropy of random walks.

To make analytic progress, we focus on unweighted, undirected networks. In this case, Eq. (\ref{S(k)}) shows that the entropy is determined by the degree sequence of the network. If we consider a random network ensemble with node degrees independently distributed according to a degree distribution $\mathcal{P}(k)$, then the average entropy rate is given by\cite{Gomez-03}
\begin{equation}
\label{S_avg}
\begin{aligned}
\langle  S\rangle  &= \frac{1}{2E}\sum_i\langle  k_i \log k_i \rangle  \\
&= \frac{\langle  k \log k\rangle }{\langle k\rangle },
\end{aligned}
\end{equation}
where the averages are taken over $\mathcal{P}(k)$.

\subsection{High-degree expansion}

Since $k\log k$ is convex in $k$, it is clear that $\langle k\log k\rangle  \ge \langle k\rangle \log \langle k\rangle $, and we arrive at a simple lower bound for the entropy,
\begin{equation}
\label{lower}
\langle S\rangle  \ge \log\langle k\rangle .
\end{equation}
In fact, one can show that $\log\langle k\rangle $ is the zeroth-order term in an expansion of $\langle S\rangle $ in the limit of large average degree $\langle k\rangle  \gg 1$. Expanding $k\log k$ around $\langle k\rangle $, we find
\begin{equation}
\label{high_deg}
\begin{aligned}
\langle S\rangle  &= \frac{1}{\langle k\rangle } \langle \langle k\rangle \log \langle k\rangle  + \left(1 + \log \langle k\rangle \right)\left(k - \langle k\rangle \right) + \frac{\left(k - \langle k\rangle \right)^2}{2\langle k\rangle } + O\left(\frac{1}{\langle k\rangle ^2}\right)\rangle  \\
&= \log \langle k\rangle  + \frac{\text{Var}(k)}{2\langle k\rangle ^2} + O\left(\frac{1}{\langle k\rangle ^3}\right),
\end{aligned}
\end{equation}
where $\text{Var}(k)$ is the variance of $k$. We therefore find that, in addition to increasing logarithmically with the average degree, the entropy of random walks grows with increasing degree variance. In turn, this result further supports the conclusion that networks with heterogeneous degrees produce random walks with higher entropy. In what follows, we derive analytic results for the entropy of random walks on various canonical network families.

\subsection{$k$-regular network}

We begin by studying $k$-regular networks, wherein each node $i$ has constant degree $k_i = k$. In this case, we arrive at the simple relation $\langle S\rangle  = \log k$, which saturates the lower bound in Eq. (\ref{lower}).\cite{Cover-01} This result shows that $k$-regular networks achieve the lowest possible entropy among networks of a given density.

\subsection{Poisson distributed network}

\begin{figure}
\centering
\includegraphics[width = \textwidth]{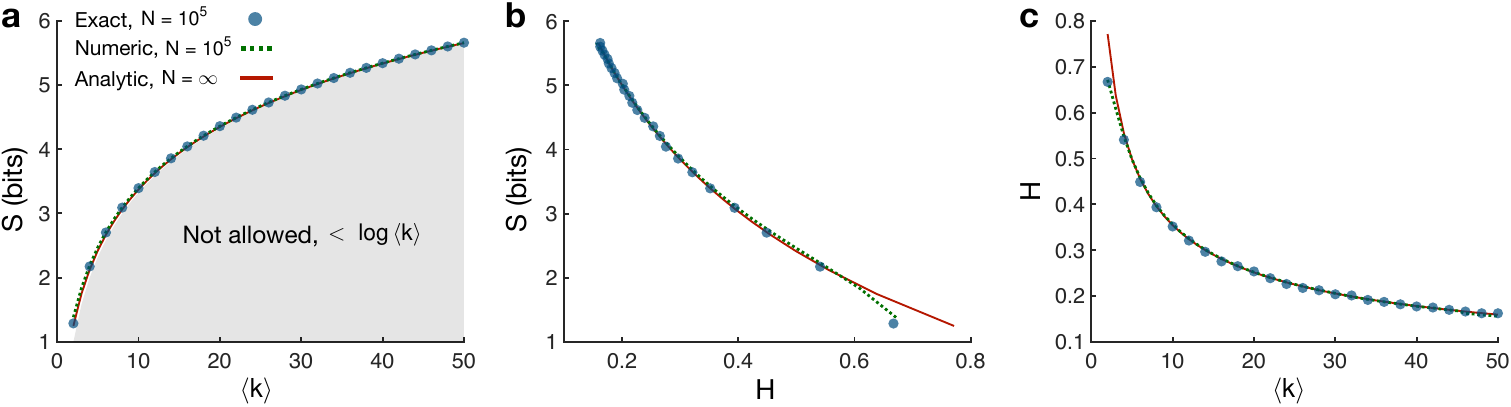} \\
\raggedright
\myfont\textbf{Fig. S\ref{entropy_poisson} $|$ Entropy of random walks in Poisson distributed networks.}
\captionsetup{labelformat=empty}
{\spacing{1.25} \caption{\label{entropy_poisson} \myfont \textbf{a}, Entropy of random walks as a function of the average degree $\langle k\rangle $ for Poisson distributed networks. Data points are exact calculations using the degree sequences of randomly-generated Erd\"{o}s-R\'{e}nyi networks of size $N = 10^4$. Dashed lines are numerical results for $N=10^4$, calculated using the Poisson degree distribution. Solid lines are analytic results for the thermodynamic limit $N\rightarrow\infty$. \textbf{b}, Entropy as a function of the degree heterogeneity $H$ for variable $\langle k\rangle $. \textbf{c}, Degree heterogeneity as a function of the average degree.}}
\end{figure}

While $k$-regular networks maintain a lattice-like structure, many real networks display random organization.\cite{Albert-02} The simplest model for generating random networks, known as the Erd\"{o}s-R\'{e}nyi model,\cite{Erdos-01} places $E$ edges uniformly at random between pairs of $N$ nodes. In the thermodynamic limit $N\rightarrow \infty$, Erd\"{o}s-R\'{e}nyi networks follow a Poisson degree distribution $\mathcal{P}(k) = e^{-\langle k\rangle } \langle k\rangle ^k/k!$. In this case, the degree variance is given by $\text{Var}(k) = \langle k\rangle $, and applying Eq. (\ref{high_deg}), we find that
\begin{equation}
\label{S_poisson}
\langle S\rangle  = \log \langle k\rangle  + \frac{1}{2\langle k\rangle } + O\left(\frac{1}{\langle k\rangle ^2}\right).
\end{equation}
Therefore, in the high-$\langle k\rangle $ limit, the entropy of random walks on an Erd\"{o}s-R\'{e}nyi network approaches the lower-bound $\langle S\rangle  \approx \log \langle k\rangle $. We find that the analytic prediction in Eq. (\ref{S_poisson}) accurately approximates the true entropy of randomly-generated Erd\"{o}s-R\'{e}nyi networks across all values of the average degree (Fig. S\ref{entropy_poisson}a).

To investigate the relationship between the entropy and the heterogeneity of degrees in a network, we defined the degree heterogeneity to be the relative average difference in degrees,
\begin{equation}
H = \frac{\langle |k_i - k_j|\rangle }{\langle k\rangle } = \frac{1}{\langle k\rangle }\sum_{k_i,k_j}|k_i-k_j|\mathcal{P}(k_i)\mathcal{P}(k_j).
\end{equation}
$H$ is a well-studied measure of the dispersion of a distribution, with range $[0,2]$. We note that other often used measures of degree heterogeneity, such as $\langle k^2\rangle /\langle k\rangle ^2$ and $\text{Var}(k)/\langle k\rangle ^2$, cannot be used to study the impact of degree heterogeneity on entropy for scale-free networks since $\langle k^2\rangle $ diverges for $\gamma \le 3$ in the limit $N\rightarrow\infty$. For Poisson distributed networks, one can show that
\begin{equation}
H = 2e^{-2\langle k\rangle }\big(I_0(2\langle k\rangle ) + I_1(2\langle k\rangle )\big),
\end{equation}
where $I_{\nu}(x)$ is the modified Bessel function of the first kind.\cite{Liu-01} For other degree distributions, however, it is generally difficult to derive an analytic form for $H$. We find that the entropy of random walks on Poisson distributed networks decreases with increasing degree heterogeneity as we vary $\langle k\rangle $ (Fig. S\ref{entropy_poisson}b), seemingly contradicting our conclusion in the main text that entropy increases with heterogeneity. However, this effect is driven by the monotonic decrease in $H$ with increasing $\langle k\rangle $ in Poisson distributed networks (Fig. S\ref{entropy_poisson}c). In the following subsections, we show that entropy does in fact increase with degree heterogeneity for other network models, confirming the results in the main text.

\subsection{Power-law distributed network}

Compared to random networks, real networks often contain a number of hub nodes with unusually high degree, leading to a heavy-tailed distribution of node degrees.\cite{Albert-02} Often this heavy-tailed distribution is associated with scale-free organization,\cite{Barabasi-01} which is characterized by a power-law degree distribution $\mathcal{P}(k) \sim k^{-\gamma}$, where $\gamma$ is the scale-free exponent. In the limit $N\rightarrow\infty$, we can approximate the averages in Eq. (\ref{S_avg}) as integrals, and one can show that\cite{Gomez-03}
\begin{equation}
\label{S_PL}
\langle S\rangle  = \frac{1}{\gamma - 2}.
\end{equation}
We see that the entropy diverges as $\gamma\rightarrow 2$, while for $\gamma > 2$ the entropy of scale-free networks is well-defined. We remark that this critical exponent is different from $\gamma = 3$, which is the critical exponent for many other network phenomena driven by the divergence of $\langle k^2\rangle $.\cite{Pastor-01,Albert-01} Instead, as $\gamma\rightarrow 2$, super-hubs emerge that connect to almost all of the nodes in the network, causing the average degree $\langle k\rangle $ to diverge.\cite{Albert-02} Each time a random walk arrives at one of these super-hubs, the entropy of the ensuing transition, roughly $-\log\frac{1}{N}$, diverges as $N\rightarrow\infty$.

\begin{figure}
\centering
\includegraphics[width = \textwidth]{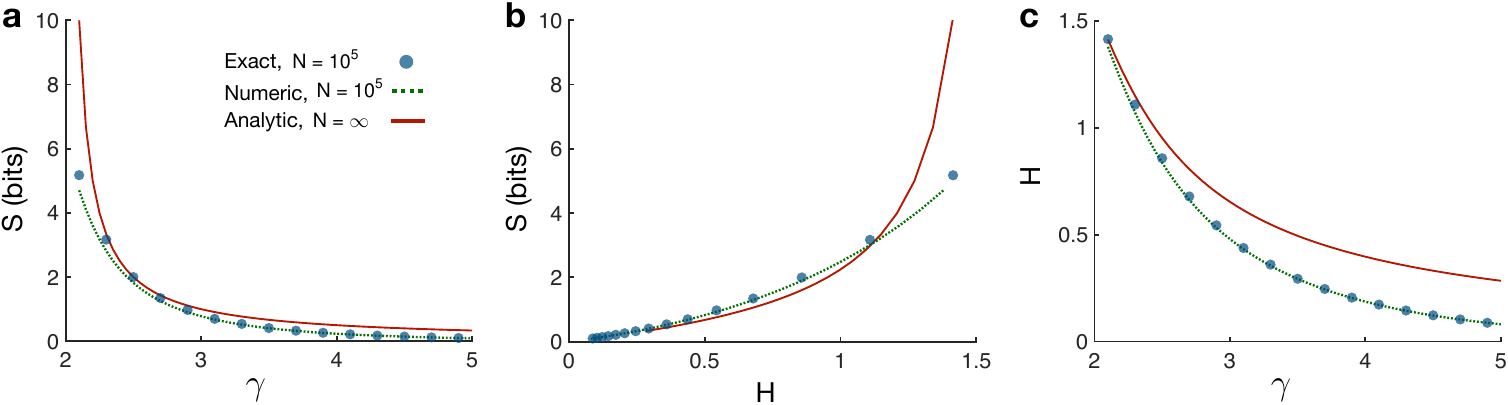} \\
\raggedright
\myfont\textbf{Fig. S\ref{entropy_PL} $|$ Entropy of random walks in power-law distributed networks.}
\captionsetup{labelformat=empty}
{\spacing{1.25} \caption{\label{entropy_PL} \myfont \textbf{a}, Entropy of random walks as a function of the scale-free exponent $\gamma$ for power-law distributed networks. Data points are exact calculations from networks of size $N = 10^4$ generated using the configuration model.\cite{Molloy-01} Dashed lines are numerical results for $N=10^4$, calculated using the power-law degree distribution. Solid lines are analytic results for the thermodynamic limit $N\rightarrow\infty$. \textbf{b}, Entropy as a function of the degree heterogeneity $H$ for variable $\gamma$. \textbf{c}, Degree heterogeneity as a function of the scale-free exponent.}}
\end{figure}

We compare the analytic prediction in Eq. (\ref{S_PL}) with exact calculations from both power-law distributed networks generated using the configuration model\cite{Molloy-01} and from numerical calculations of the averages in Eq. (\ref{S_avg}), finding that the numerical estimates agree well with the exact values (Fig. S\ref{entropy_PL}a). Moreover, we find that the entropy increases with degree heterogeneity as we sweep over $\gamma$ (Fig. S\ref{entropy_PL}b), confirming our conclusions in the main text. This increase in entropy is related to the corresponding increase in heterogeneity as $\gamma\rightarrow 2$ (Fig. S\ref{entropy_PL}c).

\subsection{Static model}

In order to test the effects of network density and degree heterogeneity independently, we turn to the static model, which is commonly used to generate scale-free networks of a given density.\cite{Goh-01} Beginning with $N$ disconnected nodes, we assign each node $i$ a weight $w_i = \, i^{-\alpha}$, where $\alpha \in [0,1)$ is a real number. Then, we randomly select a pair of nodes $i$ and $j$ with probabilities proportional to their weights, and we connect them if they have not already been connected. This process is repeated until $E = \frac{1}{2}N\langle k\rangle $ edges have been added. A number of analytic properties have been derived for the static model,\cite{Catanzaro-01,Lee-01} including the fact that, in the thermodynamic limit, the degree distribution is given by $\mathcal{P}(k) = \frac{1}{\alpha}\big(\frac{\langle k\rangle }{2}(1-\alpha)\big)^{1/\alpha} \frac{\Gamma\left(k-\frac{1}{\alpha},\, \frac{\langle k\rangle }{2}(1-\alpha)\right)}{\Gamma(k + 1)}$, where $\Gamma(\cdot)$ is the gamma function and $\Gamma(\cdot,\,\cdot)$ is the upper incomplete gamma function. In the large-$k$ limit, one can show that the degree distribution drops off as a power law $\mathcal{P}(k)\sim k^{-\gamma}$, where $\gamma = 1 + \frac{1}{\alpha}$.

We are interested in deriving an analytic form for the entropy. Using a hidden variables method,\cite{Catanzaro-01} one can show that the average degree of node $i$ is given by
\begin{equation}
\label{k_bar}
\bar{k}(i) = \langle k\rangle (1-\alpha)\left(\frac{i}{N}\right)^{-\alpha}\left(1 - N^{\alpha - 1}\right).
\end{equation}
Approximating the numerator in Eq. (\ref{S_avg}) by $\langle k\log k\rangle  \approx \frac{1}{N} \int_1^{N} \bar{k}(i)\log\bar{k}(i) \, di$, and taking the limit $N\rightarrow\infty$, we find that the entropy is given by
\begin{equation}
\label{S_SF}
\langle S\rangle  = \log\langle k\rangle  + \frac{1}{\gamma - 2} - \log\frac{\gamma - 1}{\gamma - 2}.
\end{equation}
We note that the average degree of a pure power-law network is $\langle k\rangle  = \frac{\gamma - 1}{\gamma - 2}$. Plugging this average degree into Eq. (\ref{S_SF}), we recover the entropy of power-law distributed networks in Eq. (\ref{S_PL}), as expected. Interestingly, we notice that, even for finite $\langle k\rangle $, the entropy in the static model diverges as $\gamma\rightarrow 2$ in the thermodynamic limit.

\begin{figure}[t!]
\centering
\includegraphics[width = \textwidth]{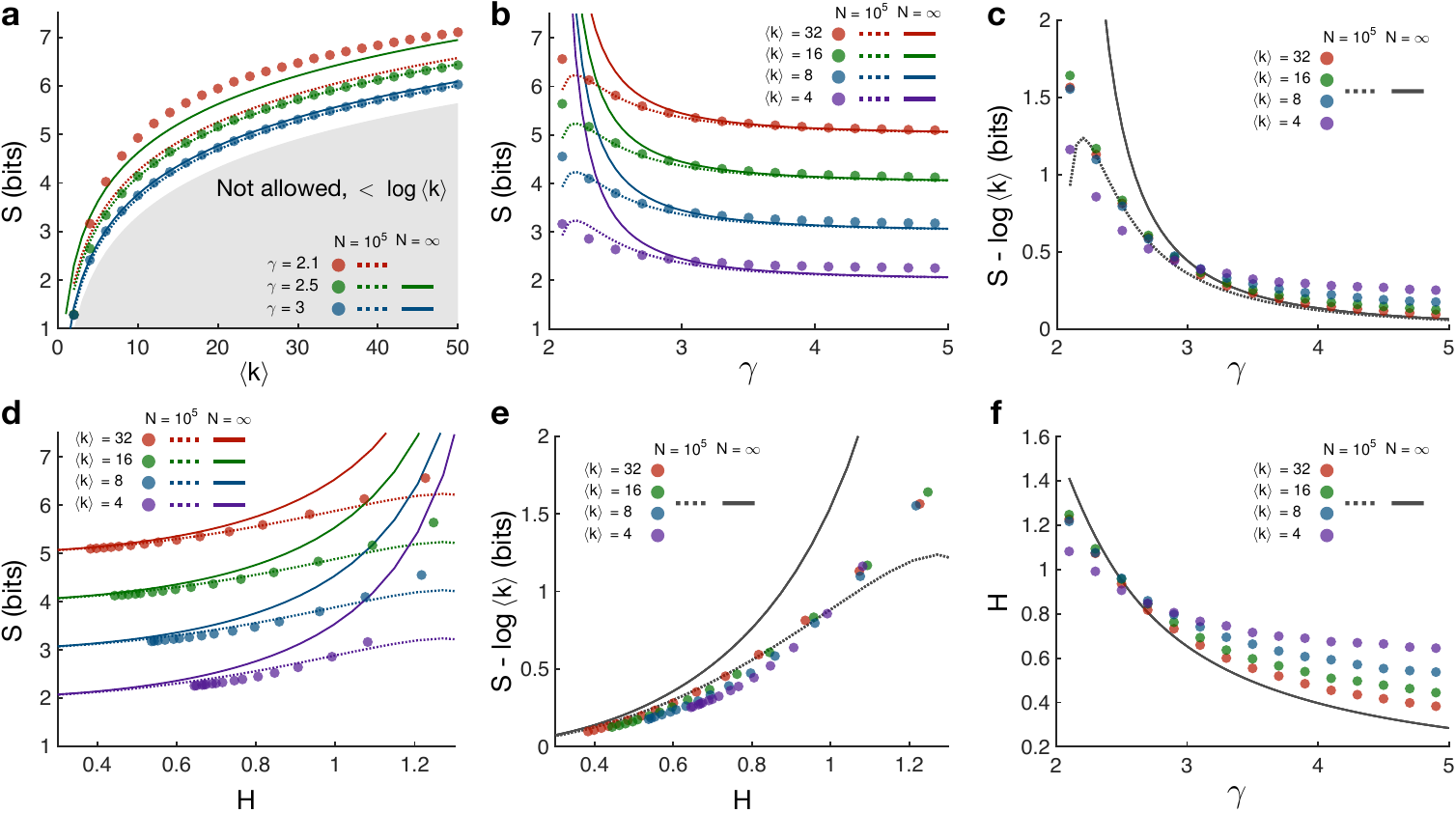} \\
\raggedright
\myfont\textbf{Fig. S\ref{entropy_SF} $|$ Entropy of random walks in static model networks.}
\captionsetup{labelformat=empty}
{\spacing{1.25} \caption{\label{entropy_SF} \myfont \textbf{a}, Entropy of random walks as a function of the average degree $\langle k\rangle $ for various values of the scale-free exponent $\gamma$ in the static model. Data points are exact calculations using the degree sequences of networks with $N = 10^4$ generated using the static model. Dashed lines are numerical results for $N=10^4$, calculated using the average degree relationship in Eq. (\ref{k_bar}). Solid lines are analytic results for the thermodynamic limit $N\rightarrow\infty$. \textbf{b}, Entropy as a function of $\gamma$ for various values of $\langle k\rangle $. \textbf{c}, The quantity $S - \log\langle k\rangle $ collapses to a single function of $\gamma$ across various values of $\langle k\rangle $. \textbf{d}, Entropy as a function of the degree heterogeneity $H$ for varying $\gamma$. \textbf{e}, The quantity $S - \log\langle k\rangle $ increases with $H$ for varying $\gamma$. \textbf{f}, Degree heterogeneity increases as $\gamma$ decreases toward the critical value $\gamma = 2$.}}
\end{figure}

We find that the entropy increases as $\langle k\rangle $ increases (Fig. S\ref{entropy_SF}a) and also as $\gamma$ decreases (Fig. S\ref{entropy_SF}b).
The thermodynamic result in Eq. (\ref{S_SF}) is accurate for $\gamma \ge 3$, while numerical calculations using Eq. (\ref{k_bar}) and including finite network size yield accurate predictions for $\gamma \ge 2.5$. We note that the only effect of $\langle k\rangle $ on the entropy in Eq. (\ref{S_SF}) is in the logarithmic lower bound, suggesting that the quantity $S - \log\langle k\rangle $ should depend exclusively on the scale-free exponent $\gamma$. Indeed, subtracting $\log\langle k\rangle $ from our entropy calculations, we find that networks of varying density collapse onto a single line (Fig. S\ref{entropy_SF}c). This result is made even more clear by considering how the quantity $S - \log \langle k\rangle $ varies with degree heterogeneity as we sweep over $\gamma$ (Fig. S\ref{entropy_SF}e). Finally, we note that $H$ increases with decreasing $\gamma$ (Fig. S\ref{entropy_SF}f), thereby explaining the monotonic relationship between entropy and degree heterogeneity in the static model (Fig. S\ref{entropy_SF}d).

\subsection{Exponentially distributed network}

\begin{figure}
\centering
\includegraphics[width = \textwidth]{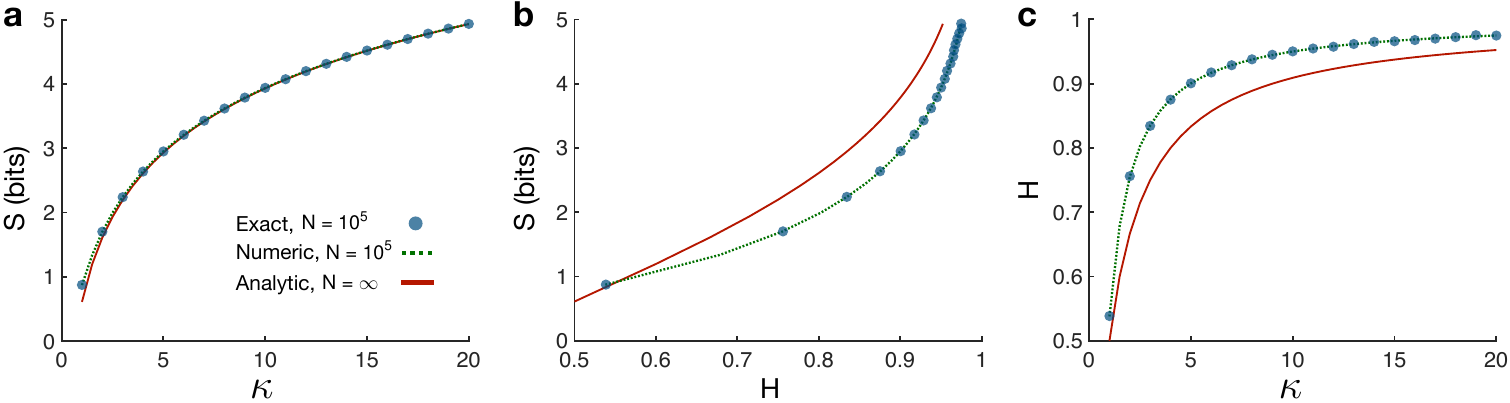} \\
\raggedright
\myfont\textbf{Fig. S\ref{entropy_exp} $|$ Entropy of random walks in exponentially distributed networks.}
\captionsetup{labelformat=empty}
{\spacing{1.25} \caption{\label{entropy_exp} \myfont \textbf{a}, Entropy of random walks as a function of the degree cutoff $\kappa$ for exponentially distributed networks. Data points are exact calculations from networks of size $N = 10^4$ generated using the configuration model.\cite{Molloy-01} Dashed lines are numerical results for $N=10^4$, calculated using the exponential degree distribution. Solid lines are analytic results for the thermodynamic limit $N\rightarrow\infty$. \textbf{b}, Entropy as a function of the degree heterogeneity for variable $\kappa$. \textbf{c}, Degree heterogeneity as a function of the exponential cutoff.}}
\end{figure}

Many real networks exhibit degree distributions with exponential cutoffs for large values of $k$.\cite{Newman-02,Albert-02} In pure exponentially distributed networks, the degree distribution follows the form $\mathcal{P}(k) \sim e^{-k/\kappa}$, where $\kappa\ge 0$ is the degree cutoff. In the thermodynamic limit, approximating the averages in Eq. (\ref{S_avg}) as integrals, we find that the entropy is given by
\begin{equation}
\langle S\rangle  = \log\langle k\rangle  + \frac{1 - \gamma_{\text{e}}}{\ln 2},
\end{equation}
where $\gamma_{\text{e}}$ is Euler's constant. We see in Fig. S\ref{entropy_exp}a that this analytic prediction accurately describes the entropy of randomly-generated exponential networks. Moreover, we find that the entropy increases with increasing degree heterogeneity (Fig. S\ref{entropy_exp}b) and that the heterogeneity increases with the degree cutoff $\kappa$ (Fig. S\ref{entropy_exp}c).

\section{KL divergence between random walks and human expectations}
\label{KL_divergence}

The results of the previous section demonstrate that, generally, the entropy of random walks is larger for networks with heterogeneous degrees, a feature that has been found in many real networks.\cite{Cancho-01, Barabasi-01, Barabasi-02, Newman-02} But what are the structural features that allow a network to maintain a low divergence from human expectations? Here, we answer this question by studying the KL divergence $D_{\text{KL}}(P||\hat{P})$ between a network's transition structure $P$ and the expectations of an observer $\hat{P}$.

\subsection{Upper bound}

For expectations $\hat{P}$ of the form in Eq. (\ref{model1}), the KL divergence is given by
\begin{equation}
\label{upper1}
\begin{aligned}
D_{\text{KL}}(P||\hat{P}) &= - \sum_i \pi_i \sum_j P_{ij} \log \frac{\hat{P}_{ij}}{P_{ij}} \\
&= - \sum_i \pi_i \sum_j P_{ij} \log \left( C \sum_{t = 0}^{\infty} f(t) \frac{(P^{t+1})_{ij}}{P_{ij}}\right),
\end{aligned}
\end{equation}
where $(P^{t+1})_{ij}/P_{ij}$ is the relative probability of transitioning from node $i$ to node $j$ in $t+1$ steps versus one step. Keeping only the first term inside the logarithm, we arrive at an upper bound for the KL divergence,
\begin{equation}
\label{upper2}
D_{\text{KL}}(P||\hat{P}) \le - \sum_i \pi_i \sum_j P_{ij} \log \left(Cf(0)\right) = -\log\left(Cf(0)\right).
\end{equation}
Eq. (\ref{upper2}) allows us to make a number of simple predictions for the KL divergence. For example, if the expectations are defined by $f(t) = \eta^t$, as presented in the main text, then $C = 1-\eta$ and so $D_{\text{KL}} \le -\log (1-\eta)$. In this case, we see that the KL divergence tends to zero as $\eta\rightarrow 0$ and that the upper bound diverges as $\eta\rightarrow 1$. In contrast, if $f(t) = (t+1)^{-\alpha}$ then $C = \zeta(\alpha)^{-1}$, where $\zeta(\cdot)$ is the Riemann zeta function, and we have $D_{\text{KL}} \le \log \zeta(\alpha)$. As a final example, if $f(t) = 1/ t!$ then $C = e^{-1}$, and so $D_{\text{KL}} \le \log e$ (which we remark is not equal to one since we use $\log$ base two).

\subsection{Relationship to clustering}

While Eq. (\ref{upper2}) provides a simple relationship between the KL divergence and parameters in the model for $\hat{P}$, we are ultimately interested in understanding the effects of network structure. To gain an intuition for the role of topology, it helps to focus on a particular model for the expectations. For example, considering $f(t) = \eta^t$, in the low-$\eta$ limit the KL divergence takes the form
\begin{equation}
\label{DKL_approx1}
\begin{aligned}
D_{\text{KL}}(P||\hat{P}) &= -\log (1-\eta) - \sum_i \pi_i \sum_j P_{ij} \log \left(1 +  \eta\frac{(P^2)_{ij}}{P_{ij}} + O(\eta^2)\right) \\
&= -\log (1-\eta) - \frac{\eta}{\ln 2}\sum_i \pi_i \sum_j P_{ij} \frac{(P^2)_{ij}}{P_{ij}} + O(\eta^2).
\end{aligned}
\end{equation}
We note that, when calculating information measures such as entropy or KL divergence, one only considers terms with non-zero probability, such that, for each node $i$, the sum on $j$ in Eq. (\ref{DKL_approx1}) implicitly runs over all nodes for which $P_{ij} = \frac{1}{k_i}G_{ij}$ is non-zero. Therefore, for undirected networks, recalling that $\pi_i = \frac{k_i}{2E}$, we have
\begin{equation}
\label{DKL_approx2}
D_{\text{KL}}(P||\hat{P}) = -\log (1-\eta) - \frac{\eta}{2E\ln 2}\sum_i k_i \sum_j G_{ij} \sum_{\ell} \left(\frac{1}{k_i}G_{i\ell}\right)\left(\frac{1}{k_{\ell}}G_{\ell j}\right) + O(\eta^2).
\end{equation}
Switching the $i$ and $\ell$ indices and canceling terms, we arrive at the concise approximation
\begin{equation}
D_{\text{KL}}(P||\hat{P}) = -\log (1-\eta) - \frac{\eta}{E\ln 2}\sum_i \frac{1}{k_i} \bigtriangleup_i + O(\eta^2),
\end{equation}
where $\bigtriangleup_i = (G^3)_{ii}/2$ is the number of (possibly weighted) triangles involving node $i$. We therefore find that the KL divergence is lower for networks with a lager number of triangles or, equivalently, a higher clustering coefficient. In the following subsections, we investigate the relationship between KL divergence and clustering in Erd\"{o}s-R\'{e}nyi and stochastic block networks.

\subsection{Erd\"{o}s-R\'{e}nyi network}

We wish to derive an analytic approximation for the KL divergence of an Erd\"{o}s-R\'{e}nyi network. Considering human expectations defined by $f(t) = \eta^t$, for undirected networks Eq. (\ref{upper1}) becomes
\begin{equation}
\begin{aligned}
D_{\text{KL}}(P||\hat{P}) &= -\sum_i \frac{k_i}{2E}\sum_j \frac{1}{k_i}G_{ij} \log\left((1-\eta)\sum_{t=0}^{\infty} \eta^t \frac{(P^{t+1})_{ij}}{P_{ij}}\right) \\
&= -\log(1-\eta) - \frac{1}{2E}\sum_{ij} G_{ij} \log\left(\sum_{t=0}^{\infty} \eta^t \frac{(P^{t+1})_{ij}}{P_{ij}}\right).
\end{aligned}
\end{equation}
We note that the second term above is an average of the logarithm over the edges in the network. Approximating this average of logarithms by a logarithm of the average, we have
\begin{equation}
\label{DKL_ER1}
D_{\text{KL}}(P||\hat{P}) \approx -\log(1-\eta) - \log\left[\frac{1}{2E}\sum_{ij} k_iG_{ij} \sum_{t=0}^{\infty} \eta^t (P^{t+1})_{ij}\right].
\end{equation}
For unweighted networks, we recognize that $\sum_j G_{ij} (P^{t+1})_{ij}$ is the probability of transitioning from node $i$ to one of $i$'s neighbors in $t+1$ steps. For $t=0$ this probability is one. For $t>0$, we consider two cases: (i) dense networks with high $\langle k\rangle $, and (ii) sparse networks with low $\langle k\rangle $.

\begin{figure}[t!]
\centering
\includegraphics[width = .67\textwidth]{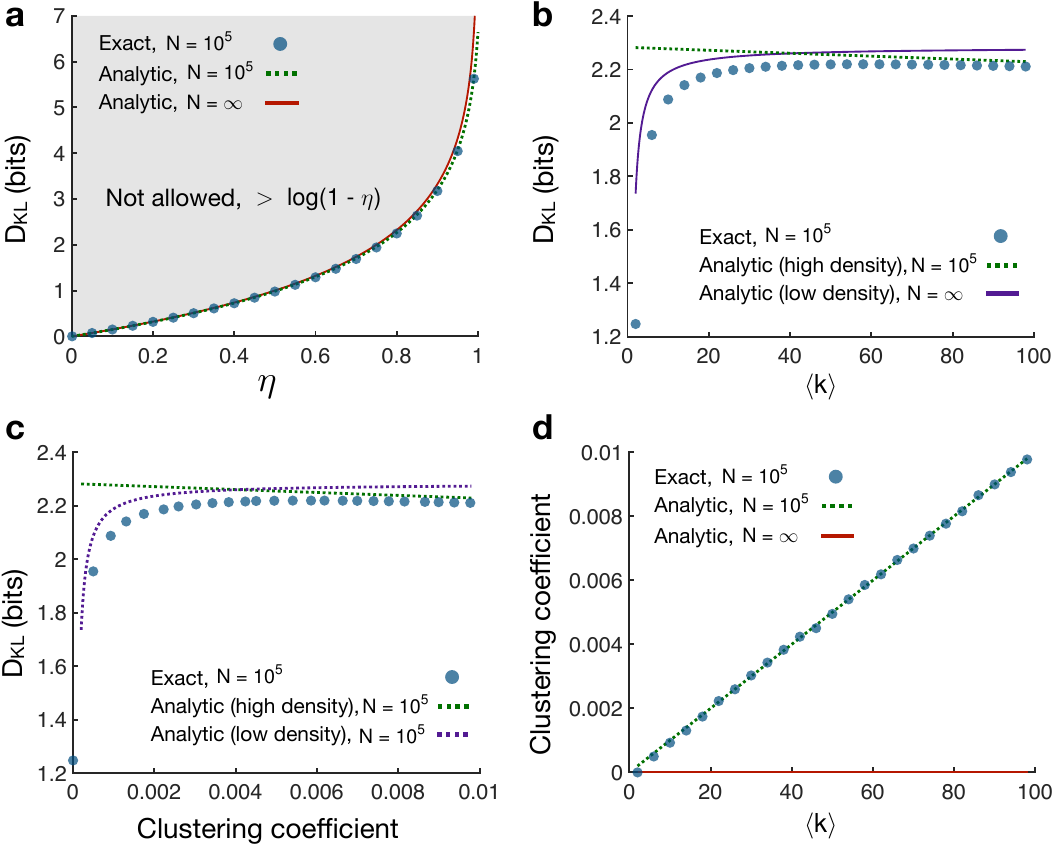} \\
\raggedright
\myfont\textbf{Fig. S\ref{DKL_ER} $|$ KL divergence from human expectations in Erd\"{o}s-R\'{e}nyi networks.}
\captionsetup{labelformat=empty}
{\spacing{1.25} \caption{\label{DKL_ER} \myfont \textbf{a}, KL divergence between random walks and human expectations as a function of the inaccuracy parameter $\eta$ for Erd\"{o}s-R\'{e}nyi networks. Data points are exact calculations for networks of size $N = 10^4$  with average degree $\langle k\rangle  = 100$. Dashed line is the analytic prediction using Eq. (\ref{DKL_ER3}) with $N=10^4$. Solid line is the analytic result for the thermodynamic limit $N\rightarrow\infty$. \textbf{b}, KL divergence as a function of the average degree $\langle k\rangle $ for $\eta$ equal to the value $0.80$ measured in the serial response experiments. Dashed line represents the high-density analytic approximation in Eq. (\ref{DKL_ER3}) with $N=10^4$, while the solid line is the low-density approximation in Eq. (\ref{DKL_ER4}). \textbf{c}, KL divergence as a function of the average clustering coefficient for variable $\langle k\rangle $. \textbf{d}, Average clustering coefficient as a function of $\langle k\rangle $. In the thermodynamic limit the clustering tends toward zero for all values of $\langle k\rangle $ (solid line).}}
\end{figure}

For dense networks, we approximate the probability of transitioning from node $i$ to one of node $i$'s neighbors in $t+1>1$ steps as $k_i/N$, the probability of randomly selecting one of the $k_i$ neighbors from all $N$ nodes. Plugging this approximation for $\sum_j G_{ij} (P^t)_{ij}$ into Eq. (\ref{DKL_ER1}), we have
\begin{equation}
\label{DKL_ER2}
\begin{aligned}
D_{\text{KL}}(P||\hat{P}) &\approx -\log(1-\eta) - \log\left[\frac{1}{2E} \sum_i k_i\left(1 + \sum_{t = 1}^{\infty} \eta^t \frac{k_i}{N}\right)\right] \\
&= -\log(1-\eta) - \log\left[1 + \frac{1}{2EN}\frac{\eta}{1- \eta}\sum_i k_i^2\right].
\end{aligned}
\end{equation}
We have now reduced the KL divergence to a function of the degree sequence $\bm{k}$. For large Erd\"{o}s-R\'{e}nyi networks, the node degrees follow a Poisson distribution, and, for large $\langle k\rangle $, we have $\langle k^2\rangle  \approx \langle k\rangle ^2$. Thus, the average KL divergence for a dense Erd\"{o}s-R\'{e}nyi network can be approximated by
\begin{equation}
\label{DKL_ER3}
\begin{aligned}
\langle D_{\text{KL}}\rangle  &\approx -\log(1-\eta) - \langle \log\left[1 + \frac{1}{2E}\frac{\eta}{1- \eta} k^2\right]\rangle  \\
&\approx -\log(1-\eta) - \log\left[1 + \frac{1}{2E}\frac{\eta}{1- \eta} \langle k^2\rangle \right] \\
&\approx -\log(1-\eta) - \log\left[1 + \frac{1}{2E}\frac{\eta}{1- \eta} \langle k\rangle ^2\right] \\
&= -\log\left[1 - \eta\left(1 - \frac{\langle k\rangle }{N}\right)\right],
\end{aligned}
\end{equation}
where the averages are taken over the degree distribution $\mathcal{P}(k)$. We find that this approximation accurately predicts the KL divergence of Erd\"{o}s-R\'{e}nyi networks as a function of the integration parameter $\eta$ (Fig. S\ref{DKL_ER}a). We also see that in the thermodynamic limit $N\rightarrow \infty$, $D_{\text{KL}}$ approaches the upper bound $-\log(1-\eta)$.

For sparse Erd\"{o}s-R\'{e}nyi networks, the number of loops is small and thus the network is locally treelike.\cite{Karrer-01} In a tree, the probability $\sum_j G_{ij} (P^t)_{ij}$ of transitioning from a given node $i$ to one of node $i$'s neighbors in $t+1$ steps is zero if $t$ is odd. For $t$ even, setting node $i$ to be the root of the tree, if we assume all nodes have the same degree $\langle k\rangle$, then the probability of moving down the tree on any given step is $1 - 1/\langle k\rangle $ and the probability of moving up the tree is $1/\langle k\rangle $. Approximating $1 - 1/\langle k\rangle  \approx 1$, then the probability of moving down the tree $t/2 + 1$ steps and back up the tree $t/2$ steps is roughly $1/\langle k\rangle ^{\frac{t}{2}}$. Plugging this expression into Eq. (\ref{DKL_ER1}), we have
\begin{equation}
\begin{aligned}
D_{\text{KL}}(P||\hat{P}) &\approx -\log(1-\eta) - \log\left[\frac{1}{2E} \sum_i k_i\sum_{t \text{ even}}\eta^t \langle k\rangle ^{-\frac{t}{2}}\right] \\
&= -\log(1-\eta) - \log\left[\frac{1}{2E} \sum_i k_i\sum_{t = 0}^{\infty} \eta^{2t} \langle k\rangle ^{-t}\right] \\
&= -\log(1-\eta) - \log\left[\frac{1}{2E} \frac{\langle k\rangle }{\langle k\rangle  - \eta^2}\sum_i k_i\right].
\end{aligned}
\end{equation}
Averaging over the Poisson degree distribution, we have
\begin{equation}
\label{DKL_ER4}
\begin{aligned}
\langle D_{\text{KL}}\rangle  &\approx -\log(1-\eta) - \left< \log\left[\frac{N}{2E} \frac{\langle k\rangle }{\langle k\rangle  - \eta^2} k\right]\right>  \\
&\approx -\log(1-\eta) - \log\left[\frac{\langle k\rangle }{\langle k\rangle  - \eta^2}\right].
\end{aligned}
\end{equation}
We find that the above approximation provides a decent estimate of the KL divergence for low $\langle k\rangle $, while the high-density approximation in Eq. (\ref{DKL_ER3}) accurately predicts the KL divergence for $\langle k\rangle  > 50$ (Fig. S\ref{DKL_ER}b).

In addition to the dependence of $D_{\text{KL}}$ on $\eta$ and $\langle k\rangle $, we are also interested in the effect of clustering. The clustering coefficient of a given node $i$ is the number of triangles $\bigtriangleup_i$ involving node $i$ divided by the number of possible triangles $\binom{k_i}{2} = k_i(k_i - 1)/2$. For Erd\"{o}s-R\'{e}nyi networks, averaging over all nodes $i$, the clustering coefficient is approximately $\langle k\rangle /N$. We find that, for small $\langle k\rangle $, the KL divergence increases with increasing clustering, while, for large $\langle k\rangle $, the KL divergence decreases (Fig. S\ref{DKL_ER}c). Given that the clustering is directly proportional to $\langle k\rangle $ in Erd\"{o}s-R\'{e}nyi networks (Fig. S\ref{DKL_ER}d), the effects of clustering on $D_{\text{KL}}$ are driven by the density of the network. To disambiguate the effects of clustering and density, in the following subsection, we study a stochastic block model in which these properties can be varied independently.

\subsection{Stochastic block network}

In order to test the effects of clustering on the KL divergence without the confounding impact of edge density, we consider the stochastic block model.\cite{Decelle-01} Specifically, the $N$ nodes are divided into $N/N_c$ communities of $N_c$ nodes each. Then, a prescribed fraction $f$ of the $E = \langle k\rangle  N/2$ edges are placed between pairs of nodes within the same community, and the remaining fraction $1-f$ of edges are placed between nodes in different communities.

We wish to understand the dependence of the KL divergence on the fraction $f$ of within-community edges. Beginning with Eq. (\ref{DKL_ER1}), we once again consider the probability $\sum_j G_{ij} (P^{t+1})_{ij}$ of transitioning from node $i$ to one of node $i$'s neighbors in $t+1$ steps. As before, for $t=0$ this probability is one. For $t > 0$, we approximate
\begin{equation}
\label{p_in}
\sum_j G_{ij} (P^{t+1})_{ij} \approx p^{\text{in}}(t+1) \frac{k_i^{\text{in}}}{N_c-1} + p^{\text{out}}(t+1)\frac{k_i^{\text{out}}}{N-N_c},
\end{equation}
where $p^{\text{in}}(t+1)$ is the probability of ending up in the same community as node $i$ after $t+1$ steps, $p^{\text{out}}(t+1)$ is the probability of ending up in a different community from node $i$ after $t+1$ steps, $k_i^{\text{in}} \approx f k_i$ is the number of edges connecting node $i$ to nodes within the same community, and $k_i^{\text{out}} \approx (1-f)k_i$ is the number of edges connecting node $i$ with nodes in different communities. We model the transitions in and out of node $i$'s community as a two-state Markov process with probability matrix
\begin{equation}
A = \left(\begin{array}{cc} P(\text{in } | \text{ in}) & P(\text{in } | \text{ out}) \\ P(\text{out } | \text{ in}) & P(\text{out } | \text{ out}) \end{array} \right) = \left(\begin{array}{cc} f & 1-f \\ \frac{N_c}{N-N_c}(1-f) & f + \frac{N - 2N_c}{N-N_c}(1-f) \end{array} \right).
\end{equation}
Using this representation, one can show that
\begin{equation}
\label{p_in_out}
\begin{aligned}
p^{\text{in}}(t+1) &= (A^{t+1})_{11} \approx \frac{1}{N}\left(N_c + (N-N_c)f^{t+1}\right), \\
\text{and} \quad p^{\text{out}}(t+1) &= (A^{t+1})_{12} \approx \frac{N-N_c}{N}(1-f^{t+1}),
\end{aligned}
\end{equation}
where the approximations follow from the assumption that $\left(\frac{fN-N_c}{N-N_c}\right)^{t+1} \approx f^{t+1}$. Plugging Eq. (\ref{p_in_out}) into Eq. (\ref{p_in}), we have
\begin{equation}
\begin{aligned}
\sum_j G_{ij} (P^{t+1})_{ij} &\approx \frac{fk_i}{N}\left(1 + f^{t+1}\left(\frac{N}{N_c} - 1\right)\right) + \frac{(1-f)k_i}{N}\left(1-f^{t+1}\right) \\
&= \frac{k_i}{N} \left(1 + f^{t+1}\left(\frac{N}{N_c}f - 1\right)\right) \\
&\approx \frac{k_i}{N} \left(1 + \frac{N}{N_c}f^{t+2} \right),
\end{aligned}
\end{equation}
where the final approximation follows from the assumption that $\frac{N}{N_c}f \gg 1$. We substitute this result into Eq. (\ref{DKL_ER1}), finding that
\begin{equation}
\begin{aligned}
D_{\text{KL}}(P||\hat{P}) &\approx -\log(1-\eta) - \log\left[\frac{1}{2E}\sum_i k_i \left(1 + k_i\sum_{t=1}^{\infty} \eta^t \left(\frac{1}{N}  + \frac{f^{t+2}}{N_c}\right)\right)\right] \\
&= -\log(1-\eta) - \log\left[1 + \frac{1}{2E}\sum_i k^2_i \sum_{t=1}^{\infty} \eta^t \left(\frac{1}{N}  + \frac{f^{t+2}}{N_c}\right)\right] \\
&= -\log(1-\eta) - \log\left[1 + \frac{1}{2E}\left(\frac{1}{N}\frac{\eta}{1-\eta} + \frac{1}{N_c}\frac{\eta f^3}{1-\eta f}\right)\sum_i k^2_i\right] \\
&= - \log\left[1- \eta + \frac{\eta}{2E}\left(\frac{1}{N} + \frac{1}{N_c}\frac{(1-\eta)f^3}{1-\eta f}\right)\sum_i k^2_i\right].
\end{aligned}
\end{equation}
For stochastic block models in the thermodynamic limit $N\rightarrow \infty$, the degree distribution is Poisson, and for large $\langle k\rangle $ we have $\langle k^2\rangle  \approx \langle k\rangle ^2$. Averaging over the Poisson degree distribution, the average KL divergence can be approximated by
\begin{equation}
\label{DKL_block1}
\begin{aligned}
\langle D_{\text{KL}}\rangle  &\approx -\left< \log\left[1- \eta + \frac{\eta}{2E}\left(\frac{1}{N} + \frac{1}{N_c}\frac{(1-\eta)f^3}{1-\eta f}\right)\sum_i k^2_i\right]\right>  \\
&\approx -\log\left[1- \eta + \frac{\eta N}{2E}\left(\frac{1}{N} + \frac{1}{N_c}\frac{(1-\eta)f^3}{1-\eta f}\right)\langle k^2\rangle \right] \\
&\approx -\log\left[1- \eta\left(1 -  \frac{\langle k\rangle }{N} - \frac{\langle k\rangle }{N_c}\frac{(1-\eta)f^3}{1-\eta f}\right)\right]. \\
\end{aligned}
\end{equation}

\begin{figure}[t!]
\centering
\includegraphics[width = .67\textwidth]{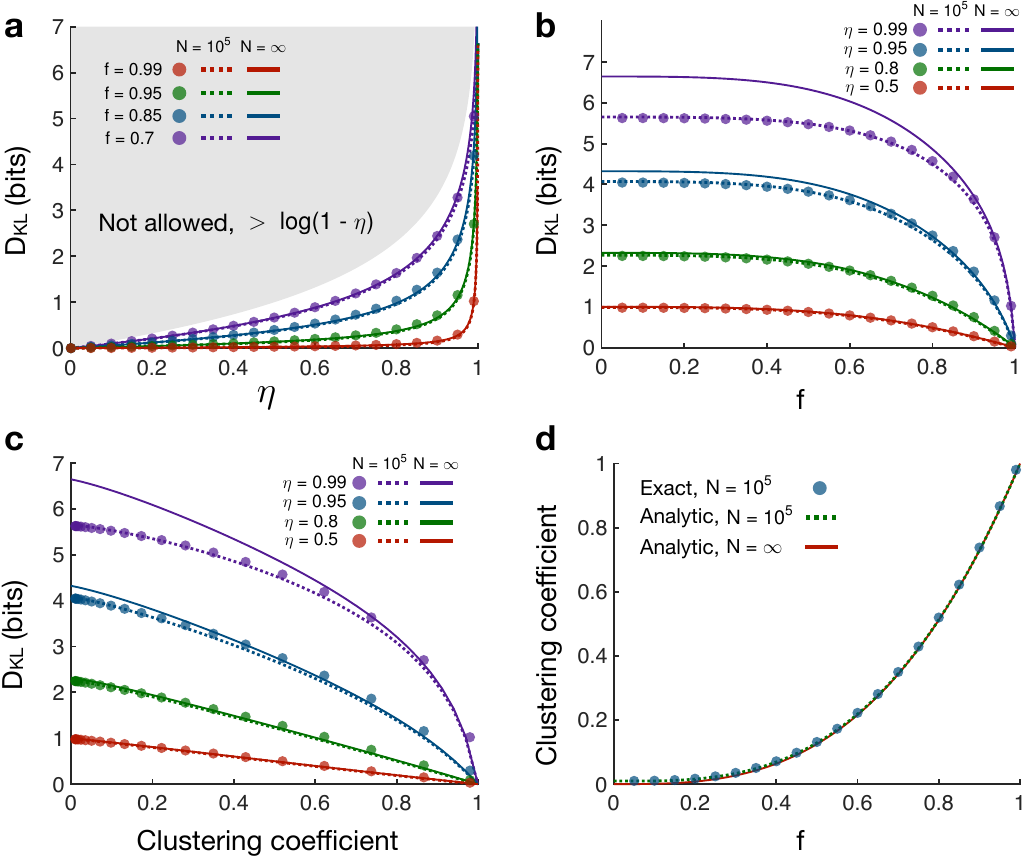} \\
\raggedright
\myfont\textbf{Fig. S\ref{DKL_block} $|$ KL divergence from human expectations in stochastic block networks.}
\captionsetup{labelformat=empty}
{\spacing{1.25} \caption{\label{DKL_block} \myfont \textbf{a}, KL divergence as a function of the integration parameter $\eta$ for stochastic block networks with average degree $\langle k\rangle  = 100$ and communities of size $N_c = 100$. Data points are exact calculations for networks of size $N = 10^4$. Dashed lines are analytic predictions using Eq. (\ref{DKL_block1}) with $N=10^4$. Solid lines are analytic results for the thermodynamic limit $N\rightarrow\infty$. \textbf{b}, KL divergence as a function of the fraction of within-community edges $f$ for different values of $\eta$. \textbf{c}, KL divergence as a function of the average clustering coefficient for variable $f$ and different values of $\eta$. \textbf{d}, Average clustering coefficient as a function of $f$. Dashed line is the analytic prediction in Eq. (\ref{cc}) with $N = 10^4$. Solid line is the analytic result in the limit $N\rightarrow\infty$.}}
\end{figure}

We remark that the first three terms inside the logarithm in Eq. (\ref{DKL_block1}) are identical to the Erd\"{o}s-R\'{e}nyi result in Eq. (\ref{DKL_ER3}), and thus the final term can be regarded as a correction resulting from the modular structure of the stochastic block model. Interestingly, this third term does not vanish in the thermodynamic limit $N\rightarrow \infty$; however, it does vanish in the limit $f\rightarrow 0$, as the network loses its block structure. We find that the analytic prediction in Eq. (\ref{DKL_block1}) is accurate across all values of $\eta$ and all fractions $f$ (Fig. S\ref{DKL_block}a,b). Furthermore, we find that the KL divergence decreases monotonically with increasing $f$ for fixed average degree $\langle k\rangle $ (Fig. S\ref{DKL_block}a,b).

In order to predict the effect of clustering, it is helpful to have an analytic approximation for the average clustering coefficient in a stochastic block network. We recall that the clustering coefficient for a node $i$ is given by $2\bigtriangleup_i/(k_i(k_i-1))$, where $\bigtriangleup_i$ is the number of triangles involving node $i$. For a stochastic block network, we define the probability of an edge existing between two nodes in the same community as $p^{\text{in}} = f\langle k\rangle /N_c$ and the probability of an edge between two nodes in different communities as $p^{\text{out}} = (1-f)\langle k\rangle /(N-N_c)$. We then arrive at the following approximation,
\begin{equation}
\begin{aligned}
\langle \bigtriangleup_i\rangle  &= \frac{k_i^{\text{in}}(k_i^{\text{in}} - 1)}{2} p^{\text{in}} + k_i^{\text{in}}k_i^{\text{out}}p^{\text{out}} + \frac{k_i^{\text{out}}(k_i^{\text{out}} - 1)}{2}\left[\frac{N_c-1}{N-N_c-1}p^{\text{in}} + \left(\frac{N-2N_c}{N-N_c-1}\right)p^{\text{out}}\right] \\
&\approx \frac{p^{\text{in}}}{2} (k_i^{\text{in}})^2 + p^{\text{out}} k_i^{\text{out}} \left( k_i^{\text{in}} + \frac{k_i^{\text{out}}}{2}\right),
\end{aligned}
\end{equation}
where the approximation follows from the assumptions that $N\gg N_c$ and $k_i^{\text{in}},k_i^{\text{out}} \gg 1$. Plugging in for $p^{\text{in}}$, $p^{\text{out}}$, $k_i^{\text{in}} = fk_i$, and $k_i^{\text{out}} = (1-f)k_i$, we have
\begin{equation}
\langle \bigtriangleup_i\rangle  \approx \frac{\langle k\rangle  k_i^2}{2}\left(\frac{f^3}{N_c} + \frac{(1+f)(1-f)^2}{N-N_c}\right).
\end{equation}
Thus the average clustering coefficient is given by
\begin{equation}
\label{cc}
\frac{1}{N}\sum_i\frac{2\langle \bigtriangleup_i\rangle }{k_i(k_i - 1)} \approx \langle k\rangle \left(\frac{f^3}{N_c} + \frac{(1+f)(1-f)^2}{N-N_c}\right),
\end{equation}
where the approximation follows from the assumption that $k_i \gg 1$. We see that this analytic result accurately predicts the increase in the average clustering coefficient with increasing modularity $f$ (Fig. S\ref{DKL_block}d). More importantly, we find that the KL divergence decreases with increasing clustering for fixed $\eta$ and $\langle k\rangle $ (Fig. S\ref{DKL_block}c). This final result indicates that increased modularity helps human observers maintain accurate representations, thereby reducing their inefficiency when processing information.

\section{Hierarchically modular networks}
\label{hierarchically_modular}

The combination of high entropy and low KL divergence exhibited by real networks is driven by heterogeneous degrees and modular structure. Interestingly, degree heterogeneity and modularity are ubiquitous in natural and human-made systems,\cite{Cancho-01, Barabasi-01, Barabasi-02, Girvan-01, Rosvall-02, Motter-01, Eriksen-01} and together they define hierarchically modular organization.\cite{Ravasz-01} In order to simultaneously study entropy and KL divergence, it is helpful to have a model for generating networks with variable heterogeneity and modularity. One of the earliest models of hierarchical systems was developed to understand metabolic networks.\cite{Ravasz-02, Ravasz-01} Yet this model is deterministic, generating fractal networks in which it is difficult to tune the heterogeneity or modularity. Another common model is the nested stochastic block model,\cite{Arenas-01, Arenas-02} wherein small modules are nested inside larger modules. However, this model does not include heterogeneous degrees (a heavy-tailed degree distribution). Perhaps the closest model to what we require was recently developed to study the emergence of complex dynamics in the brain.\cite{Zamora-01} In this model, the nested stochastic block model is combined with a preferential attachment rule to generate a rich club of hub nodes.

Here we propose a model that directly combines the static model\cite{Goh-01,Catanzaro-01,Lee-01} and the stochastic block model.\cite{Decelle-01} Beginning with $N$ disconnected nodes, we first assign each node $i$ a weight $w_i = i^{-\alpha}$, where $\alpha\in[0,1]$ is related to the scale-free exponent by $\gamma = 1 + \frac{1}{\alpha}$. We also assign each node $i$ to a community. Then, we randomly select pairs of nodes $i$ and $j$ within the same community with probabilities proportional to their weights, and we connect them if they have not already been connected. This process is repeated until $fE = \frac{1}{2}f\langle k\rangle N$ edges have been added within communities. We then repeat this process again until $(1-f)E = \frac{1}{2}(1-f)\langle k\rangle  N$ edges have been added between communities. The resulting network has a degree distribution that drops off as a power law $\mathcal{P}(k) \sim k^{-\gamma}$ and also has the same community structure as a stochastic block model.

\begin{figure}[t!]
\centering
\includegraphics[width = .7\textwidth]{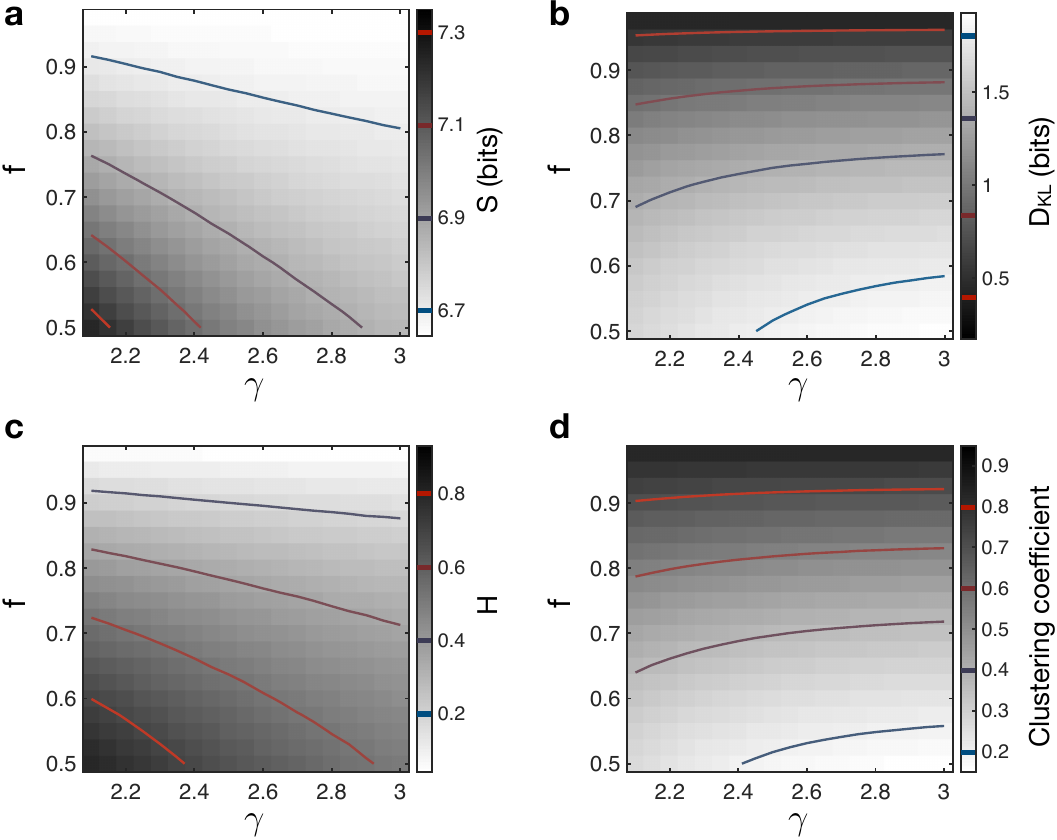} \\
\raggedright
\myfont\textbf{Fig. S\ref{HM_sweep} $|$ Information and structural properties of hierarchically modular networks.}
\captionsetup{labelformat=empty}
{\spacing{1.25} \caption{\label{HM_sweep} \myfont \textbf{a}, Entropy as a function of the scale-free exponent $\gamma$ and the fraction of within-community edges $f$ for hierarchically modular networks with average degree $\langle k\rangle  = 100$ and communities of size $N_c = 100$. Each point is an exact calculation for a network of size $N = 10^4$. \textbf{b}, KL divergence as a function of $\gamma$ and $f$ in the same networks with $\eta$ fixed to the average value $0.80$ from our experiments. \textbf{c}, Degree heterogeneity $H$ varies as a function of $\gamma$ and $f$ in a similar fashion to the entropy (\textbf{a}). \textbf{d}, Average clustering coefficient varies as a function of $\gamma$ and $f$ much like the KL divergence (\textbf{b}).}}
\end{figure}

Sweeping over the two parameters $\gamma$ and $f$, while fixing the average degree $\langle k\rangle  = 100$ and community size $N_c = 100$, we see that our hierarchically modular model exhibits a variety of entropies (Fig. S\ref{HM_sweep}a) and KL divergences (Fig. S\ref{HM_sweep}b). Additionally, we verify that the model can attain a wide range of degree heterogeneities (Fig. S\ref{HM_sweep}c) and clustering coefficients (Fig. S\ref{HM_sweep}d). Notably, the variation in the degree heterogeneity and clustering coefficient with $\gamma$ and $f$ appears almost identical to the variation in the entropy and KL divergence, respectively, once again indicating that entropy is primarily driven by heterogeneity and KL divergence is primarily driven by clustering or modularity.

\begin{figure}[t!]
\centering
\includegraphics[width = .5\textwidth]{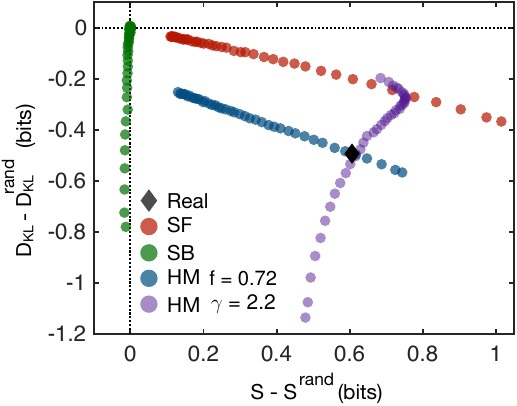} \\
\raggedright
\myfont\textbf{Fig. S\ref{HM_comp} $|$ Comparing the information properties of real and model networks.}
\captionsetup{labelformat=empty}
{\spacing{1.25} \caption{\label{HM_comp} \myfont Entropies and KL divergences of real and model networks compared to fully randomized versions. For each model network in Table 1, we generate SF networks with variable $\gamma$ (red), SB networks with communities of size $N_c \approx \sqrt{N}$ and variable $f$ (green), and HM networks with $N_c \approx \sqrt{N}$ and variable $\gamma$ (fixed $f = 0.72$; blue) or variable $f$ (fixed $\gamma = 2.2$; purple), all with the same number of nodes $N$ and edges $E$ as the real network. Each real and model network is then compared with 100 randomized versions; data points are first averaged over the 100 randomized networks and then averaged over the set of real networks in Table 1. HM networks with $\gamma = 2.2$ and $f = 0.72$ match the average entropy and KL divergence of real networks.}}
\end{figure}

Given our investigation of the information properties of different network models, it is ultimately important to compare against real communication networks. For each network listed in Table S12, we generate series of scale-free networks with various exponents $\gamma$, stochastic block networks with various within-community fractions $f$, and hierarchically modular networks with various exponents $\gamma$ (for fixed $f$) and various $f$ (for fixed $\gamma$). Each model network maintains the same number of nodes $N$ and edges $E$ as the corresponding real network. For the stochastic block and hierarchically modular networks, we choose community sizes that are roughly the square root of the network size $N_c \approx \sqrt{N}$ for the purpose of remaining consistent with our model-based analysis (wherein $N=10^4$ and $N_c = \sqrt{10^4} = 100$). Comparing each real and model network with completely randomized versions of the same networks (Fig. S\ref{HM_comp}), we find that: (i) scale-free networks cannot attain the low KL divergence displayed by real networks and (ii) stochastic block networks cannot attain the high entropy displayed by real networks, but (iii) hierarchically modular networks can achieve both with a parameter combination of $\gamma \approx 2.2$ and $f \approx 0.72$. Thus, we confirm that both heterogeneous degrees and modular structure are required (that is, hierarchical organization is required) to match the information properties of real networks.

\section{Network datasets}
\label{network_datasets}

The real-world networks analyzed in the main text are listed and briefly described in Table S12. While the semantic, web, citation, and social networks are gathered from online network repositories, the language and music networks are novel. For the language networks, we developed code to (i) remove punctuation and white space, (ii) filter words by their part of speech, and (iii) record the transitions between the filtered words. Here we focus on networks of transitions between nouns, noting that the same methods could be used to record transitions between other parts of speech. The raw text was gathered from Project Gutenberg (\texttt{gutenberg.org/wiki/Main\_Page}).

For the music networks, we read in audio files in MIDI format using the \texttt{readmidi} function in MATLAB (R2018a). For each song, we split the notes by their channel, which represents the different instruments. For each channel, we created a network of note transitions. We then create a transition network representing the entire song by aggregating the transitions between notes across the different channels. The MIDI files were gathered from \texttt{midiworld.com} and from \texttt{kunstderfuge.com}. Our code and data are available upon request from the corresponding author.

\addtocounter{figure}{-1}
\begin{figure}
\myfont
\centering
{\fontsize{9.5}{9}\selectfont
\begin{tabular}{l l l l l}
\hline
\textbf{Type} & Name & $N$ & $E$ & Description \\
\hline
\hline
\textcolor{MyBlue}{\textbf{Language}} & Shakespeare\cite{Shakespeare-01}$^{*+}$ & 11,234 & 97,892 & Noun transitions in Shakespeare's work. \\
& Homer\cite{Homer-01}$^{*+}$ & 3,556 & 23,608 & Same as above (Homer's Iliad). \\
& Plato\cite{Plato-01}$^{*+}$ & 2,271 & 9,796 & Same as above (Plato's Republic). \\
& Jane Austen\cite{Austen-01}$^{*+}$ & 1,994 & 12,120 & Same as above (Pride and Prejudice). \\
& William Blake\cite{Blake-01}$^{*+}$ & 370 & 781 & Same as above (Songs of Innocence...). \\
& Miguel de Cervantes\cite{Cervantes-01}$^{*+}$ & 6,090 & 43,682 & Same as above (Don Quixote). \\
& Walt Whitman\cite{Whitman-01}$^{*+}$ & 4,791 & 16,526 & Same as above (Leaves of Grass). \\
\hline
\textcolor{MyTurquoise}{\textbf{Semantic}} & Bible\cite{Kunegis-01} & 1,707 & 9,059 & Pronoun co-occurrences in Bible verses. \\
& Les Miserables\cite{Kunegis-01} & 77 & 254 & Character co-occurrences. \\
& Edinburgh Thesaurus\cite{Kiss-01,Batagelj-01}$^{*}$ & 7,754 & 226,518 & Word similarities in human experiments. \\
& Roget Thesaurus\cite{Roget-01, Batagelj-01}$^{*}$ & 904 & 3,447 & Linked semantic categories. \\
& Glossary terms\cite{Batagelj-01} & 60 & 114 & Words used in definitions of other words. \\
& FOLDOC\cite{Howe-01, Batagelj-01}$^{*}$ & 13,274 & 90,736 & Same as above (computing terms). \\
& ODLIS\cite{Reitz-01,Batagelj-01}$^{*}$ & 1,802 & 12,378 & Same as above (information science terms). \\
\hline
\textcolor{MyGreen}{\textbf{Web}} & Google internal\cite{Palla-01, Kunegis-01}$^{*}$ & 12,354 & 142,296 & Hyperlinks between Google's own cites. \\
& Education\cite{Gleich-01,Rossi-01} & 2,622 & 6,065 & Hyperlinks between education webpages. \\
& EPA\cite{DeNooy-01,Rossi-01} & 2,232 & 6,876 & Pages linking to www.epa.gov. \\
& Indochina\cite{Boldi-01,Rossi-01} & 9,638 & 45,886 & Hyperlinks between pages in Indochina. \\
& 2004 Election blogs\cite{Adamic-01,Kunegis-01}$^{*}$ & 793 & 13,484 & Hyperlinks between blogs on US politics. \\
& Spam\cite{Castillo-01, Rossi-01} & 3,796 & 36,404 & Hyperlinks between spam pages. \\
& WebBase\cite{Boldi-01,Rossi-01} & 6,843 & 16,374 & Hyperlinks gathered by web crawler. \\
\hline
\textcolor{MyOrange}{\textbf{Citations}} & arXiv Hep-Ph\cite{Leskovec-01, Kunegis-01}$^{*+}$ & 12,711 & 139,500 & Citations in Hep-Ph section of the arXiv. \\
& arXiv Hep-Th\cite{Leskovec-01, Kunegis-01}$^{*+}$ & 7,464 & 115,932 & Citations in Hep-Th section of the arXiv. \\
& Cora\cite{Subelj-01, Kunegis-01}$^{*}$ & 3,991 & 16,621 & Citation network between scientific papers.  \\
& DBLP\cite{Ley-01,Kunegis-01}$^{*}$ & 240 & 858 & Citation network between scientific papers. \\
\hline
\textcolor{MyRed}{\textbf{Social}} & Facebook\cite{Viswanath-01, Kunegis-01}$^{+}$ & 13,130 & 75,562 & Subset of the Facebook network. \\
& arXiv Astr-Ph\cite{Leskovec-01, Kunegis-01} & 17,903 & 196,972 & Coauthorships in Astr-Ph section of arXiv. \\
& Adolescent health\cite{Moody-01, Kunegis-01}$^{*}$ & 2,155 & 8,970 & Friendships between students. \\
& Highschool\cite{Coleman-01, Kunegis-01}$^{*}$ & 67 & 267 & Friendships between highschool students. \\
& Jazz\cite{Gleiser-01, Kunegis-01} & 198 & 2,742 & Collaborations between jazz musicians. \\
& Karate club\cite{Zachary-01, Kunegis-01} & 34 & 78 & Interactions between karate club members. \\
\hline
\textcolor{MyPurple}{\textbf{Music}} & Thriller -- Michael Jackson\cite{Jackson-01}$^{*+}$ & 67 & 446 & Network of note transitions. \\
& Hard Day's Night -- Beatles\cite{Beatles-01}$^{*+}$ & 41 & 212 & Same as above. \\
& Bohemian Rhapsody -- Queen\cite{Queen-01}$^{*+}$ & 71 & 961 & Same as above. \\
& Africa -- Toto\cite{Toto-01}$^{*+}$ & 39 & 163 & Same as above. \\
& Sonata No 11 -- Mozart\cite{Mozart-01}$^{*+}$ & 55 & 354 & Same as above. \\
& Sonata No 23 -- Beethoven\cite{Beethoven-01}$^{*+}$ & 69 & 900 & Same as above. \\
& Nocturne Op 9-2 -- Chopin\cite{Chopin-01}$^{*+}$ & 59 & 303 & Same as above. \\
& Clavier Fugue 13 -- Bach\cite{Bach-01}$^{*+}$ & 40 & 143 & Same as above. \\
& Ballade Op 10-1 -- Brahms\cite{Brahms-01}$^{*+}$ & 69 & 670 & Same as above. \\
\hline
\end{tabular}} \\[-.75em]
\raggedright
\captionsetup{labelformat=empty}
{\spacing{1.25} \caption{\myfont \textbf{Table S12 $|$ Real networks analyzed in the main text.} For each network we list its type; name, reference, whether it has a directed version (denoted by $*$), and whether it has a temporally evolving version (denoted by $+$); number of nodes $N$; number of edges $E$; and a brief description.}}
\end{figure}

\newpage

\section*{References}

\bibliographystyle{naturemag}
\bibliography{GraphLearningBib}

\end{document}